\documentclass[twocolumn]{aastex7}
\usepackage[nomain,acronyms,nohypertypes={acronym}]{glossaries}

\makeatletter
\DeclareFontFamily{OMX}{MnSymbolE}{}
\DeclareSymbolFont{MnLargeSymbols}{OMX}{MnSymbolE}{m}{n}
\SetSymbolFont{MnLargeSymbols}{bold}{OMX}{MnSymbolE}{b}{n}
\DeclareFontShape{OMX}{MnSymbolE}{m}{n}{
    <-6>  MnSymbolE5
   <6-7>  MnSymbolE6
   <7-8>  MnSymbolE7
   <8-9>  MnSymbolE8
   <9-10> MnSymbolE9
  <10-12> MnSymbolE10
  <12->   MnSymbolE12
}{}
\DeclareFontShape{OMX}{MnSymbolE}{b}{n}{
    <-6>  MnSymbolE-Bold5
   <6-7>  MnSymbolE-Bold6
   <7-8>  MnSymbolE-Bold7
   <8-9>  MnSymbolE-Bold8
   <9-10> MnSymbolE-Bold9
  <10-12> MnSymbolE-Bold10
  <12->   MnSymbolE-Bold12
}{}

\let\llangle\@undefined
\let\rrangle\@undefined
\DeclareMathDelimiter{\llangle}{\mathopen}%
                     {MnLargeSymbols}{'164}{MnLargeSymbols}{'164}
\DeclareMathDelimiter{\rrangle}{\mathclose}%
                     {MnLargeSymbols}{'171}{MnLargeSymbols}{'171}
\makeatother

\usepackage{mathtools}
\usepackage{dsfont}
\usepackage[nolist]{acronym} 
\usepackage[nameinlink,capitalise,noabbrev]{cleveref}

\makeatletter
\usepackage{etoolbox}
\patchcmd\H@refstepcounter{\protected@edef}{\protected@xdef}{}{}
\makeatother

\newcommand{\REV}[1]{{{#1}}}

\newacronym{BE}{BE}{Bonnor-Ebert}
\newacronym{TES}{TES}{turbulent equilibrium sphere}
\newacronym{ISM}{ISM}{interstellar medium}
\newacronym{MHD}{MHD}{magnetohydrodynamic}
\newacronym{GMC}{GMC}{giant molecular cloud}
\newacronym{VL2}{VL2}{second-order van Leer}
\newacronym{LP}{LP}{Larson-Penston}
\newacronym{CMF}{CMF}{core mass function}
\newacronym{SMF}{SMF}{sink particle mass function}
\newacronym{IMF}{IMF}{initial mass function}
\newacronym{PDF}{PDF}{probability distribution function}
\newacronym{YSO}{YSO}{young stellar object}
\newacronym{RMS}{RMS}{root mean square}
\newacronym{AMR}{AMR}{adaptive mesh refinement}
\newacronym{SFE}{SFE}{star formation efficiency}

\defcitealias{moon24}{Paper~I}
\defcitealias{paperII}{Paper~II}

\shorttitle{Turbulent Core Collapse}
\shortauthors{Moon \& Ostriker}

\begin{document}

\title{Prestellar Cores in Turbulent Clouds: Numerical Modeling and Evolution to Collapse}

\author[0000-0002-6302-0485]{Sanghyuk Moon}
\affiliation{Department of Astrophysical Sciences, Princeton University,
  Princeton, NJ 08544, USA}
\email[show]{sanghyuk.moon@princeton.edu}
\author[0000-0002-0509-9113]{Eve C.\ Ostriker}
\affiliation{Department of Astrophysical Sciences, Princeton University,
  Princeton, NJ 08544, USA}
\email[show]{eco@astro.princeton.edu}

\begin{abstract}
A fundamental issue in star formation is understanding the precise mechanisms leading to the formation of prestellar cores, and their subsequent gravitationally unstable evolution.
To address this question, we carefully construct a suite of turbulent, self-gravitating numerical simulations, and analyze the development and collapse of individual prestellar cores.
We show that the numerical requirements for resolving the sonic scale and internal structure of anticipated cores are essentially the same in self-gravitating clouds, calling for the number of cells per dimension to increase quadratically with the cloud's Mach number.
In our simulations, we follow evolution of individual cores by tracking the region around each gravitational potential minimum over time.
Evolution in nascent cores is towards increasing density and decreasing turbulence, and there is a wide range of critical density for initiating collapse.
At given spatial scale the turbulence level also varies widely, and tends to be correlated with density.
By directly measuring the radial forces acting within cores, we identify a distinct transition to a state of gravitational runaway.
We use our new theory for turbulent equilibrium spheres to predict the onset of each core's collapse.
Instability is expected when the critical radius becomes smaller than the tidal radius; we find good agreement with the simulations.
Interestingly, the imbalance between gravity and opposing forces is only $\sim 20\%$ during core collapse, meaning that this is a quasi-equilibrium rather than a free-fall process.
For most of their evolution, cores exhibit both subsonic contraction and transonic turbulence inherited from core-building flows; supersonic radial velocities accelerated by gravity only appear near the end of the collapse.
\end{abstract}

\section{Introduction}\label{sec:intro}

Stars form in the coldest and densest regions of \glspl{GMC}, systems that are pervaded by supersonic turbulence \citep[e.g.][]{andre14,heyer15}.
The turbulent velocity field in GMCs creates structures at a range of scales (from negative divergences), while also dispersing structures (from positive divergences and shear) and contributing to support against gravity \citep[e.g.][]{mckee07,hennebelle12}.

The perturbations in density are organized by the velocity field into spatially correlated, hierarchical structures, which are sometimes characterized by the density power spectrum \citep{jskim05,kritsuk07,2016MNRAS.460.4483K}.
If the density power spectrum approximately follows a power law, the structures in configuration space are expected to display some degree of self-similarity, often motivating fractal descriptions of density structures \citep{stutzki98,thiesset23}.
When the gas density is so high that the self-gravity starts to affect the dynamics, however, self-similarity is no longer expected.

Observations of nearby molecular clouds indicate that in the dense regions where self-gravity is clearly important and stars form, the gas is organized into roughly spherical, compact ($\lesssim 0.1\,\mathrm{pc}$), centrally-concentrated objects called dense (starless) cores \citep[see][for dedicated reviews on dense cores]{bergin07,difrancesco07}.
Cores' radial density profiles are characterized by a flat central plateau and outer envelope approximately following $\rho\propto r^{-2}$, suggestive of gravitational stratification where self-gravity is roughly balanced by pressure gradients.
The widths of observed molecular emission lines are somewhat (but not much) broader than the thermal value, indicating that motions within dense cores exhibit subsonic or transonic turbulence \citep[e.g.,][]{goodman98,choudhury21}.
The turbulence within cores presumably is a legacy of their turbulent formation environments.
Observed statistics suggest that dense cores are likely transient objects that live no more than a few free-fall times (see below), and their mass distributions have a similar shape to the \gls{IMF} but shifted toward higher masses by a factor $2\text{--}3$ \citep[see][and references therein]{andre14}.

It is generally understood that dense cores form in regions where turbulent flows happen to locally converge, and under certain conditions undergo runaway gravitational collapse to form a single star or a multiple system \citep{mckee07,andre14,offner14,padoan14}.
To investigate this process in a highly idealized setup, \citet{gong09} performed spherically symmetric simulations for supersonic converging flows at a range of Mach number, to study how cores form and evolve in dense post-shock regions.
As mass is added, cores in those simulations initially evolve in a quasi-equilibrium fashion with increasing density stratification, until the onset of
outside-in collapse dramatically (and rapidly) increases the central density.
In less idealized three-dimensional simulations focused on post-shock regions in converging, turbulent flows, a similar transition from core building to rapid internal collapse was identified \citep{gong15}. 
Recently, \citet[see also \citealt{collins23}]{collins24} utilized tracer particles to analyze evolution of individual cores self-consistently formed in three-dimensional simulations of self-gravitating isothermal turbulence.
They found that cores are formed by converging flows that sweep up low-density gas and then undergo gravitational collapse once the density becomes moderately high and internal turbulent motions decay to transonic or slightly supersonic level.

Although the details differ, a common theme in characterizing evolution of simulated cores is that there exists a critical stage at which a core transitions to a state of runaway gravitational collapse (e.g., from the \emph{core building} to the \emph{core collapse} stage in \citet{gong09}; from the \emph{hardening} to the \emph{singularity} stage as described in \citet{collins24}; the first step of the inertial-inflow scenario of \citet{padoan20}).
A flip side of this is the possible existence of failed cores that do not satisfy the physical conditions associated with the critical stage and disperse back into the \gls{ISM} \citep[e.g.,][]{vazquez-semadeni05,smullen20,offner22}.

While the transition from core building to collapse stages has been identified in simulations, it has not been explained physically exactly what \emph{triggers} the onset of collapse.
Traditionally, the \gls{BE} sphere \citep{bonnor56,ebert55,ebert57} has been regarded as the most relevant theoretical model for distinguishing critical conditions: spherical isothermal equilibria exist for all radii, but beyond a certain radius the equilibrium is unstable and collapse would be expected.
However, the \gls{BE} solution assumes a completely isolated spherical equilibrium supported entirely by thermal pressure, while in contrast real cores (1) are affected by internal velocity structure as they form from the supersonically turbulent \gls{ISM}, and (2) do not exist in isolation but are surrounded and gravitationally affected by neighboring structures.

To address the first of the above limitations, in \citet{moon24} (hereafter \citetalias{moon24}) we developed a new theoretical model of quasi-equilibrium isothermal spheres supported by both thermal and turbulent pressure, with solutions obtained by directly solving the time-steady, angle-averaged equations of hydrodynamics.
A salient feature of this model, termed the \gls{TES}, is that the turbulent pressure naturally arises from a power-law velocity structure function (see \cref{sec:tes}) rather than from a phenomenological equation of state.
The \gls{BE} solutions are recovered in the limit of vanishing turbulent velocity dispersion.
\citetalias{moon24} found that, for a radially stratified \gls{TES} solution, there exists a critical radius $r_\mathrm{crit}$ at which the equilibrium becomes unstable to radial perturbations\footnote{Alternatively, for a given confining pressure at the edge, there exists a maximum mass above which no equilibrium solution exists.}, and that $r_\mathrm{crit}$ increases with the strength of the turbulence for a given density.

One of the main conclusions of \citetalias{moon24} is that a quasi-equilibrium core will collapse when its maximum radius exceeds $r_\mathrm{crit}$.
However, it is not obvious what determines the ``maximum radius'' of a real core in a real cloud, or if an outer radius exists at all.
In an approximately isothermal medium such as a GMC, a core does not possess a well-defined boundary, but instead continuously blends into the surrounding gas.
This might seem to imply that no core can remain stable because every core extends indefinitely beyond $r_\mathrm{crit}$.
In reality, however, a core is generally surrounded by neighboring structures, such that there is an effective maximum radius beyond which the core cannot be considered an isolated object from the point of view of the gravitational potential.

In order to identify the critical conditions for collapse that fully take into account the hierarchical structure in which cores live, and include the effects of the velocity field, careful analyses of numerical simulations are needed.
In this paper, we present results from investigations of dynamical evolution of individual cores forming in a suite of three-dimensional numerical simulations of self-gravitating isothermal turbulence.
Thus, the present work takes on the second of the limitations of the traditional \gls{BE} stability analysis mentioned above, armed with the \gls{TES} solutions of \citetalias{moon24}.
In a companion paper (\citealt{paperII}; hereafter \citetalias{paperII}), we will present detailed properties of the critical cores (defined at the onset of collapse) and compare their structure with the \gls{TES} model.

Another important observational and theoretical issue concerns the dynamical status of prestellar cores as they evolve to reach a high degree of central concentration.
At one extreme, prestellar cores are viewed as quasi-static objects slowly evolving under magnetic support to reach the singular isothermal profile, which then undergoes near-pressureless collapse with a rarefaction wave propagating outward from the innermost region \citep{shu77}.
The other extreme is to treat the pressureless free-fall stage as beginning from a state with a flat density profile near the center, rather than a power law \citep{whitworth01,myers05}.
The pressure-modified ``outside-in'' dynamical collapse models \citep{larson69,penston69,hunter77,foster93} lie somewhere in between these two extremes, often viewed as being qualitatively more similar to the latter.
The measured infall speeds of starless cores based on the ``blue asymmetry'' in the molecular line profiles are too fast and spatially extended to be explained by quasi-static contraction driven by ambipolar diffusion \citep{cwlee99,cwlee01}.
Observed inflows are sometimes considered too slow for dynamical collapse \citep{campbell16}, although in fact supersonic speeds only appear toward the end of the collapse \citep[e.g.,][]{foster93,myers05}.
Alternatively, observed infall motions may simply reflect the initial momentum of converging, core-building flows (e.g., \citealt{cwlee01}; \citealt{gong09}; \citealt{padoan20}; \citealt{collins24}) rather than representing gravitational collapse \citep[see also][for discussion of filament-forming flows]{chen20}.
Direct number counting of prestellar cores relative to \glspl{YSO} indicates that the core collapse typically takes $2\text{--}5$ times the free-fall time \citep{ward-thompson07,konyves15}, consistent with neither free-fall nor quasi-static contraction.
Directly measuring the forces acting on simulated cores as a function of time is helpful to quantitatively characterize the dynamics of the collapse process.
Measurements of this kind are part of the analysis we present in this paper, leading to the new physical concept of ``quasi-equilibrium collapse.''

The remainder of this paper is organized as follows.
In \cref{sec:tes}, we briefly summarize selected content from \citetalias{moon24} that will be referenced throughout this work.
In \cref{sec:simulations}, we outline the hydrodynamic equations we solve and describe numerical methods and resolution requirements.
We also describe our algorithm for tracking cores through successive snapshots in our simulations.
\cref{sec:results} presents our main results; we describe the overall time evolution of our models, analyze dynamical evolution of individual cores, and identify the critical conditions for collapse.
We discuss implications of our results in \cref{sec:discussion} and conclude in \cref{sec:conclusion}.

\section{Review of the Turbulent Equilibrium Sphere Model}\label{sec:tes}

In \citetalias{moon24}, we developed a semi-analytic model of isothermal spheres supported by thermal and turbulent pressure, in which the latter naturally arises from the radius-dependent velocity dispersion rather than from a phenomenological equation of state.
We refer to members of this family of solutions as a Turbulent Equilibrium Sphere (TES).
In this section, we briefly summarize relevant features of the \gls{TES} model that will be used throughout this work.
We refer the reader to \citetalias{moon24} for a comprehensive presentation.

By averaging the continuity and momentum equation over the full solid angle, \citetalias{moon24} derived the equation of motion governing the dynamics of fluid parcels distributed over a spherical shell at radius $r$:
\begin{equation}\label{eq:lagrangian_eom}
  \frac{\partial \left<v_r \right>_\rho}{\partial t} + \left<v_r \right>_\rho \frac{\partial \left<v_r \right>_\rho}{\partial r} = f_\mathrm{net}.
\end{equation}
Here, $v_r$ is the radial velocity measured with respect to the origin of the local spherical coordinate system at the potential minimum, $f_\mathrm{net}$ is the net force per unit mass, and the subscripted angled bracket denotes the mass-weighted angle-averaging operator defined by
\begin{equation}\label{eq:mass-weighted-angle-average}
  \left<Q \right>_\rho \equiv \frac{\oint_{4\pi} \rho Q\,d\Omega}{\oint_{4\pi} \rho\,d\Omega},
\end{equation}
where $\rho$ is gas density and $Q$ is a physical quantity to be averaged.
It is related to the volume-weighted average
\begin{equation}\label{eq:volume-weighted-angle-average}
  \left<Q \right> \equiv \frac{1}{4\pi} \oint_{4\pi} Q\, d\Omega
\end{equation}
by $\left<Q \right>_\rho \equiv \left<\rho Q \right>/ \left<\rho \right>$.
\REV{We note that one can derive an integral version of \cref{eq:lagrangian_eom} (\autoref{app:angle-averaged-equations}), which is useful for analyzing internal radial dynamics of cores.}

In \cref{eq:lagrangian_eom}, the net specific force 
\begin{equation}\label{eq:fnet}
f_\mathrm{net} = f_\mathrm{thm} + f_\mathrm{trb} + f_\mathrm{grv} + f_\mathrm{cen} + f_\mathrm{ani}
\end{equation}
is comprised of the sum of the thermal pressure gradient force $f_\mathrm{thm}$, turbulent pressure gradient force $f_\mathrm{trb}$, gravitational force $f_\mathrm{grv}$, centrifugal force $f_\mathrm{cen}$, and the residual force due to the anisotropic turbulence $f_\mathrm{ani}$, which are given by
\begin{align}
  f_\mathrm{thm} &= -\frac{1}{\left<\rho \right>} \frac{\partial \left( \left<\rho \right>c_s^2 \right)}{\partial r},\label{eq:def_fthm} \\
  f_\mathrm{trb} &= -\frac{1}{\left<\rho \right>} \frac{\partial \left( \left<\rho \right> \left<\delta v_r^2 \right>_\rho \right)}{\partial r},\label{eq:def_ftrb} \\
  f_\mathrm{grv} &= \left<g_r \right>_\rho,\label{eq:def_fgrv}\\
  f_\mathrm{cen} &= \frac{\left<v_\theta \right>_\rho^2 + \left<v_\phi \right>_\rho^2}{r},\label{eq:def_fcen}\\
  f_\mathrm{ani} &= \frac{\left<\delta v_\theta^2 \right>_\rho + \left<\delta v_\phi^2 \right>_\rho - 2 \left<\delta v_r^2 \right>_\rho}{r}\label{eq:def_fani}.
\end{align}
Here, $c_s$ is the isothermal sound speed, $g_r$ is the radial component of the gravitational acceleration, and $v_\theta$ and $v_\phi$ are the meridional and azimuthal components of velocity.
\REV{The turbulent velocity components are defined by
\begin{equation}\label{eq:def_delta_v}
    \delta v_i \equiv v_i - \left< v_i \right>_\rho,
\end{equation}
where $i = r$, $\theta$, and $\phi$.}
We note that in \citetalias{moon24} and the present work, we neglect the contribution of magnetic stresses.  These may in fact be significant at earlier stages of evolution, but by the time cores begin to collapse magnetic terms are sub-dominant, with cores that are thermally supercritical also magnetically supercritical \citep[e.g.][]{chen15}.

Equilibrium solutions are obtained by assuming the left hand side of \cref{eq:lagrangian_eom} is zero, requiring $f_\mathrm{net}=0$.
\citetalias{moon24} focused on the particular equilibrium solutions for the case when the rotation is negligible ($f_\mathrm{cen}=0$) and turbulence is isotropic ($f_\mathrm{ani} = 0$), such that $f_\mathrm{thm} + f_\mathrm{trb} = -f_\mathrm{grv} $.
Coupled with the Poisson equation, the equilibrium equation becomes
\begin{equation}\label{eq:dimensional_steady_equilibrium}
  \frac{1}{r^2} \frac{\partial}{\partial r} \left[ \frac{r^2}{\left<\rho \right>} \frac{\partial}{\partial r} \left( \left<\rho \right> c_s^2 + \left<\rho \right> \left<\delta v_r^2 \right>_\rho \right) \right] = -4\pi G \left<\rho \right>
\end{equation}

In observations of GMCs and their substructures, the turbulent velocity field is spatially correlated such that the observed linewidth increases as a power law with the size scale of structures \citep{larson81,solomon87,goodman98,jijina99,heyer09}.
Additionally, high-resolution numerical simulations of supersonic turbulence have found that within a certain range of scales, the 
\gls{RMS} velocity difference $\Delta v(l)$ between a pair of points separated by a distance $l$ closely follows a power-law
\begin{equation}\label{eq:structure_function}
  \Delta v(l) \propto l^p
\end{equation}
with $p \approx 0.5$ \citep{kritsuk07,federrath21}, which is consistent with observed velocity structures within Galactic \glspl{GMC} \citep{heyer04}.

Motivated by the above observational and theoretical results regarding the power-law scaling of the velocity structure function, the \gls{TES} model assumes that the \REV{shell-averaged radial velocity dispersion $\left<\delta v_r^2\right>_\rho^{1/2}$ appearing in \cref{eq:dimensional_steady_equilibrium} increases with radial distance from the center} as a power law:
\begin{equation}\label{eq:linewidth_size}
  \left<\delta v_r^2 \right>_\rho^{1/2} = c_s \left( \frac{r}{r_s} \right)^p ,
\end{equation}
where $r_s$ is the sonic radius and $p$ is the power-law exponent.
A closely related quantity is the volume-averaged one-dimensional velocity dispersion \REV{$\sigma_\mathrm{1D}$ defined by 
\begin{equation}\label{eq:sigma1d}
    \sigma_\mathrm{1D} \equiv \frac{1}{\sqrt{3}}\left(\frac{\int \rho \vert\mathbf{v} - \mathbf{v}_\mathrm{com}\vert^2 d\mathcal{V}}{\int \rho\, d\mathcal{V}}\right)^{1/2} \equiv  c_s \left( \frac{r}{\lambda_s} \right)^p,
\end{equation}
where the integrals are taken over a ball of radius $r$ and $\mathbf{v}_\mathrm{com} \equiv (\int \rho \mathbf{v}d\mathcal{V})/(\int \rho\,d\mathcal{V})$ is the center-of-mass velocity.
The second equality of \cref{eq:sigma1d} defines the sonic scale $\lambda_s$ when $\sigma_\mathrm{1D}$ is assumed to follow a power-law in radius.}
\REV{If the turbulent velocities are statistically isotropic and considerably larger than bulk velocities, then $\sigma_\mathrm{1D}$ and $\left<\delta v_r^2\right>_\rho^{1/2}$ are related by 
\begin{equation}\label{eq:sigma_vol_to_dv}
    \sigma_\mathrm{1D} = \eta_d\left(\frac{3}{2p+3}\right)^{1/2}\left<\delta v_r^2\right>_\rho^{1/2},
\end{equation}
(see \autoref{app:sigma_r}), where the order unity factor $\eta_d$ would be equal to unity if the density is uniform, and more generally depends on internal density stratification.
For the \gls{TES} solutions, we find $\eta_d$ is almost constant at $\eta_d \approx 0.9$, with a very weak dependence on $p$ and $\sigma_\mathrm{1D}/c_s$.
The sonic scale $\lambda_s$ and the sonic radius $r_s$ are related by}
\begin{equation}\label{eq:lambda_s_to_rs}
    \lambda_s = \eta_d^{-\frac{1}{p}}\left(\frac{2p+3}{3}\right)^{\frac{1}{2p}} r_s.
\end{equation}
A sphere with diameter $2 \lambda_s$ would have equal thermal and nonthermal contributions to the velocity dispersion observed on a pencil beam through its center. 
We note that for $p=0.5$ \REV{and $\eta_d = 0.9$, $\sigma_\mathrm{1D} = 0.779\left<\delta v_r^2\right>_\rho^{1/2}$ and $\lambda_s = 1.65r_s$.}

For given central density $\rho_c\equiv \rho(r=0)$, sound speed $c_s$, and prescribed turbulent velocity field (i.e., choices for $r_s$ and $p$), \cref{eq:dimensional_steady_equilibrium} can be integrated outward to yield the equilibrium density structure.
In obtaining solutions, it is convenient to nondimensionalize the variables and equations, dividing the density by $\rho_c$, velocities by $c_s$, and lengths by
\begin{equation}\label{eq:rc}
    r_c \equiv \frac{c_s}{\sqrt{4\pi G \rho_c}},
\end{equation}
\REV{so that the normalized radial coordinate is 
\begin{equation}
\xi \equiv \frac{r}{r_c}.
\end{equation}
}
The dimensionless sonic radius that parameterizes solutions is   
\begin{equation}\label{eq:xis}
  \begin{split}
    \xi_s &\equiv \frac{r_s}{r_c}\\
          &= 5.14 \left( \frac{r_s}{0.05\,\mathrm{pc}} \right) \left( \frac{n_{\mathrm{H}_2,c}}{10^5\,\mathrm{cm^{-3}}} \right)^{1/2}\left( \frac{T}{10\,\mathrm{K}} \right)^{-1/2} .
  \end{split}
\end{equation}
In the limit of $\xi_s\to \infty$, the \gls{TES} becomes identical to the \gls{BE} sphere.

\citetalias{moon24}\REV{, following \citet{bonnor56}, calculated for each equilibrium solution with a given $\xi_s$ the bulk modulus $K\equiv -\partial \ln P_\mathrm{eff} / \partial \ln V$ as a function of the dimensionless radius $\xi = r/r_c$, and found the values $\xi = \xi_\mathrm{crit}$ at which $K$ first becomes negative (i.e. the radius at which the effective surface pressure begins to {\it increase} with increasing volume). 
  A \gls{TES} with outer radius $\xi_\mathrm{max}$ is unstable if $\xi_\mathrm{max} > \xi_\mathrm{crit}$.
\citetalias{moon24} numerically found that the critical radius $\xi_\mathrm{crit}$ decreases with increasing $\xi_s$, with the exact dependence varying for different values of $p$.}
For given $p$, the \emph{dimensional} critical radius $r_\mathrm{crit} = \xi_\mathrm{crit}r_c$ is therefore a function of $c_s$, $\rho_c$, and $r_s$.
The related critical mass is then defined by $M_\mathrm{crit} = M_\mathrm{enc}(r_\mathrm{crit})$, where $M_\mathrm{enc}$ is the enclosed mass
\begin{equation}\label{eq:menc}
  M_\mathrm{enc}(r) \equiv \int_0^r 4\pi r'^2 \left<\rho\right>\,dr'.
\end{equation}
For \gls{TES} solutions, the critical mass and radius (normalized to the mass and radius scales based on a given central density) as well as the center-to-edge density contrast all decrease with increasing $\xi_s$ (i.e., weaker turbulence), 
as shown in Fig. 4 of \citetalias{moon24}.

\citetalias{moon24} showed that \REV{for some parameter choices, turbulence can completely prevent instability: when $\xi_s < \xi_{s,\mathrm{min}}$, the \gls{TES} 
has 
$\xi_\mathrm{crit} \to \infty$}.
When $p=0.5$, this minimum value is numerically found to be $\xi_{s,\mathrm{min}} = 2.42$.
For a given local sonic radius $r_s$, the existence of $\xi_{s,\mathrm{min}}$ implies that collapse cannot occur when the central density is below \REV{ $\rho_{c,\mathrm{min}}=[\xi_{s,\mathrm{min}}c_s/(\sqrt{4\pi G} r_s)]^2$, corresponding to} 
\begin{equation}\label{eq:rhocmin}
    \frac{\rho_{c,\mathrm{min}}}{\rho_0} = \frac{\xi_{s,\mathrm{min}}^2}{4\pi^2} \left(\frac{r_s}{L_{J,0}}\right)^{-2},
\end{equation}
\REV{where $L_{J,0} = c_s[\pi/(G\rho_0)]^{1/2}$ is the Jeans length at a reference density $\rho_0$ (this could be, e.g., the mean density in a GMC).}
We emphasize that $\rho_{c,\mathrm{min}}$ is \emph{not} a critical density for collapse: when $\rho_c$ is very close to $\rho_{c,\mathrm{min}}$, the critical radius would exceed typical sizes of any realistic core forming region; prestellar core collapse is expected to occur at densities at least a factor of a few higher than this minimum (see Fig. 5 in \citetalias{moon24} and related discussion following \cref{eq:sigma_to_xis} below).

Because the parametric dependency of the critical quantities on $\xi_s$ becomes extremely steep as $\xi_s \to \xi_{s,\mathrm{min}}$, in \citetalias{moon24} we also presented $r_\mathrm{crit}$ and $M_\mathrm{crit}$ (and some other quantities) in terms of the mass-weighted velocity dispersion $\sigma_\mathrm{1D}$ defined in \cref{eq:sigma1d}.
In particular, the critical mass and radius for $p=0.5$ were found to be well-approximated by
\begin{equation}\label{eq:mcrit_fit}
    M_\mathrm{crit} \approx M_\mathrm{BE}(\overline{\rho}) \left( 1 + \frac{1}{2}\frac{\sigma_\mathrm{1D}^2}{c_s^2}\right),
\end{equation}
\begin{equation}\label{eq:rcrit_fit}
    r_\mathrm{crit} \approx R_\mathrm{BE}(\overline{\rho}) \left( 1 + \frac{1}{2}\frac{\sigma_\mathrm{1D}^2}{c_s^2}\right)^{1/3},
\end{equation}
within a relative error of $5\%$ for $\sigma_\mathrm{1D} < 9.5 c_s$ and $\sigma_\mathrm{1D} < 13 c_s$.
\REV{Here,
\begin{equation}\label{eq:rbe}
  R_\mathrm{BE}(\overline{\rho}) = 0.762 \frac{c_s}{G^{1/2}\overline{\rho}^{1/2}},
\end{equation}
\begin{equation}\label{eq:mbe}
    M_\mathrm{BE}(\overline{\rho})\equiv \frac{4\pi}{3} \overline{\rho}R_\mathrm{BE}^3(\overline{\rho}) = 1.86 \frac{c_s^3}{G^{3/2}\overline{\rho}^{1/2}}
\end{equation}
are the radius and mass of the critical \gls{BE} sphere with the average density $\overline{\rho}$. The numerical coefficients in \cref{eq:rbe,eq:mbe} are adjusted if $R_\mathrm{BE}$ and $M_\mathrm{BE}$ are defined in terms of edge densities (rather than mean densities); see e.g.~Equations (18)--(19) of \citet{gong09}.}
Equations (60)--(61) of \citetalias{moon24} also provide the critical density contrast and mean density in terms of $\sigma_\mathrm{1D}/c_s$.
We also note that, for $p=0.5$, an approximate relation
\begin{equation}\label{eq:sigma_to_xis}
    \xi_s \approx \xi_{s,\mathrm{min}} + 4\left(\frac{\sigma_\mathrm{1D}}{c_s}\right)^{-2}
\end{equation}
holds for $\sigma_\mathrm{1D} < 10 c_s$, with the maximum relative error of $1.85\%$.
This expression for $\xi_s$ may be 
used 
to determine the central density required, 
\REV{ $\rho_{c}=[\xi_{s}c_s/(\sqrt{4\pi G} r_s)]^2$},
for collapse of a core with velocity dispersion $\sigma_\mathrm{1D}$ for a given average density $\overline{\rho}$,  as in \cref{eq:mcrit_fit,eq:rcrit_fit}.

\begin{figure}[htpb]
  \epsscale{1.1}
  \plotone{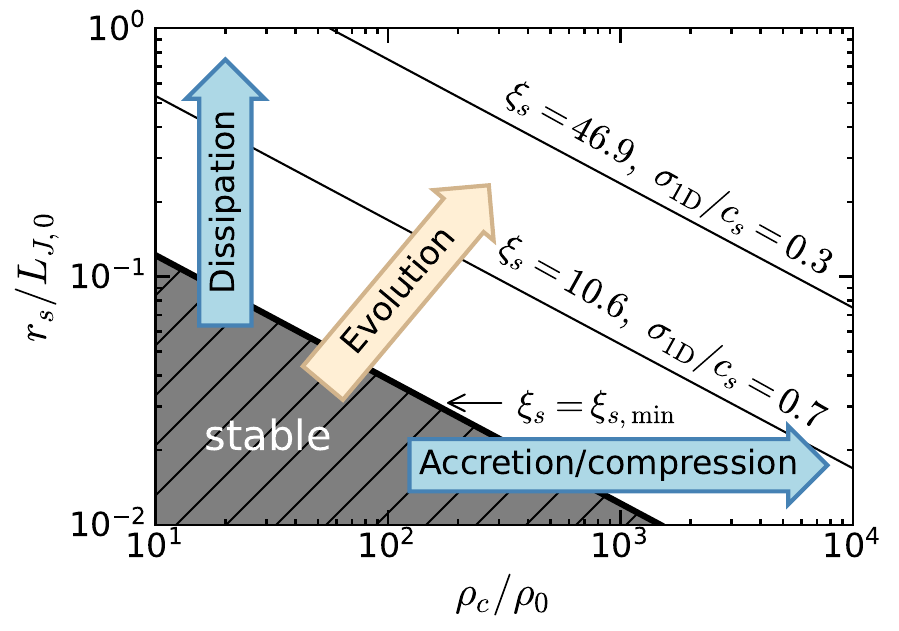}
  \caption{Schematic diagram showing the overall direction of evolution for prestellar cores that undergo collapse.
    \REV{Other than block arrows, all the lines plotted are quantitative results from the \gls{TES} model with $p=0.5$.}
  The thick black line marks $\xi_s = \xi_{s,\mathrm{min}}$, equal to $2.42$ for $p=0.5$; all cores with $\xi_s < \xi_{s,\mathrm{min}}$ are stable.  
  For structures originating in the stable regime, local dissipation of turbulence would lead to increase in \REV{the sonic radius} $r_s$ (upward arrow), whereas accretion (or compression in the Lagrangian sense) leads to higher \REV{central density} $\rho_c$ (rightward arrow).
  In typical core-forming regions, both processes are expected to occur at the same time and therefore forming prestellar cores would evolve diagonally in this diagram.
  The thin solid lines mark loci of constant $\xi_s = 10.6$ and $46.9$, respectively corresponding to $\sigma_\mathrm{1D}/c_s = 0.7$ and $0.3$ for $p=0.5$ (\cref{eq:sigma_to_xis}).}
  \label{fig:tes_schematics}
\end{figure}

Considering application of the \gls{TES} analysis to core formation and collapse in GMCs, it is generally expected that cores evolve in the direction of increasing $\xi_s \propto r_s\rho_c^{1/2}$ as turbulence decays and the central density increases.
This is schematically illustrated in \cref{fig:tes_schematics}.
Initially, structures originate in the stable regime where $\rho_c < \rho_{c,\mathrm{min}}$ given by \cref{eq:rhocmin}.
With density increasing and turbulence decreasing as a core develops, \cref{eq:rcrit_fit} indicates that $r_\mathrm{crit}$ tends to decrease over time.
When the compression driven by converging flows is strong enough, $r_\mathrm{crit}$ can become small enough compared to the effective core size $r_\mathrm{max}$ that the core will become unstable to collapse.
Otherwise, if $r_\mathrm{crit}$ remains larger than $r_\mathrm{max}$, a core would remain stable and may disperse back into the ambient medium (a schematic drawing of this scenario is provided in Fig. 11 of \citetalias{moon24}).
We note that cores will appear in different regions of the parameter space shown in \cref{fig:tes_schematics}, depending on the internal and cloud-scale Mach numbers.
In particular, transonic ($\sigma_\mathrm{1D} \sim c_s$; see \citetalias{paperII}), cores would appear somewhere near the diagonal line marked $\sigma_\mathrm{1D}/c_s =0.7$.
Cores forming in a high Mach number cloud would generally be expected to have high $\rho_c$ and small $r_s$, towards the bottom right, while cores forming in a low Mach number cloud would have low  $\rho_c$ and large $r_s$, toward the upper left.

\section{Numerical Simulations}\label{sec:simulations}

\subsection{Governing Equations and Numerical Methods}

We model the star-forming \gls{ISM} as an isothermal, self-gravitating fluid.
While the gas temperature in real \glspl{GMC} is clearly not uniform in \REV{either space or time, the actual variations are relatively small, and} the isothermal assumption allows us to compare the simulation results to simpler semi-analytic models, which aids understanding of the physical processes governing the collapse.

The governing equations we solve are
\begin{equation}\label{eq:continuity}
  \frac{\partial\rho}{\partial t} + \boldsymbol{\nabla}\cdot(\rho \mathbf{v}) = 0,
\end{equation}
\begin{equation}\label{eq:momentum}
  \frac{\partial(\rho \mathbf{v})}{\partial t} + \boldsymbol{\nabla}\cdot(\rho \mathbf{v}\mathbf{v} + P\mathds{I}) = -\rho\boldsymbol{\nabla}\Phi,
\end{equation}
\begin{equation}\label{eq:poisson}
  \boldsymbol{\nabla}^2\Phi = 4\pi G (\rho + \rho_*),
\end{equation}
where $\rho$ and $\rho_*$ are, respectively, the volume density of gas and (smoothed) sink particles that form during the simulations, $\mathbf{v}$ is the gas velocity, $P = c_s^2\rho$ is the gas pressure, $c_s$ is the isothermal sound speed, $\mathds{I}$ is the identity matrix, $\Phi$ is the gravitational potential associated with both gas and sink particles, and $G$ is the Newton's gravitational constant.

\cref{eq:continuity,eq:momentum,eq:poisson} are numerically solved under periodic boundary conditions using the \textit{Athena++} code \citep{stone20}.
We use the HLLE Riemann solver with piecewise linear reconstruction to calculate the fluxes and apply first-order flux correction \citep[see the Appendix of][]{lemaster09} when needed.
We use the \gls{VL2} integrator \citep{stone09} to advance the conserved variables $\rho$ and $\rho \mathbf{v}$ in time using the Riemann fluxes.
In both the predictor and corrector steps of the \gls{VL2} integrator, \cref{eq:poisson} is approximately solved by one execution of the full-multigrid algorithm followed by 3 additional iterations of multigrid V-cycles \citep{tomida23}.
We use the triangle-shaped cloud interpolation scheme \citep{hockney88} to deposit the sink particle masses onto the grid to evaluate $\rho_*$.

The actual numerical computations are carried out in terms of dimensionless hydrodynamic variables.
We adopt the mean density $\rho_0 = 1.4 m_\mathrm{H}n_\mathrm{H,0}$ in the simulation box as the unit of density, where $m_\mathrm{H}$ is the mass of a hydrogen atom and $n_\mathrm{H,0}$ is the hydrogen number density averaged over the box.
We take the isothermal sound speed
\begin{equation}\label{eq:sound_speed}
  c_s = 0.266\,\mathrm{km}\,\mathrm{s}^{-1} \left( \frac{T}{20\,\mathrm{K}} \right)^{1/2}
\end{equation}
as the unit of velocity and use the Jeans length
\begin{equation}\label{eq:ljeans}
  L_{J,0} \equiv \left(\frac{\pi c_s^2}{G\rho_0}\right)^{1/2} = 3.86\,\mathrm{pc}\left( \frac{T}{20\,\mathrm{K}} \right)^{1/2} \left( \frac{n_\mathrm{H,0}}{100\,\mathrm{cm}^{-3}} \right)^{-1/2}
\end{equation}
as the unit of length.
The corresponding mass and time units are the Jeans mass
\begin{equation}\label{eq:mjeans}
  M_{J,0} \equiv \rho_0 L_{J,0}^3 = 200\,\mathrm{M}_\odot\,\left( \frac{T}{20\,\mathrm{K}} \right)^{3/2}\left( \frac{n_\mathrm{H,0}}{100\,\mathrm{cm}^{-3}} \right)^{-1/2} 
\end{equation}
and the Jeans time
\begin{equation}\label{eq:tjeans}
  t_{J,0} \equiv \frac{L_{J,0}}{c_s} = 14.2\,\mathrm{Myr}\,\left( \frac{n_\mathrm{H,0}}{100\,\mathrm{cm}^{-3}} \right)^{-1/2}
\end{equation}
at the mean density of the cloud.
The free-fall time at the mean density is related to $t_{J,0}$ by
\begin{equation}\label{eq:tff0}
\begin{split}
  t_\mathrm{ff,0} &\equiv \left( \frac{3\pi}{32 G\rho_0} \right)^{1/2} = 0.306t_{J,0}\\
                  &= 4.35\,\mathrm{Myr} \left( \frac{n_\mathrm{H,0}}{100\,\mathrm{cm}^{-3}} \right)^{-1/2}.
\end{split}
\end{equation}

\subsection{Initial Conditions and Model Physical Parameters}

Our computational domain is a Cartesian cube with volume $L_\mathrm{box}^3$ uniformly divided into $N^3$ cells, each having side length $\Delta x \equiv L_\mathrm{box}/N$.
The computational domain is initialized with constant density $\rho = \rho_0$ and random velocity perturbations $\mathbf{v} = \delta \mathbf{v}_0$.
The initial velocity field $\delta \mathbf{v}_0$ is characterized by a power spectrum $P(k) \propto k^{-2}$ (corresponding to a linewidth-size relation $\Delta v(l) \propto l^{1/2}$), with two thirds of the total power in solenoidal modes and the remaining one third in compressive modes.
The amplitude is normalized such that the one-dimensional mass-weighted \gls{RMS} velocity dispersion is $\sigma_\mathrm{1D,box}$, and the three-dimensional RMS Mach number on the largest scale (i.e. the velocity dispersion in code units) is $\mathcal{M}_\mathrm{3D} = \sqrt{3}\sigma_\mathrm{1D,box}/c_s$. With our adopted code units, for decaying self-gravitating turbulence simulations we must specify $L_\mathrm{box}/L_{J,0}$ and  $\mathcal{M}_\mathrm{3D}$.

\REV{We define the effective virial parameter\footnote{In \citetalias{moon24}, we defined the virial parameter for an unstratified spherical cloud as $\alpha_\mathrm{vir,cloud} = 5 (R_\mathrm{cloud}/r_s)^{2p} c_s^2 R_\mathrm{cloud}/G M_\mathrm{cloud}$. From \cref{eq:linewidth_size,eq:sigma_vol_to_dv}, these definitions are related by  $\alpha_\mathrm{vir,cloud} = (1 + 2p/3) \alpha_\mathrm{vir,box}$ \REV{if we take $R_\mathrm{cloud} = L_\mathrm{box}/2$ and $M_\mathrm{cloud} = M_\mathrm{box}$}.}
\begin{equation}\label{eq:cloud_virial_parameter}
  \alpha_\mathrm{vir,box} \equiv \frac{5 \sigma_\mathrm{1D,box}^2 L_\mathrm{box}}{2G M_\mathrm{box}},
\end{equation}
by treating $L_\mathrm{box}/2$ as an effective ``radius'' of the computational domain\footnote{Although $\sigma_\mathrm{1D,box}$ is defined as a box-average, we find that the difference between $\sigma_\mathrm{1D,box}$ and the spherically averaged velocity dispersion is only $2.7\%$.} containing $M_\mathrm{box} = L_\mathrm{box}^3 \rho_0$.
Although $\alpha_\mathrm{vir,box}$ cannot be directly interpreted as the ratio of kinetic to gravitational potential energies due to the assumed periodic boundary conditions, it still provides a convenient dimensionless parameter characterizing the relative importance of turbulence to gravity.
As discussed in \autoref{app:initial_conditions}, for example, the flow crossing time and the gravitational free-fall time are comparable to each other when $\alpha_\mathrm{vir,box} \sim 2$. The value of $\alpha_\mathrm{vir,box}$ also allows us to connect to observed clouds that have properties comparable to those in our simulation.
}
If we specify the two dimensionless parameters $\alpha_\mathrm{vir,box}$ and $\mathcal{M}_\mathrm{3D}$, then the side length $L_\mathrm{box}$ of our computational domain is determined by
\begin{equation}\label{eq:lbox}
  \begin{split}
    L_\mathrm{box} &= \left( \frac{5}{6\pi} \right)^{1/2}L_{J,0}\alpha_\mathrm{vir,box}^{-1/2} \mathcal{M}_\mathrm{3D} \\
                   &= 3.64L_{J,0} \left( \frac{\alpha_\mathrm{vir,box}}{2} \right)^{-1/2} \left( \frac{\mathcal{M}_\mathrm{3D}}{10} \right).
  \end{split}
\end{equation}

For the simulations presented in this paper, we consider two models with $\mathcal{M}_\mathrm{3D} = 5$ (hereafter model \texttt{M5}) and $\mathcal{M}_\mathrm{3D} = 10$ (hereafter model \texttt{M10}).
In order to have $\alpha_\mathrm{vir,box} \sim 2$, we set the box size to $L_\mathrm{box} = 2L_{J,0}$ and $4L_{J,0}$ for model \texttt{M5} and \texttt{M10}, respectively, leading to $\alpha_\mathrm{vir,box} = 1.66$.
The total mass within the domain is therefore $8 M_{J,0}$ for model \texttt{M5} and $64M_{J,0}$ for model \texttt{M10}, corresponding respectively to $1.60\times 10^3 M_\odot$ and $1.28 \times 10^4  M_\odot$ for the fiducial density and temperature in \cref{eq:mjeans}.
\cref{tb:models} lists the parameters adopted for both models. 
Column (1) gives the model name.
Columns (2) and (3) give $\mathcal{M}_\mathrm{3D}$ and $L_\mathrm{box}$, respectively.
Column (4) gives $\alpha_\mathrm{vir,box}$.
In addition to the above physical parameters, \cref{tb:models} lists numerical parameters for both models; the resolution requirements are discussed in  \cref{sec:resolution} and the termination condition at the end of \cref{sec:sink_particles}.

\begin{deluxetable*}{lccccccccccc}
    \tablecaption{Model parameters\label{tb:models}}
    \tablehead{
\colhead{Model} &
\colhead{$\mathcal{M}_\mathrm{3D}$} &
\colhead{$L_\mathrm{box}/L_{J,0}$} &
\colhead{$\alpha_\mathrm{vir,box}$} &
\colhead{$N$} &
\colhead{$\Delta x/L_{J,0}$} &
\colhead{$\overline{\rho}_\mathrm{max}/\rho_0$} &
\colhead{$M_\mathrm{min}/M_{J,0}$} &
\colhead{$1-a$} &
\colhead{\# of sims.} &
\colhead{$t_\mathrm{final}/t_\mathrm{ff,0}$}\\
\colhead{(1)} &
\colhead{(2)} &
\colhead{(3)} &
\colhead{(4)} &
\colhead{(5)} &
\colhead{(6)} &
\colhead{(7)} &
\colhead{(8)} &
\colhead{(9)} &
\colhead{(10)} &
\colhead{(11)} &
    }
    \startdata
    \texttt{M10} & 10 & 4 & 1.66 & 1024 & $3.91 \times 10^{-3}$ & 189 & $2.42\times 10^{-2}$ & $1.15 \times 10^{-2}$ & 7 & $1.1 \pm 0.1$\\
    \texttt{M5}  & 5  & 2 & 1.66 & 512  & $3.91 \times 10^{-3}$ &  189 & $2.42\times 10^{-2}$ & $2.33\times 10^{-4}$ & 40 & $1.3 \pm 0.2$
    \enddata
    \tablecomments{Columns (7) and (8) are based on \cref{eq:resolution_requirement} and \cref{eq:Mmin} assuming the core radius is resolved by $N_\mathrm{core,res}=8$ cells. For $N_\mathrm{core,res}=4$,  $\bar{\rho}_\mathrm{max}$ and $M_\mathrm{min}$ are increased and decreased by a factor 4 or 2, respectively. The unresolved mass fraction in Column (9) would then become $1-a= 2.19\times 10^{-6}$ and $9.76\times 10^{-4}$ for \texttt{M5} and \texttt{M10}, respectively.
    }
\end{deluxetable*}

\begin{figure}[htpb]
  \epsscale{1.1}
  \plotone{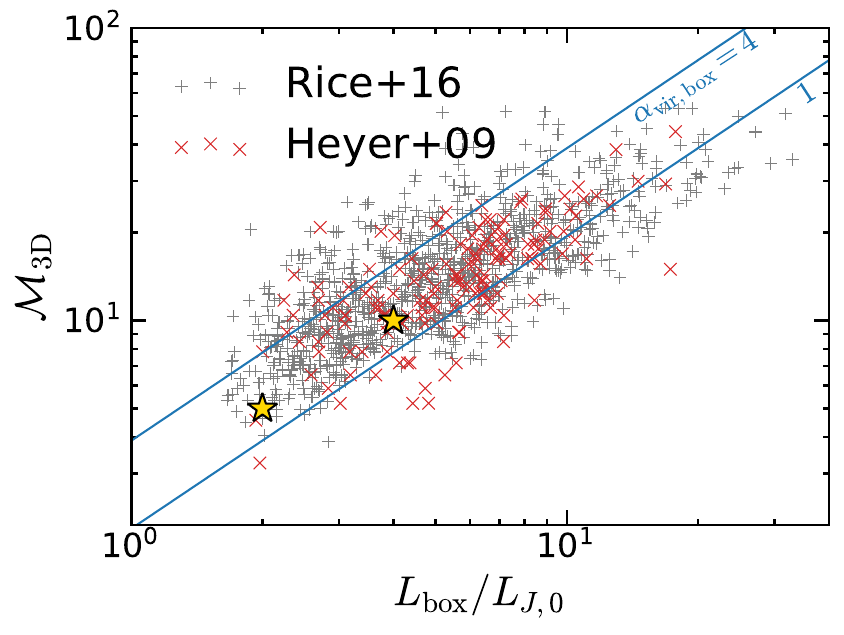}
  \caption{Dimensionless linewidth-size relation of the Galactic \gls{GMC} sample of \citet{heyer09} (red cross symbols) and \citet{rice16} (gray plus symbols).
  We assume $T=20\,\mathrm{K}$ \citep{yoda10} to derive $c_s$ and $L_{J,0}$.
  Because both $\mathcal{M}_\mathrm{3D}$ and \REV{$L_\mathrm{box}/L_{J,0}$ scale $\propto T^{-1/2}$, the observed points would move diagonally upward/rightward along the constant $\alpha_\mathrm{vir,box}$ lines (plotted in blue solid lines for $\alpha_\mathrm{vir,box} = 1$ and $4$)} if a lower temperature were assumed.
  The loci of our models \texttt{M5} and \texttt{M10} are marked with star symbols.}
  \label{fig:clouds}
\end{figure}

To compare our models with observed clouds, in \cref{fig:clouds} we plot the Galactic \gls{GMC} sample of \citet{heyer09} and \citet{rice16} in the $\mathcal{M}_\mathrm{3D}$--$L_\mathrm{box}/L_{J,0}$ plane.
\REV{We take $L_\mathrm{box} = (4\pi R_\mathrm{cl}^3/3)^{1/3}$ for the observed cloud radius $R_\mathrm{cl}$ and multiply the observed line-of-sight velocity dispersion by $\sqrt{3}/c_s$ to calculate $\mathcal{M}_\mathrm{3D}$.}
Since neither \citet{heyer09} nor \citet{rice16} provide temperature measurements, we simply assume $c_s = 0.266\,\mathrm{km}\,\mathrm{s}^{-1}$ appropriate for $T = 20\,\mathrm{K}$ \citep{yoda10}\footnote{Based on the near-constantness of the brightness temperature ratio between $^{12}\text{CO}(J=2\text{--}1)$ and $^{12}\text{CO}(J=1\text{--}0)$ transitions, \citet{yoda10} placed a lower limit of $19\,\mathrm{K}$ on the kinetic temperature of a moderate density ($n_\mathrm{H} \sim \text{a few }10^2\,\mathrm{cm}^{-3}$) gas.}.

The loci of our models are marked with yellow star symbols in \cref{fig:clouds}, where our chosen $\alpha_\mathrm{vir,box}=1.66$ value lies between the diagonal lines marking $\alpha_\mathrm{vir,box}=4$ and $1$.
Model \texttt{M10} represents a typical moderate-mass cloud, similar e.g. to Taurus \citep{pineda10}, while model \texttt{M5} lies near the lower edge of the observed distribution.
Alternatively, our models may be regarded as local overdense patches within a larger mass \gls{GMC} (with higher mass, size, and velocity dispersion than our box).
Because such a local region would have higher $n_\mathrm{H,0}$ than the average over a whole GMC, if one takes the latter viewpoint the dimensional length, mass, and time scales given in \cref{eq:ljeans,eq:mjeans,eq:tjeans} would be appropriately scaled down, e.g.~by a factor $0.3$ if $n_\mathrm{H,0}\sim 10^3 \, \mathrm{cm}^{-3}$.
If the temperature is lower in overdense gas, it would also reduce $T$ and therefore the dimensional values of $L_{J,0}$ and $M_{J,0}$.

We note that while the simulations we present here all have initial $\alpha_\mathrm{vir,box}=1.66$, we have also tested a range of initial virial parameters.
Since observed \glspl{GMC} form via condensation of more diffuse, more turbulent gas, and the distribution shown \cref{fig:clouds} in fact extends above $\alpha_\mathrm{vir,box}=4$, the most relevant regime for further exploration would be towards higher virial parameter.
We have tested simulations with initially larger $\alpha_\mathrm{vir,box}$, and we find that after a period of initial turbulent decay, evolution is quite similar to that of the models which are the primary focus of our presentation; we show one such model, \texttt{M15L2}, in \autoref{app:initial_conditions}.
The case with low initial $\alpha_\mathrm{vir,box}$ is less relevant as a model of a star-forming \gls{GMC}, but we have also tested this case, with model \texttt{M3L4} included in \autoref{app:initial_conditions}.
When the initial virial parameter is very low, evolution is quite different from our standard simulations, with smooth filaments forming and fragmenting primarily due to the action of self-gravity.

\subsection{Turbulent, Self-Gravitating Cloud Structure and Numerical Resolution Requirements}\label{sec:resolution}

\REV{Motivated by observations and previous numerical results for the properties of turbulent, self-gravitating clouds, in this section we discuss numerical requirements for resolving dense prestellar cores and their diffuse progenitors. While a rigorous evaluation of the necessary resolution criteria can only be substantiated by a thorough convergence study, the analysis provided here quantitatively explains the motivation for the numerical choices made at the outset of the present study.  
Readers primarily interested in our results, rather than numerical details, may wish to skip this section. 
For the convenience of those readers, key takeaways are:
\begin{itemize}
\item[1.] For the purpose of the present study, we define cores as ``resolved'' when the criterion in \cref{eq:resolvedness_criterion} is met, where $N_\mathrm{core,res}$ is an arbitrary parameter (our fiducial choice is 8). This leads to \cref{eq:Nmin_99} as a provisional resolution criterion for dense cores in turbulent clouds. 
\item[2.] \cref{eq:lmb_sonic} provides a useful reference point for the sonic scale relative to the Jeans scale, based on cloud-scale averages. This leads to \cref{eq:nmin_sonic} as a provisional resolution criterion in diffuse gas. 
\end{itemize}
}

\subsubsection{Resolving dense, self-gravitating cores}


To correctly follow internal dynamics of \REV{self-gravitating} cores, it is very important to resolve \REV{the critical radius (see \cref{eq:rcrit_fit})} $r_\mathrm{crit}$ with a sufficient number of cells.
We therefore introduce the resolution criterion
\begin{equation}\label{eq:resolvedness_criterion}
  \Delta x\le \frac{r_\mathrm{crit}}{N_\mathrm{core,res}}
\end{equation}
in which $N_\mathrm{core,res}$ is an arbitrary threshold, such that only those cores satisfying \cref{eq:resolvedness_criterion} are considered as being fully ``resolved''.
In this work, we adopt a fiducial choice $N_\mathrm{core,res} = 8$ unless otherwise stated.
This choice is conservative, in the sense that cores with e.g.  $N_\mathrm{core,res} = 4$ would still be marginally resolved.

In \citetalias{moon24}, we show (see Figure 8(b) there) that for given mean core density $\bar\rho$ (where $\bar\rho$ is a factor 1.5-2.5 above the density at the edge for critical cores), the smallest possible $r_\mathrm{crit}$ is that for the case without turbulence, i.e. the radius of a critical Bonnor-Ebert sphere $R_\mathrm{BE}$ (\cref{eq:rbe}).
Because $R_\mathrm{BE}/L_{J,0}=0.43 (\bar{\rho}/\rho_0)^{-1/2}$ decreases when the core density is larger compared to the average in the box, it becomes more difficult to resolve cores forming in higher density regions.
Together, \cref{eq:rbe} and \cref{eq:resolvedness_criterion} imply that $\Delta x$ must vary as the inverse square root of the density in order to satisfy a fixed resolution criterion.
\REV{As we point out below (see \cref{eq:lognormal_var})}, the increase of density variance with Mach number  implies the numerical resolution requirements become more stringent in simulations of more highly turbulent clouds.

In many simulations of gravoturbulent fragmentation, resolution is controlled via \gls{AMR}, with the typical criterion that the local Jeans length must always be resolved by a fixed number of cells ($\Delta x = L_J/N_J \equiv c_s[\pi/(G\rho)]^{1/2}/N_J$; $N_J = 4$ was originally recommended by \citealt{truelove97} to avoid artificial fragmentation, while more stringent $N_J = 30$ was suggested by \citealt{federrath11} to resolve local turbulence).
Since the mass resolution under this criterion varies as $\Delta m\propto \rho^{-1/2}$, if strictly applied it could imply that the precursor to a given core (in a Lagrangian sense) is not necessarily resolved before it becomes self-gravitating.
An alternative approach of \gls{AMR} with fixed mass resolution \citep[or Lagrangian methods such as that of ][]{hopkins16}, i.e. $\Delta x \propto \rho^{-1/3}$ instead of  $\Delta x \propto \rho^{-1/2}$, would impose greater resolution demands in a larger fraction of the domain, but would ensure that the precursor material of a given core is resolved.
However, even with fixed mass resolution as set by a target mass to be resolved, substructure due to turbulence introduced at early, low density stages, can potentially be missed depending on the sonic scale in a simulation; we return to this issue below.

In the present work, which adopts a fixed uniform mesh rather than \gls{AMR}, we control the resolution by introducing a target density parameter $\overline{\rho}_\mathrm{max}$ \REV{below} which cores would be resolved with at least $N_\mathrm{core,res}$ cells.
One can place a conservative upper limit on the mesh cell size by substituting $R_\mathrm{BE}$ for $r_\mathrm{crit}$ in \cref{eq:resolvedness_criterion}, leading to
\begin{equation}\label{eq:resolution_requirement}
  \begin{split}
    \Delta x_\mathrm{max} &\equiv \frac{R_\mathrm{BE}(\overline{\rho}_\mathrm{max})}{N_\mathrm{core,res}}\\
                          &= \frac{0.430L_{J,0}}{N_\mathrm{core,res}} \left( \frac{\overline{\rho}_\mathrm{max}}{\rho_0} \right)^{-1/2}.
  \end{split}
\end{equation}
\cref{eq:resolution_requirement} implies that, for a desired $\overline{\rho}_\mathrm{max}$ and $N_\mathrm{core,res}$, the cell size must be smaller than $\Delta x_\mathrm{max}$ to satisfy \cref{eq:resolvedness_criterion}.
Alternatively, at a given numerical resolution $\Delta x/L_{J,0}$, \cref{eq:resolution_requirement} yields the density parameter $\overline{\rho}_\mathrm{max}$ that is related to the resolution parameter $N_\mathrm{core,res}$ by
\begin{equation}\label{eq:rhomax_res}
    \frac{\overline{\rho}_\mathrm{max}}{\rho_0} = 0.185\left(\frac{\Delta x}{L_{J,0}}\right)^{-2} N_\mathrm{core,res}^{-2}.
\end{equation}
We note that the target density $\overline{\rho}_\mathrm{max}$ is a conservative limit: turbulent cores have $r_\mathrm{crit} > R_\mathrm{BE}$ and therefore can be resolved at densities higher than $\overline{\rho}_\mathrm{max}$.

The minimum resolvable core mass depends on the target density or alternatively on the resolution parameter for a given numerical resolution as
\begin{subequations}\label{eq:Mmin}
\begin{align}
    M_\mathrm{min}&\equiv M_\mathrm{BE}(\overline{\rho}_\mathrm{max})\nonumber\\
                  &= 2.36\times 10^{-2} M_{J,0} \left( \frac{\overline{\rho}_\mathrm{max}}{200\rho_0} \right)^{-1/2}\label{eq:Mmin_given_rhomax}\\
                  &= 2.43 \frac{c_s^2}{G} N_\mathrm{core,res}\Delta x\label{eq:Mmin_given_dx}\\
                  &= \frac{4\pi N_\mathrm{core,res}^3}{3} \Delta m_\mathrm{eff}\label{eq:Mmin_given_dm}
\end{align}
\end{subequations}
\REV{where $M_\mathrm{BE}$ is the \gls{BE} mass defined in \cref{eq:mbe} and \cref{eq:mjeans} provides a conversion of $M_{J,0}$ to physical units.}
In \cref{eq:Mmin_given_dm}, $\Delta m_\mathrm{eff} = \overline{\rho}_\mathrm{max}\Delta x^3$ is the effective mass resolution at the target density.
We note that each form in \cref{eq:Mmin} serves different purposes: \cref{eq:Mmin_given_rhomax} provides the minimum resolvable mass in terms of the target density parameter, where the resolution must be self-consistently determined to satisfy \cref{eq:resolution_requirement}; when the resolution of the simulation is already given, \cref{eq:Mmin_given_dx,eq:Mmin_given_dm} yield the minimum mass of cores whose radii would be resolved with $N_\mathrm{core,res}$ cells.

In order to resolve critical cores with density $\bar{\rho}_\mathrm{max}$, the minimum required number of cells (per dimension) in the simulation must satisfy
\begin{subequations}\label{eq:minimum_number_of_cells}
\begin{align}
  N_\mathrm{min} &\equiv \frac{L_\mathrm{box}}{\Delta x_\mathrm{max}}
  \nonumber\\
                   &= 1052
                   \frac{N_\mathrm{core,res}}{8}
                   \frac{L_\mathrm{box}}{4L_{J,0}}
                   \left( \frac{\overline{\rho}_\mathrm{max}}{200\rho_0} \right)^{1/2}\label{eq:minimum_number_of_cells_a}\\
                    &= 958
                    \frac{N_\mathrm{core,res}}{8}
                     \left(\frac{\alpha_\mathrm{vir,box}}{2} \right)^{-1/2}
                    \frac{\mathcal{M}_\mathrm{3D}}{10}
                    \left( \frac{\overline{\rho}_\mathrm{max}}{200\rho_0} \right)^{1/2},\label{eq:minimum_number_of_cells_b}
\end{align}
\end{subequations}
where for \cref{eq:minimum_number_of_cells_b} we used \cref{eq:lbox}.

\cref{eq:Mmin_given_rhomax} and \cref{eq:minimum_number_of_cells} show that at higher $\overline{\rho}_\mathrm{max}$, cores at lower mass can be resolved, but only at the expense of an increasingly costly simulation, given that the total computational cost scales $\propto N_\mathrm{min}^4$ in the ideal case.
In \cref{eq:minimum_number_of_cells_b}, there is both an explicit dependence of numerical resolution $N_\mathrm{min}$ on $\mathcal{M}_\mathrm{3D}$ (from assuming the parent cloud is self-gravitating), and an implicit dependence through $\rho_\mathrm{max}$, because higher Mach number turbulence produces denser gas.
However, it is not necessary to increase the resolution to allow for arbitrarily large $\bar{\rho}_\mathrm{max}$, because the amount of mass actually present at extremely high densities would be vanishingly small.
In practice, combining the expected abundance of high-density gas with limits on computational resources provides a constraint on what values of $\mathcal{M}_\mathrm{3D}$ may be studied.
We quantify this next.  

\subsubsection{Resolution requirement for cores in turbulent clouds}

It is empirically known from turbulent, isothermal numerical simulations that the logarithm of gas density approximately follows a normal distribution in which the mean is related to the variance by
\begin{equation}\label{eq:lognormal_mean}
  \mu = \pm \sigma_\rho^2/2,
\end{equation}
where plus and minus signs correspond to mass- and volume-weighting, respectively (e.g., \citealt{vazquez-semadeni94,padoan97,ostriker99,klessen00pdf}; see \citealt{rabatin23a} for deviations from the log-normal form).
The measured variance increases with the amplitude of turbulence \citep[and is somewhat modulated by magnetic fields, e.g.][]{ostriker01,lemaster08}.  The functional form for quasi-steady driven turbulence proposed by \citet{padoan97}, 
\begin{equation}\label{eq:lognormal_var}
  \sigma_\rho^2 = \ln(1 + b^2\mathcal{M}_\mathrm{3D}^2)
\end{equation}
with $b\sim 1/2$, has been  
found to be consistent with many simulations \citep[e.g.][]{kritsuk07,2019ApJ...881..155P}.   
In detail, the fitting parameter $b$ has been found to be related to the dimensionality and compressiveness of turbulence \citep{federrath08}.

We assume the density \gls{PDF} prior to the core formation follows a log-normal distribution with the mean $\mu$ and variance $\sigma_\rho^2$ given by \cref{eq:lognormal_mean} (the positive sign) and \cref{eq:lognormal_var}, respectively.
Here we adopt $b=0.4$ appropriate for a natural mixture of the compressive and solenoidal modes in our initial conditions \citep{federrath10}.
We then consider the density $\rho_a$, defined such that a fraction $a$ ($0 < a < 1$) of the total mass is distributed below $\rho_a$.
This is given by 
\begin{equation}\label{eq:percentile_density}
  \rho_a = \rho_0\exp[\mu + \sqrt{2}\sigma_\rho \mathrm{erfc}^{-1}(2-2a)].
\end{equation}
Here, $\mathrm{erfc}^{-1}$ is the inverse of the complementary error function $\mathrm{erfc}(z) \equiv 2\pi^{-1/2}\int_z^\infty e^{-t^2} dt$.

\begin{figure}[tpb]
  \epsscale{1.1}
  \plotone{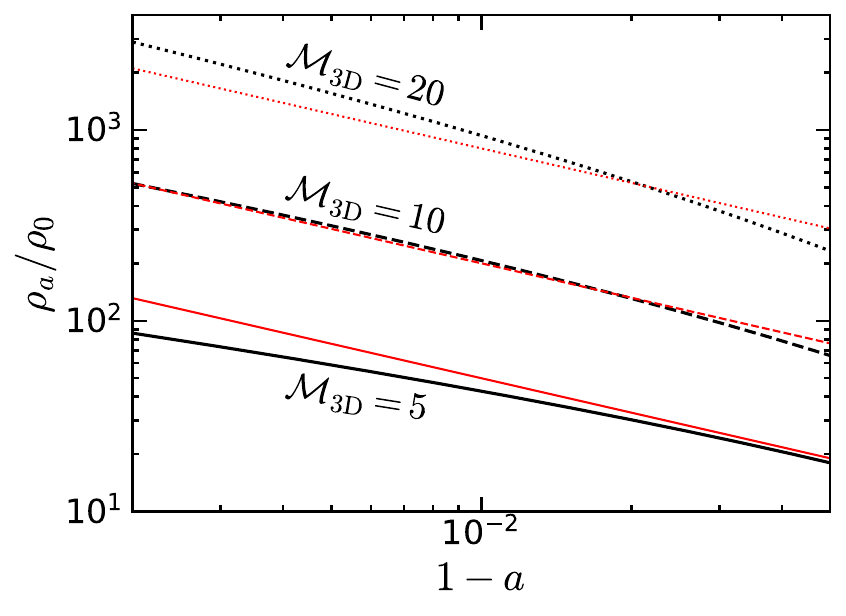}
  \caption{Density contrast such that only a fraction $1-a$ of the mass is above $\rho_a/\rho_0$, for a lognormal \gls{PDF} following \cref{eq:lognormal_mean,eq:lognormal_var}, with selected Mach numbers $\mathcal{M}_\mathrm{3D} = 5$ (black solid), $10$ (black dashed), and $20$ (black dotted).
  The corresponding red lines show the power-law approximation given in \cref{eq:rhoa_approx}.
  }
  \label{fig:rhoa}
\end{figure}

If we set $\overline{\rho}_\mathrm{max} = \rho_a$ in \cref{eq:resolution_requirement,eq:Mmin,eq:minimum_number_of_cells}, we can expect to resolve cores that form in all but a fraction $1-a$ of the mass within the simulation.  
\cref{fig:rhoa} plots $\rho_a/\rho_0$ as a function of $(1-a)$ for $\mathcal{M}_\mathrm{3D}=5$, $10$, and $20$.
If we consider the $\mathcal{M}_\mathrm{3D}=10$ curve, it shows that e.g. $a=97\%$, 99\%, or 99.7\% of the mass will be below $\rho_a/\rho_0 = 98, 207, 421$, respectively.
To resolve most of the star formation in terms of mass, $a$ must be set sufficiently high such that the ``unresolved'' mass fraction $1-a$ is much smaller than the expected net \gls{SFE} of a cloud.
If we target a value of $a \gtrsim 0.99$, which is sufficiently high for Taurus-like \glspl{GMC} having $\mathrm{SFE} \sim 5-10\%$ \citep[e.g.][]{evans09}, the use of \cref{eq:percentile_density} for $\bar{\rho}_\mathrm{max}=\rho_a$ in \cref{eq:resolution_requirement} and \cref{eq:minimum_number_of_cells} yields a Mach number dependent constraint on the required numerical resolution, such that critical cores with masses as small as that in \cref{eq:Mmin} will be resolved.
We note that while $a \sim 0.99$ is high enough to resolve most core formation \emph{by mass}, it may not be sufficient to resolve numerous low-mass cores forming below the characteristic mass of the system.
We return to this issue when we discuss the \gls{CMF} and related resolution criteria in \citetalias{paperII}.

\begin{figure}[tpb]
  \epsscale{1.1}
  \plotone{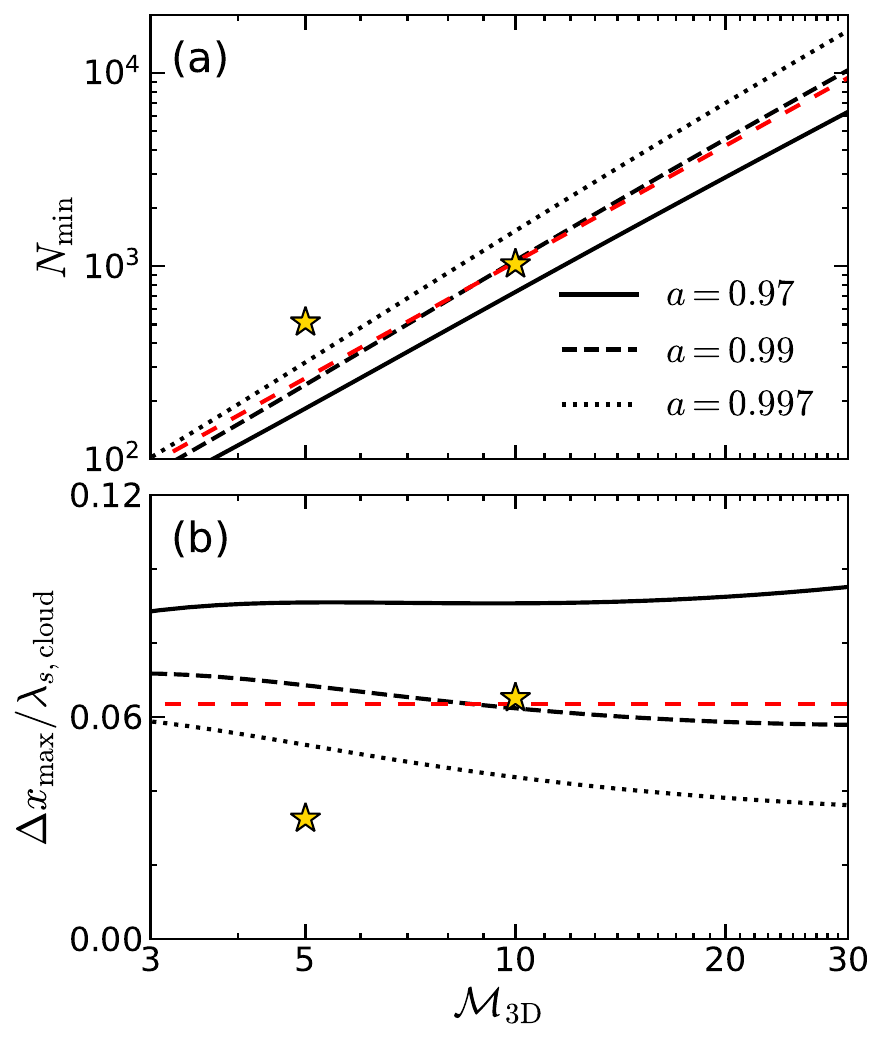}
  \caption{Numerical requirements for resolving core formation and the sonic scale. (a) Required number of cells (per dimension) $N_\mathrm{min}$ to resolve $R_\mathrm{BE}$ as a function of the box-scale Mach number $\mathcal{M}_\mathrm{3D}$.  We apply \cref{eq:minimum_number_of_cells}, and assume the core is at density $\overline{\rho}_\mathrm{max} = \rho_a$, with $a = 0.97$ (solid line), $0.99$ (dashed line), and $0.997$ (dotted line) corresponding to the fraction of mass expected to have $\rho<\rho_a$ for a lognormal distribution.
  The red dashed line plots \cref{eq:Nmin_99}, which shows $N_\mathrm{min} \propto \mathcal{M}_\mathrm{3D}^2$ approximately holds.
  (b) The ratio of the maximum allowed cell size for resolving self-gravitating cores $\Delta x_\mathrm{max}$ (\cref{eq:resolution_requirement}) to the sonic scale $\lambda_{s,\mathrm{cloud}}$ (\cref{eq:rs0}) as a function of $\mathcal{M}_\mathrm{3D}$ for  $\rho_\mathrm{max}$ as in panel (a).
  Red line uses \cref{eq:dxmax_99}.  
  In both panels, we adopt
  $\alpha_\mathrm{vir,box} = 1.66$
  and $N_\mathrm{core,res} = 8$.
The star symbols mark the position of models \texttt{M5} and \texttt{M10}.
  We note that to keep $\alpha_\mathrm{vir,box}$ constant, the box size must be scaled in proportion with $\mathcal{M}_\mathrm{3D}$ according to \cref{eq:lbox}.
}
  \label{fig:resolution}
\end{figure}

\cref{fig:resolution}(a) plots the required simulation resolution $N_\mathrm{min}$ as a function of $\mathcal{M}_\mathrm{3D}$ using $N_\mathrm{core,res} = 8$ and $\overline{\rho}_\mathrm{max} = \rho_a$ for $a=0.97$, $0.99$, and $0.997$, adopting  $\alpha_\mathrm{vir,box}=1.66$ in \cref{eq:minimum_number_of_cells_b}.
This shows that the required number of cells steeply increases with the cloud-scale Mach number approximately as $N_\mathrm{min}\propto \mathcal{M}_\mathrm{3D}^2$ for a fixed value of $a$.

For the ranges of $\mathcal{M}_\mathrm{3D}$ and $1-a$ shown in \cref{fig:rhoa}, we find an simple estimate $\rho_a \approx \widetilde{\rho}_a$ approximately holds, where
\begin{equation}\label{eq:rhoa_approx}
    \widetilde{\rho}_a \equiv 200\rho_0 \left(\frac{\mathcal{M}_\mathrm{3D}}{10}\right)^2\left(\frac{1-a}{0.01}\right)^{-0.6}
\end{equation}
is shown by red lines in \cref{fig:rhoa}.
For $a = 0.99$, $\widetilde{\rho}_{0.99} = 2\mathcal{M}_\mathrm{3D}^2\rho_0$ is a factor of two larger than the characteristic post-shock density for an isothermal shock at the large-scale \gls{RMS} Mach number of the cloud.

When we evaluate \cref{eq:resolution_requirement} and \cref{eq:minimum_number_of_cells_b} at $\overline{\rho}_\mathrm{max} = \widetilde{\rho}_{0.99}$, we obtain convenient rules for the numerical requirements to resolve cores forming in 99\% of the mass:
\begin{equation}\label{eq:dxmax_99}
\begin{split}
  \Delta x_\mathrm{max}^{[0.99]}&\equiv \frac{R_\mathrm{BE}(\widetilde{\rho}_{0.99})}{N_\mathrm{core,res}}\\ 
  &= 3.80\times 10^{-3}L_{J,0} \left(\frac{N_\mathrm{core,res}}{8}\right)^{-1}  \left( \frac{\mathcal{M}_\mathrm{3D}}{10} \right)^{-1}
\end{split}
\end{equation}
and
\begin{equation}\label{eq:Nmin_99}
\begin{split}
  N_\mathrm{min}^{[0.99]}
  &\equiv \frac{L_\mathrm{box}}{\Delta x_\mathrm{max}^{[0.99]}} \\
  &= 958 \left(\frac{N_\mathrm{core,res}}{8}\right)
\left( \frac{\alpha_\mathrm{vir,box}}{2} \right)^{-1/2}
\left( \frac{\mathcal{M}_\mathrm{3D}}{10} \right)^2
\end{split}
\end{equation}
respectively, which are shown in \cref{fig:resolution} as red dashed lines.

The results shown in \cref{fig:resolution} are based on requirements to resolve $R_\mathrm{BE}$ by $N_\mathrm{core,res}=8$ cells.  If a less restrictive resolution requirement were adopted for cores, e.g.~relaxing from $N_\mathrm{core,res}=8$ to $4$, $N_\mathrm{min}$ would shift down and $\Delta x_\mathrm{max}$ would shift up by the same factor.
We note that the effective mass resolution requirement using $\widetilde{\rho}_{0.99} = 2\mathcal{M}_\mathrm{3D}^2\rho_0$ for density is 
\begin{equation}\label{eq:dmmax_99}
\begin{split}
    \Delta m_\mathrm{max}^{[0.99]} &= \frac{3M_\mathrm{BE}(\widetilde{\rho}_{0.99})}{4\pi N_\mathrm{core,res}^3}\\
    &= 1.10\times 10^{-5}M_{J,0}\left(\frac{N_\mathrm{core,res}}{8}\right)^{-3}  \left( \frac{\mathcal{M}_\mathrm{3D}}{10} \right)^{-1},
\end{split}
\end{equation}
We also note that while \cref{eq:dxmax_99,eq:Nmin_99,eq:dmmax_99} use an approximate percentile density $\widetilde{\rho}_a$ (\cref{eq:rhoa_approx}) and assume $\overline{\rho}_\mathrm{max} = \widetilde{\rho}_{0.99}$, exact numerical results that are valid for any Mach number and $a$ can be readily obtained by setting $\overline{\rho}_\mathrm{max} = \rho_a$ using \cref{eq:percentile_density}, as shown in \cref{fig:resolution}.

\subsubsection{Resolving diffuse, turbulent gas}
Another important physical scale in turbulent systems is the sonic scale at which the characteristic turbulent velocity differences are expected to become subsonic.
If we take the cloud-scale velocity dispersion $\sigma_\mathrm{1D,box} = c_s\mathcal{M}_\mathrm{3D}/\sqrt{3}$ at the box radius $L_\mathrm{box}/2$ and assume $p=0.5$ scaling, the cloud-average sonic scale from \cref{eq:sigma1d} is
\begin{equation}\label{eq:lmb_sonic}
\begin{split}
    \lambda_{s,\mathrm{cloud}} &= \frac{3}{2}\frac{L_\mathrm{box}}{\mathcal{M}_\mathrm{3D}^2}\\
    &= 5.46\times 10^{-2}L_{J,0}\left( \frac{\alpha_\mathrm{vir,box}}{2} \right)^{-1/2} \left( \frac{\mathcal{M}_\mathrm{3D}}{10} \right)^{-1},
\end{split}
\end{equation}
where we use \cref{eq:lbox} for the second expression.
One can also use \cref{eq:lambda_s_to_rs} with $p=0.5$\REV{ and $\eta_d=1$ for uniform density} to define the cloud-average sonic radius:
\begin{equation}\label{eq:rs0}
    r_{s,\mathrm{cloud}} = \frac{3}{4}\lambda_{s,\mathrm{cloud}} = \frac{9}{8}\frac{L_\mathrm{box}}{\mathcal{M}_\mathrm{3D}^2}.
\end{equation}
At scales well below $\lambda_{s,\mathrm{cloud}}$, we expect substructure to be modest, i.e. density perturbations will be well below order unity.
But by the same token, $\lambda_{s,\mathrm{cloud}}$ should be resolved by several cells since we expect non-negligible substructure to be produced by trans-sonic motions at that scale.

\cref{fig:resolution}(b) plots the ratio of $\Delta x_\mathrm{max} = L_\mathrm{box}/N_\mathrm{min}$ to $\lambda_{s,\mathrm{cloud}}$, for the same choices of $\rho_\mathrm{max}$ as described above, $\alpha_\mathrm{vir,box} = 1.66$, and $N_\mathrm{core,res}=8$.
We note that since both $\lambda_{s,\mathrm{cloud}}$ and $\Delta x_\mathrm{max}$ vary with Mach number approximately $ \propto \mathcal{M}_\mathrm{3D}^{-1}$, their ratio is nearly constant.  \cref{fig:resolution}(b) demonstrates that with our standard choice of $N_\mathrm{core,res} = 8$, the sonic scale is resolved with more than $15$ cells when the core resolution criterion for a fraction $a \gtrsim 0.99$ of the mass is satisfied.

It is worth emphasizing the implication of \cref{fig:resolution}(b) for simulations of self-gravitating clouds: nearly the same spatial resolution is required to capture structure created by turbulence in moderate-density gas as by self-gravity in high-density gas.
Quantitatively, the ratio $\lambda_{s,\mathrm{cloud}}/R_\mathrm{BE}= (4/3)r_{s,\mathrm{cloud}}/R_\mathrm{BE}  = 2.0$ for $\bar{\rho}=\widetilde{\rho}_{0.99}$ if we adopt $\alpha_\mathrm{vir,box}=1.66$.
That is, resolving the sonic scale in ambient gas presents almost the same challenge numerically as resolving dense, self-gravitating structures that develop.
For example, if we wish to resolve $\lambda_{s,\mathrm{cloud}}$ by at least $N_\mathrm{s,res}$ elements, the required minimum number of cells per dimension would be
\begin{equation}\label{eq:nmin_sonic}
\begin{split}
    N_{\mathrm{min}} &= \frac{2}{3}N_\mathrm{s,res} \mathcal{M}_\mathrm{3D}^2\\
    &= 1067\left(\frac{N_{s,\mathrm{res}}}{16}\right)\left(\frac{\mathcal{M}_\mathrm{3D}}{10}\right)^2,
\end{split}
\end{equation}
which may be compared to \cref{eq:Nmin_99}.
Thus, for simulations with Mach numbers $\mathcal{M}_\mathrm{3D} \gtrsim 10$ comparable to those in observed GMCs, a root grid of at least $\sim 1024^3$ (or refinement based on the local turbulence level) would be required in order to resolve the structure that is created by turbulence, even before any self-gravitating cores form.  Moreover, as we shall show in \cref{sec:discussion_resolution},  local variations in $\lambda_{s}$ to values $\sim 0.1\lambda_{s,\mathrm{cloud}}$ imply that an order of magnitude larger grid than that given in \cref{eq:nmin_sonic} would actually be required to resolve turbulence essentially everywhere, although $N_{s,\mathrm{res}} \sim 16$ (corresponding to $N_\mathrm{core,res}\sim 8$ in \cref{eq:Nmin_99})  resolves $\lambda_s$ in $\sim 70\%$  of gas.      

\subsubsection{Resolution choices for this work}
In this work, we use $N=512$ and $1024$ for models \texttt{M5} (with $L_\mathrm{box}=2 L_{J,0}$) and \texttt{M10} (with $L_\mathrm{box}=4 L_{J,0}$), respectively.
If we require $N_\mathrm{core,res} = 8$, the corresponding density percentiles at which $N = N_\mathrm{min}$ are $1-a = 2.33\times 10^{-4}$ and $1.15\times 10^{-2}$ for models \texttt{M5} and \texttt{M10}.
This results in $\Delta x = 3.91 \times 10^{-3} L_{J,0}$, $\overline{\rho}_\mathrm{max} =189\rho_0$, and $M_\mathrm{min} = 2.42\times 10^{-2} M_{J,0}$ for both models.
With the adopted values of $\Delta x$ in our simulations, if we relaxed the core resolution criterion from $R_\mathrm{BE}(\bar{\rho}_\mathrm{max})/\Delta x_\mathrm{max}\equiv N_\mathrm{core,res}=8$ to $N_\mathrm{core,res}=4$, cores with $\bar{\rho}_\mathrm{max}/\rho_0 =758$ and $M_\mathrm{min}=1.21\times 10^{-2} M_{J,0}$ would be considered resolved; this density corresponds to $1-a= 2.19\times 10^{-6}$ and $9.76\times 10^{-4}$ for \texttt{M5} and \texttt{M10}, respectively. 
This means that there is an upper limit on the density of critical cores that we are able to resolve, in the range $\bar{\rho}_\mathrm{max}/\rho_0 \sim 10^2-10^3$, and a corresponding lower limit on core mass.

It is worth noting that when the Jeans criterion $L_J/\Delta x=N_J$ is applied at the central density of a critical BE sphere, it translates to $N_\mathrm{core,res}=[R_\mathrm{BE}(\bar{\rho})/L_J(\rho_c)][L_J(\rho_c)/\Delta x] = 1.03N_J$, where we have used \cref{eq:ljeans,eq:rbe} and $\rho_c/\bar\rho =5.7$.
Thus, our more ``relaxed'' resolution criterion is comparable to what would be considered resolved under the original \citet{truelove97} criterion of $N_J = 4$.

\REV{We remind the reader that, because our simulations adopt uniform resolution, our default choice of $N_\mathrm{core,res} = 8$ in \cref{eq:resolvedness_criterion} does not imply that $r_\mathrm{crit}$ is resolved with only $8$ cells for every core; most cores are resolved with more than $8$ cells, and only the smallest ones reach this limit.
Nevertheless, attaining strict numerical convergence for all cores may require even higher $N_\mathrm{core,res}$, because numerical dissipation can affect the turbulent cascade at scales much larger than $\Delta x$, depending on specific numerical methods adopted \citep{porter94,kritsuk07,federrath11}.
Using even more conservative values of $N_\mathrm{core,res}>8$ would shift the curves in \cref{fig:resolution}(a) and (b) upward and downward, respectively, by a factor $N_\mathrm{core,res}/8$.
As we shall discuss in \cref{sec:discussion_resolution}, this more stringent requirement is comparable to the resolution needed to resolve the sonic scale everywhere.
}

In \cref{tb:models}, we include the numerical parameters adopted for our simulations.
Columns (5) and (6) give $N$ and $\Delta x/L_{J,0}$.
Columns (7) and (8) give $\overline{\rho}_\mathrm{max}/\rho_0$ and $M_\mathrm{min}/M_{J,0}$ adopting $N_\mathrm{core,res}=8$.
Column (9) gives the mass fraction above the maximum resolvable density $\overline{\rho}_\mathrm{max}$, which is obtained by setting $\rho_a = \overline{\rho}_\mathrm{max}$ in \cref{eq:percentile_density}.
For each model, we run a number of simulations with different random seed $n_\mathrm{seed}$ to generate various statistical realizations of $\delta \mathbf{v}_0$.
Column (10) gives the number of simulations performed for each model.

\subsection{Sink Particles}\label{sec:sink_particles}

Gravitational collapse of isothermal flows is a runaway process that cannot be followed to later stages at fixed numerical resolution.
To overcome this difficulty and extend the simulation run time beyond the collapse of the first core, we implement a sink particle algorithm in \textit{Athena++}, based on the method described in \citet{gong13} with some modifications.
We note that the main focus of this work is the evolution of individual cores \emph{before} sink particle formation, and therefore most of the results are insensitive to the details of the sink particle implementation.

We create a sink particle when the following two conditions are met simultaneously: 1) the gas density $\rho$ exceeds a threshold $\rho_\mathrm{thr}$ and 2) the cell is at the local potential minimum.
Our choice of the density threshold is physically motivated by numerical studies of isothermal collapse, which have shown that the inner part of the gravitationally collapsing region approaches the asymptotic \gls{LP} solution \citep{larson69,penston69},
\begin{equation}\label{eq:rho_LP}
  \rho_\mathrm{LP}(r) = \frac{8.86 c_s^2}{4\pi G r^2},
\end{equation}
for a wide range of initial and boundary conditions \citep{bodenheimer68,larson69,penston69,hunter77,foster93,ogino99,hennebelle03,motoyama03,vorobyov05,gomez07,burkert09,gong09,gong11,gong15}.
Following \citet{gong13}, we take as our density threshold
\begin{equation}
  \rho_\mathrm{thr} = \rho_\mathrm{LP}(0.5\Delta x) = \frac{8.86}{\pi}\frac{c_s^2}{G\Delta x^2}.
\end{equation}
For our adopted resolution, the threshold density is high enough, $\rho_\mathrm{thr}/\rho_0 \sim 6\times 10^4 \gg \mathcal{M}_\mathrm{3D}^2$, such that random turbulent compression alone almost never leads to $\rho > \rho_\mathrm{thr}$ without involving gravitational collapse.
Although disks forming around sink particles can trigger $\rho > \rho_\mathrm{thr}$ at late times, the second condition of the local potential minimum would prevent sink particle creation unless such an event truly results from a gravitational disk fragmentation.
Once created, the position and velocity of the sink particles are updated using a drift-kick-drift variant of the leap-frog integrator that conserves the total momentum of gas and sink particles when coupled with the \gls{VL2} integrator (C.-G. Kim et al. \emph{in prep.})

Due to the gravitational attraction of a sink particle, the region around it accretes mass and momentum from the surrounding gas while the particle moves through the domain, and we use this to update the particle's mass and momentum.
We take a conservative approach of resetting the cubic control volume consisting of $27$ cells centered on the particle-containing cell at every timestep, using the average values taken from the outer adjacent cells sharing common faces \citep{cgkim20}.
This is equivalent to treating the control volume as ghost zones and applying outflow boundary conditions.
The change in the mass contained in the control volume due to the reset procedure, $\Delta M_\mathrm{reset}$, is then conservatively dumped to the sink particle, such that
\begin{equation}\label{eq:dmsink1}
  \Delta M_\mathrm{sink} = -\Delta M_\mathrm{reset}.
\end{equation}
Because the density generally increases toward the accreting sink particles, in most cases $\Delta M_\mathrm{reset} < 0$, leading to positive accretion onto the particle.
At late times, however, the flow around sink particles may depart from simple spherical accretion due to, e.g., disk or binary formation, and become chaotic, which can sometimes lead to $\Delta M_\mathrm{reset} > 0$.
Because $\Delta M_\mathrm{sink} < 0$ would be unphysical, in this case we do not update the sink particle mass and restore the fluid variables in the control volume to their original values before the reset.
Noting that the change in the total mass in the control volume, $\Delta M_\mathrm{ctrl}$ is caused by the mass flux through the control volume boundary as well as the reset procedure, \cref{eq:dmsink1} can be equivalently written as
\begin{equation}\label{eq:dmsink2}
  \Delta M_\mathrm{sink} = \Delta t \oint \mathbf{F}_\rho \cdot d \mathbf{A} - \Delta M_\mathrm{ctrl},
\end{equation}
where $\mathbf{F}_\rho$ is the mass flux returned by the Riemann solver averaged over the timestep.
\cref{eq:dmsink2} indicates that the accretion rate of the sink particle is determined by the mass flux into the control volume, modulated by the rate of change of the total mass in the control volume.
For steady flows, $\Delta M_\mathrm{ctrl} = 0$ and the accretion rate becomes identical to the mass flux.
\cref{eq:dmsink2} is by construction mass-conservative because the mass entering the control volume is distributed into the sink particle and the control volume.
We apply the same method of mass accretion described above to make sink particles accrete gas momentum as well.

When the control volumes of two sink particles overlap each other, we merge them into a single particle created at the center of mass of the two merging particles, with the total mass and momentum being conserved.
To verify that our sink particle implementation is correct, we repeat the test suites of the two-particle orbit, self-similar accretion of \citet{shu77}, and the Galilean invariance of accretion presented in \citet[Section 3.1, 3.2, 3.3]{gong13}.

Unlike in real \glspl{GMC}, there are no internal or external agents that can halt the star formation process in our simulations, and the sink particles could accrete indefinitely until they consume all the gas within the domain.
Because this would not be consistent with the observed low \glspl{SFE} in molecular clouds \citep[$\sim 1\text{--}10\%$, e.g.,][]{williams97,evans09}, we terminate the simulation when the total mass in sink particles reaches $15\%$ of the initial gas mass, i.e., $\mathrm{SFE} \equiv M_\mathrm{sink}/M_\mathrm{box} = 0.15$.
Column (10) of \cref{tb:models} gives the median and standard deviation of the termination time $t_\mathrm{final}$, in units of $t_\mathrm{ff,0}$ (see \cref{eq:tff0} for conversion to physical units), where the median is taken over different realizations of the initial velocity field.

\begin{figure}[htpb]
  \epsscale{1.1}
  \plotone{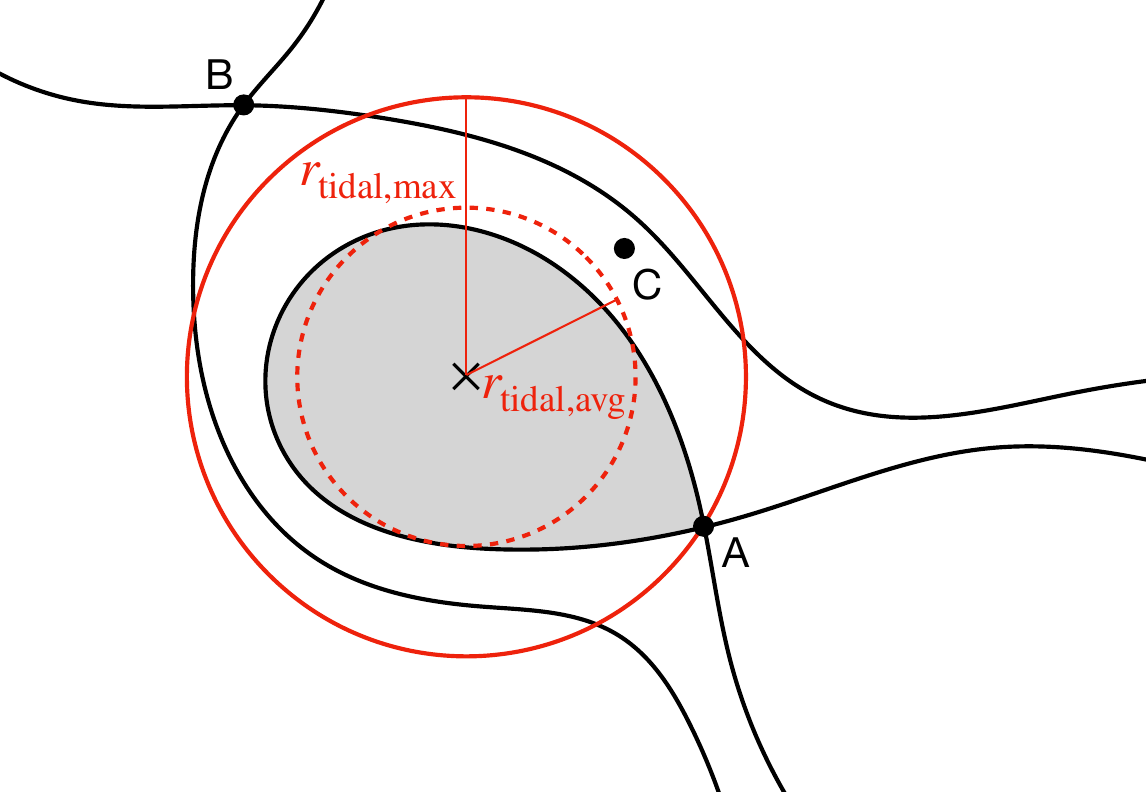}
  \caption{Schematic of the gravitational potential geography around a core.
  The black cross symbol marks the minimum of $\Phi$ corresponding to the core center, and the black solid lines draw the contours (isosurfaces in 3D) of $\Phi$.
  Points A and B are saddle points of the gravitational potential field.
  The distance to the nearest saddle point (here, point A) defines the maximum tidal radius $r_\mathrm{tidal,max}$, which sets the largest sphere (red solid circle) that can be considered as belonging to the ``core'' in our angle-averaged analysis.
  The gray shaded region corresponds to the ``leaf'' of the dendrogram of $\Phi$, whose volume $\mathcal{V}_\mathrm{leaf}$ defines the average tidal radius $r_\mathrm{tidal,avg} \equiv [3 \mathcal{V}_\mathrm{leaf} / (4\pi)]^{1/3}$ (marked with a red dashed circle).
  Point C is situated in no man's land and would be considered as a part of the core based on $r_\mathrm{tidal,max}$ but not based on $r_\mathrm{tidal,avg}$.}
  \label{fig:contours_schematics}
\end{figure}

\begin{figure*}[htpb]
  \plotone{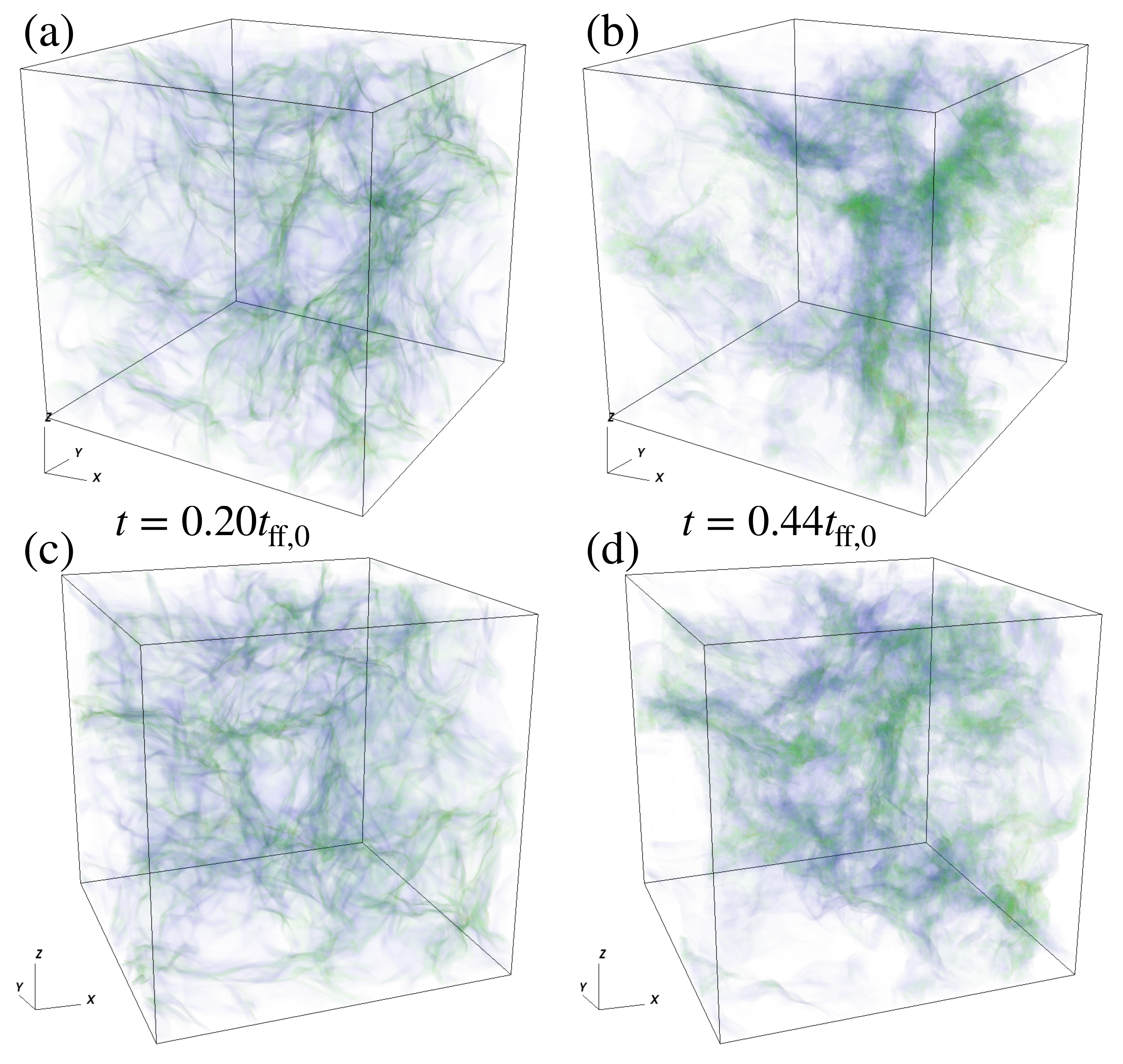}
  \caption{Volume rendering of the three-dimensional density structure of model \texttt{M10} run with $n_\mathrm{seed} = 0$.
  Panels (a) and (c) are taken at the same instant $t = 0.06 t_{J,0} = 0.20 t_\mathrm{ff,0}$, but viewed from different directions.
  Panels (b) and (d) are taken at a later time $t = 0.135 t_{J,0} = 0.44 t_\mathrm{ff,0}$.}
  \label{fig:volume}
\end{figure*}

\begin{figure*}[htpb]
  \epsscale{0.9}
  \plotone{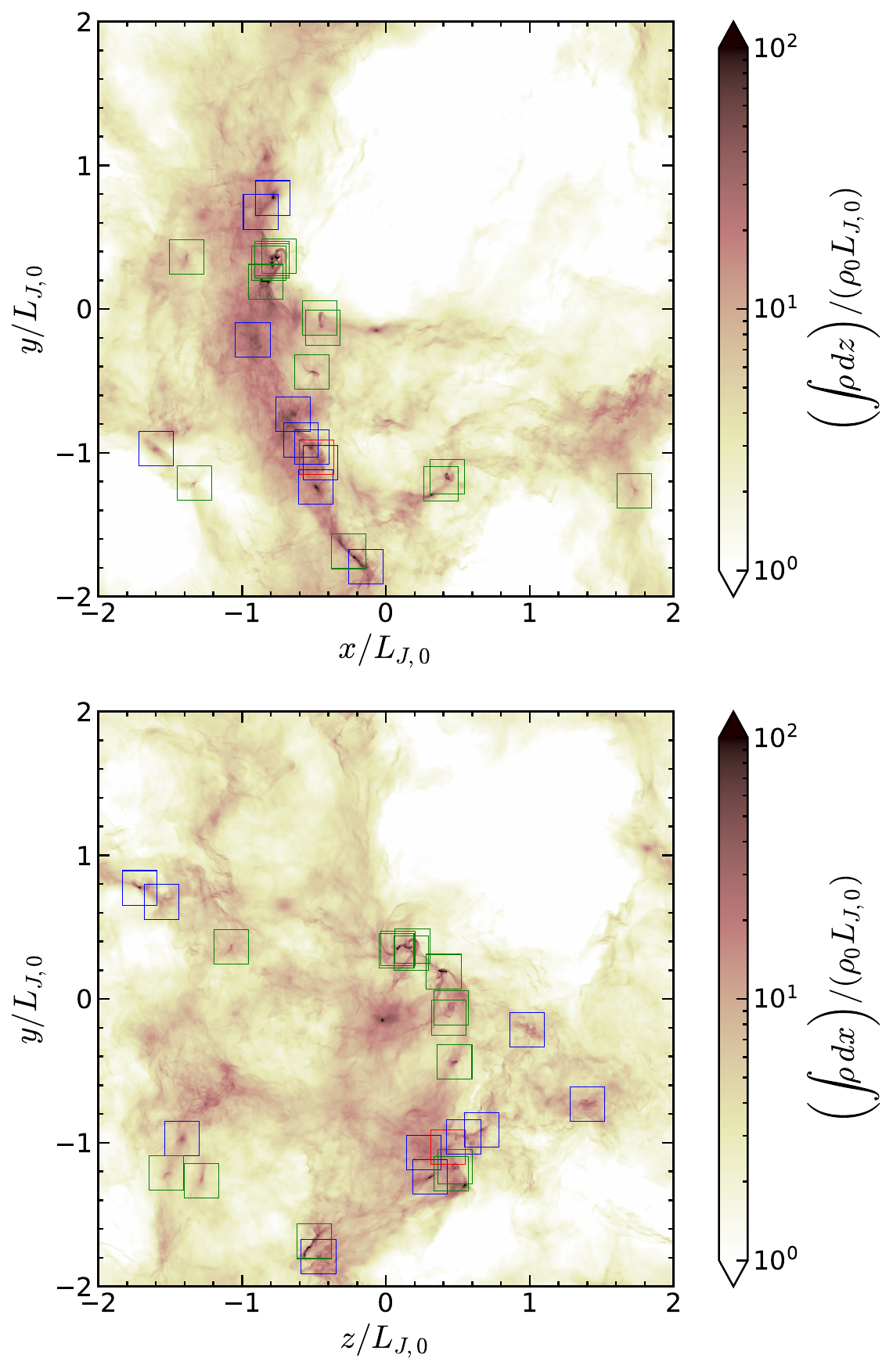}
  \caption{Gas surface density maps of model \texttt{M10} (run with $n_\mathrm{seed} = 1$) at $t = 0.267t_{J,0} = 0.872t_\mathrm{ff,0}$, projected along the $z$ (top) and $x$ (bottom) directions.
  Locations of prestellar and protostellar cores \REV{satisfying $N_\mathrm{res,core} = 4$} are highlighted with blue and green squares, respectively, where in the simulation this is equivalent to before and after sink particle formation.
  The red square marks the location of the selected prestellar core whose evolution is illustrated in \cref{fig:evolution_projected_map,fig:evolution_radial_profiles}.
  Comparison of the top and bottom panels reveals that the apparent large-scale filament seen in the top panel extending from $y = -2L_{J,0}$ to $y=L_{J,0}$ is in fact a sheet-like structure seen in projection, rather than a genuine three-dimensional filament.}
  \label{fig:projection_global}
\end{figure*}

\begin{figure*}[htpb]
  \plotone{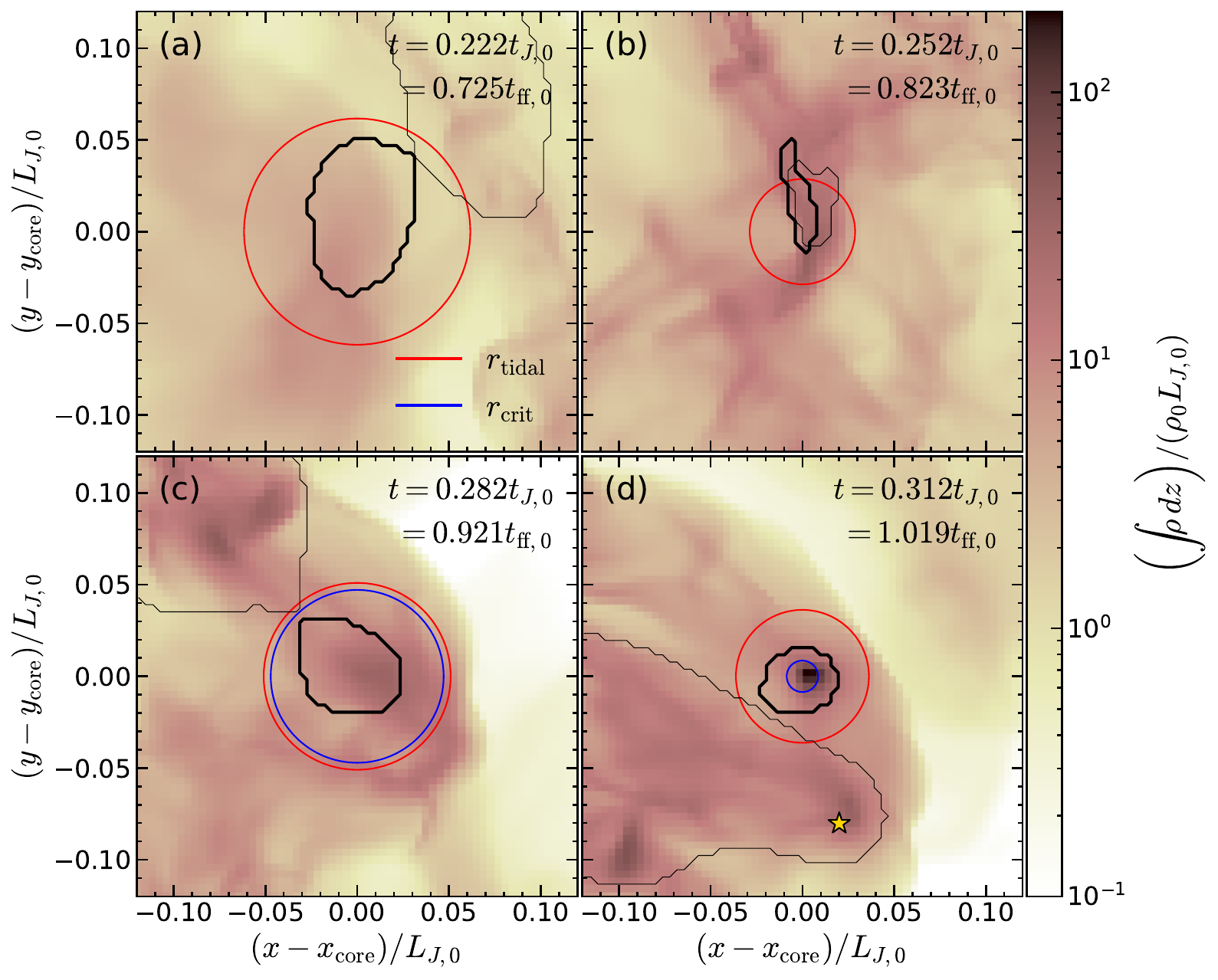}
  \caption{Zoom-in evolution of a selected core in model \texttt{M10} (highlighted with the red square in \cref{fig:projection_global}).
  The color scale shows the projected surface density along $\hat{z}$, only accounting for gas within a cube centered on the core with side length $0.24 L_{J,0}$ (identical to squares shown in \cref{fig:projection_global}).
  The panels show important epochs in physical evolution, when (a) supersonic flows (coming from the right side) are about to collide with an overdense region, (b) the flow collision leads to a dense, turbulent post-shock layer, (c) runaway collapse is initiated within the core ($t = t_\mathrm{crit}= 0.282 t_{J,0}$), and (d) a sink particle is about to form at the center (this core formed a sink particle at $t = 0.317t_{J,0}=1.04t_\mathrm{ff,0}$).
  The corresponding times in units of $t_{J,0}$ and $t_\mathrm{ff,0}$ are annotated in each panel.
  Thick and thin black lines delineate projected isosurfaces of $\Phi$ passing through the saddle point that defines $r_\mathrm{tidal,max}$ (see \cref{fig:contours_schematics} and the related text).
  The potential minimum within the thick contour defines the center of this core.
  The red and blue circles centered on the core mark $r_\mathrm{tidal}$ and $r_\mathrm{crit}$, respectively.
  The yellow star symbol in panel (d) marks the position of a sink particle from another core that collapsed earlier than the selected one.}
  \label{fig:evolution_projected_map}
\end{figure*}

\begin{figure*}[htpb]
  \plotone{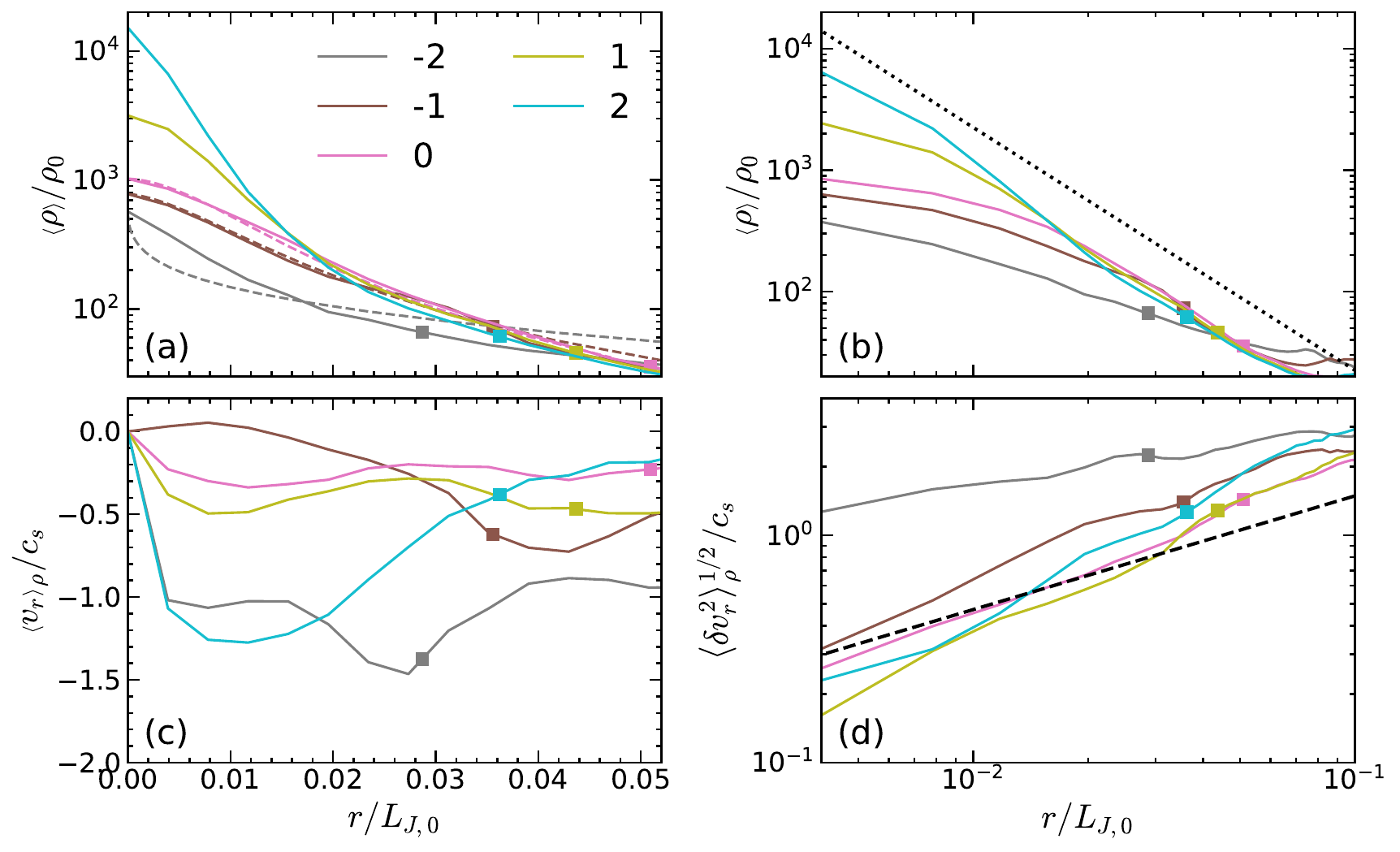}
  \caption{Time evolution of radial profiles in the selected core shown in \cref{fig:evolution_projected_map}.
  Lines with different colors represent times $t_n=t_\mathrm{crit} + n \Delta t$ at equal intervals $\Delta t = 0.015 t_{J,0}$, such that negative and positive $n$ correspond to epochs before and after the onset of the collapse, respectively.
  (a) The radial density profiles of the core directly measured from the simulation (solid lines) and of the critical \gls{TES} (dashed lines) constructed from the measured $\rho_c$, $r_s$, and $p$ for $t \le t_\mathrm{crit}$.
  (b) Similar to (a), but shown with logarithmic scale for radius.
  Dotted line plots the \gls{LP} asymptotic solution (\cref{eq:rho_LP}).
  (c) Mass-weighted average radial velocity at each radius.
  (d) Mass-weighted average radial velocity dispersion at each radius.
  Dashed line plots the average scaling $c_s(r/r_{s,\mathrm{cloud}})^{0.5}$ associated with our initial condition (see \cref{eq:rs0}).
  In all panels, square symbols mark $r = r_\mathrm{tidal,max}$ at each instant.
  Panels (a), (b), (c), (d) in \cref{fig:evolution_projected_map} correspond to $t_{-4}$, $t_{-2}$, $t_0$, and $t_2$, respectively.
  }
  \label{fig:evolution_radial_profiles}
\end{figure*}

\subsection{Reverse Core Tracking}\label{sec:core_tracking}

One of the main goals of this work is to quantitatively analyze the evolution of the prestellar cores that form in the simulations.
To accomplish this, we track the position of cores backward in time, starting from the collapse time $t_\mathrm{coll}$ at which a sink particle forms.
First, we load the snapshot immediately before $t_\mathrm{coll}$ and find the local minimum of $\Phi$ closest to the birth position of the sink, which defines the center of the prestellar core.
We then examine the gravitational potential structure around the center, schematically illustrated in \cref{fig:contours_schematics}.
As one moves further out from the center, isosurfaces of the gravitational potential form saddle points where they touch the same isosurface surrounding nearby potential minima (e.g., points A and B in \cref{fig:contours_schematics}).
We define the maximum ``tidal radius'' $r_\mathrm{tidal,max}$ as the distance to the nearest saddle point, beyond which the gravitational field is significantly affected by neighboring structures and a ``core'' cannot be considered as a single, relatively isolated entity.
We also define the average tidal radius $r_\mathrm{tidal,avg}\equiv [3 \mathcal{V}_\mathrm{leaf}/(4\pi)]^{1/3}$ where $\mathcal{V}_\mathrm{leaf}$ is the volume enclosed by the gravitational potential isosurface touching the nearest saddle point (i.e., the gray shaded region in \cref{fig:contours_schematics}).
\REV{We find that the ratio $r_\mathrm{tidal,max}/r_\mathrm{tidal,avg}$ varies from $1.5$ (10th percentile) to $2.9$ (90th percentile), with a median value $2.1$.}

Having defined the central position $\mathbf{x}_\mathrm{cen}$ and the tidal radii of the core at $t_\mathrm{coll}$, we successively load the previous snapshots to find the potential minimum that is closest to the \REV{expected position $\mathbf{x}'_\mathrm{cen} = \mathbf{x}_\mathrm{cen} - \mathbf{v}(\mathbf{x}_\mathrm{cen})\Delta t_\mathrm{out}$.}
The time step between the two output files is $\Delta t_\mathrm{out} = 0.003 t_{J,0}$, which is short enough to ensure that the distance criterion alone is sufficient to unambiguously identify previous images of a core.
Once the center is identified with the location of the potential minimum, we calculate $r_\mathrm{tidal,max}$ and $r_\mathrm{tidal,avg}$ in the same way.
We terminate the core tracking procedure when the distance between the two potential minima in the successive snapshots exceeds $r_\mathrm{tidal,max}$ (the larger among the two adjacent snapshots) which occurs when the potential well that defines $r_\mathrm{tidal,max}$ becomes too shallow and a ``core'' loses its identity.

Identifying all the saddle points and the gravitational potential isosurfaces dissected by them is equivalent to constructing a dendrogram of $\Phi$.
To accomplish this, we have developed a python package \texttt{GRID-dendro}\footnote{\url{https://sanghyukmoon.github.io/grid_dendro/intro.html}} that implements the dendrogram construction algorithm described in \citet{mao20}.
The algorithm takes the three-dimensional array $\Phi_{ijk}$ defined at Cartesian grid points and constructs a dendrogram after flattening and sorting $\Phi_{ijk}$ into a one-dimensional sequence of increasing $\Phi$.
The dendrogram construction is deterministic and involves no free parameters.
We refer the reader to \citet[Appendix A]{mao20} for a detailed description of the algorithm.
Once the dendrogram is constructed, we ``prune'' it by requiring the largest closed isosurface surrounding each potential minimum (i.e., ``leaves'' in the standard dendrogram terminology; e.g., the gray shaded region in \cref{fig:contours_schematics}) contains at least $27$ cells.
If a leaf is smaller than $27$ cells, we merge it and its ``sibling'' (i.e., the isosurface that shares a saddle point with the leaf) to their parent structure, which then becomes a new leaf.
We note that the dendrogram of the gravitational potential is less sensitive to transient density peaks compared to the density dendrogram and could provide a physically meaningful way of defining structures in simulations \citep{gong11,gong13gridcore,chen14,chen15,chen18,mao20} and observations \citep{gong13gridcore,li15}.

\section{Results}\label{sec:results}

\subsection{Overall Evolution}\label{sec:overall_evolution}

The initial supersonic turbulence creates an intricate network of filamentary and clumpy density structures (\cref{fig:volume}).
Analogous structures have been analyzed using a number of different approaches in turbulent cloud simulations as well as in observations \citep[see e.g.~review of][]{hacar23}.
At early times, the structures mostly consist of shock fronts exhibiting sharp density contrasts and dense post-shock layers behind them (\cref{fig:volume}(a), (c)), which are transient structures readily dispersed within one sound crossing time \citep{robertson18}.
As the gas experiences multiple shocks, the density \gls{PDF} approaches a log-normal distribution and the density structures become smoother (\cref{fig:volume}(b), (d)).
Comparison of the identical snapshot seen from different angles (e.g., \cref{fig:volume}(b) versus (d)) suggests the need to be careful when interpreting structures seen in projection, because physically distinct structures can happen to align along a line of sight, masquerading as a large coherent filament \citep[e.g.,][]{ostriker01,smith14,robertson18}.
\cref{fig:projection_global} also demonstrates that an apparent large-scale filament can in fact be a sheet-like structure, as in the case of the California molecular cloud \citep{rezaei22}.

Self-gravitating cores form inside the overdense structures shaped by supersonic turbulence.
To illustrate the evolution of a typical core in our simulations, in \cref{fig:evolution_projected_map} we plot the projected density distribution and the gravitational potential contours around a selected core in model \texttt{M10}, at four characteristic epochs: a) before the flow collision, b) after the flow collision, c) at the beginning of core collapse, and d) at the end of the collapse (i.e., right before the sink particle formation).
We note that the first two epochs correspond to the core building stage, while the last two to the core collapse stage, in the terminology of \citet{gong09}.

To analyze cores at different stages of evolution, we set up a local spherical coordinate system centered at the potential minimum, with the $z$-axis aligned with the angular momentum vector integrated within $r_\mathrm{tidal,max}$.
We then calculate the angle-averaged radial profiles of density $\left<\rho \right>$, velocity $\left<v_r \right>_\rho$, and velocity dispersion $\left<\delta v_r^2 \right>^{1/2}_\rho$ (see \cref{eq:volume-weighted-angle-average,eq:mass-weighted-angle-average} for the definition of the angle brackets), and plot them in \cref{fig:evolution_radial_profiles} for a few selected epochs.
For ease of description, we use the integer $n$ to label each epoch equally spaced in time, such that $t_n = t_\mathrm{crit} + n\Delta t$ for $\Delta t=0.015t_{J,0}$.
Here, $t_\mathrm{crit}$ is the time when collapse starts; a quantitative definition of $t_\mathrm{crit}$ will be given in the \cref{sec:collapse_condition}.

Initially, the core-forming region has a moderate central density $\rho_c \sim 10^2\rho_0$ and low velocity dispersion (this stage is not shown in the \cref{fig:evolution_radial_profiles}).
A supersonic flow approaches from the right and collides with the gas already in the core-forming region (between the snapshot (a) and (b) in \cref{fig:evolution_projected_map}; also manifested by large negative $\left<v_r \right>_\rho$ in \cref{fig:evolution_radial_profiles}(c) at $t_{-2}$).
The impact of the supersonic wave results in a dramatic increase in the central density and the velocity dispersion.
After the flow collision, both $\left<v_r \right>_\rho$ and $\left<\delta v_r^2 \right>^{1/2}$ decrease in time, bringing the core into a quasi-equilibrium (from $t_{-2}$ to $t_0$).
\cref{fig:evolution_projected_map}(b) and (c) illustrate that the tidal radius of the core is set by the neighboring structure that is simultaneously created by the flow collision.
As soon as the core settles into a quasi-equilibrium at the new higher density, both the infall speed and the central density start to increase in an accelerated manner (from $t_{0}$ to $t_2$), producing the highly centrally-concentrated structure seen in \cref{fig:evolution_projected_map}(d).
\cref{fig:evolution_radial_profiles}(a) shows that the increase in the central density during $t_{0}\text{--}t_{2}$ is almost an order of magnitude larger than the same time interval during $t_{-2}\text{--}t_{0}$.
As the collapse proceeds, the density profile becomes steeper, and the central part approaches the \gls{LP} solution, similar to earlier spherically symmetric, one-dimensional simulations \citep[e.g.,][]{hunter77, foster93, gong09} as well as simulations focused on core formation and evolution in post-shock layers \citep{gong11,gong15}.

\begin{figure}[htpb]
  \epsscale{1.1}
  \plotone{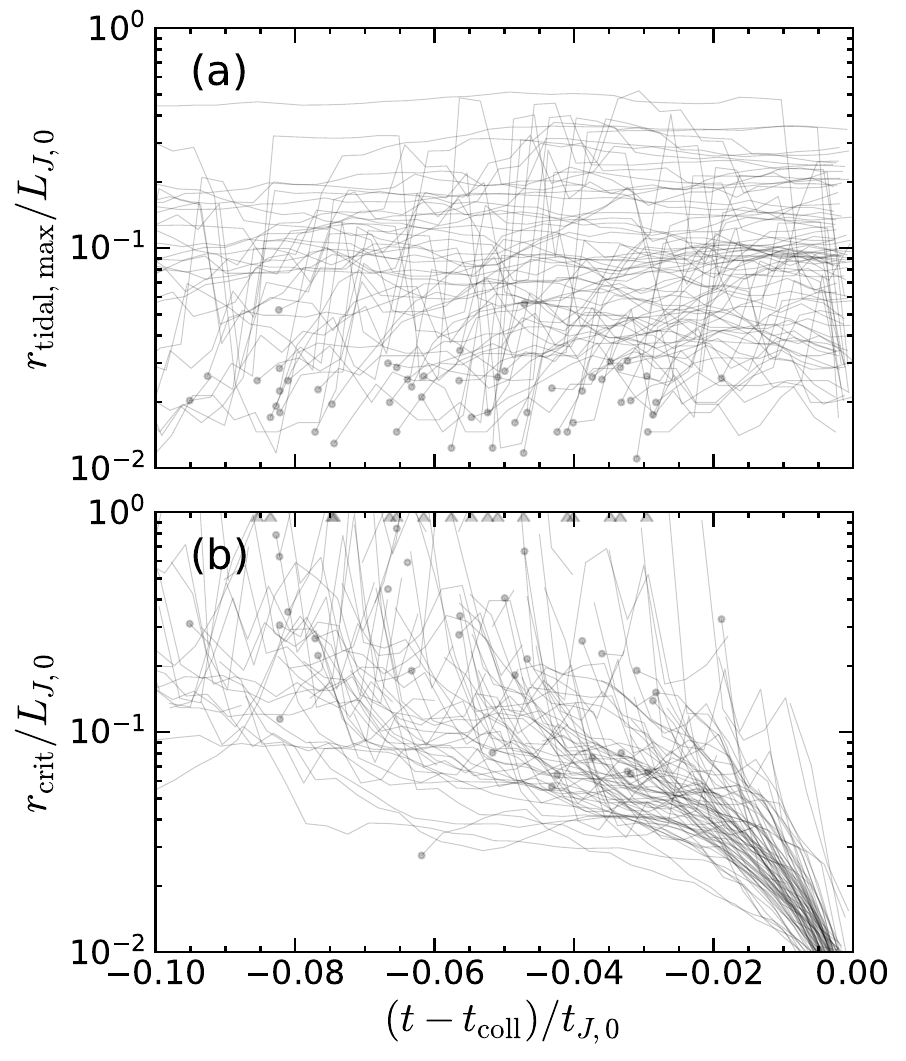}
  \caption{Temporal variations of (a) the maximum tidal radius and (b) the critical radius of cores in model \texttt{M10}.
    Each line corresponds to an individual core, where the abscissa represents the time relative to the instant of the sink particle formation.
    Circles mark the time when we terminate the reverse core tracking procedure (\cref{sec:core_tracking}).
    If $r_\mathrm{crit} = \infty$ when the core tracking is terminated, we use the upper caret symbols instead of circles in panel (b) for visualization purpose.}
  \label{fig:rcrit_rtidal}
\end{figure}

To compare the structure of our simulated cores with the \gls{TES} solutions, we measure the power-law exponent $p$ and the sonic radius $r_s$ by fitting \cref{eq:linewidth_size} to the actual profile of $\left<\delta v_r^2 \right>_\rho^{1/2}$ within $r_\mathrm{tidal,max}$.
\REV{We use these values together with the measured central density $\rho_c$ to calculate $\xi_s = r_s / r_c$ (see \cref{eq:rc,eq:xis}), and then apply the \gls{TES} model\footnote{\citet{tesphere} provides a Python pacakge implementing the \gls{TES} model. The source code is also hosted on \url{https://github.com/sanghyukmoon/tesphere}.}
to find the predicted dimensionless critical radius $\xi_\mathrm{crit}$.
The dimensional critical radius is then obtained as $r_\mathrm{crit} = r_c \xi_\mathrm{crit}$, where $r_c$ is the scale radius at a given $\rho_c$.
We solve \cref{eq:dimensional_steady_equilibrium} from $r=0$ to $r=r_\mathrm{crit}$ at each epoch prior to $t_\mathrm{crit}$ for the selected core, and the resulting \gls{TES} solutions
are overplotted in \cref{fig:evolution_radial_profiles}(a) as dashed lines.}
During the flow collision ($t=t_{-2}$), the measured density profile deviates from the \gls{TES} profile, indicating the core is undergoing dynamic compression.
As turbulence dissipates and the core approaches a quasi-equilibrium, the density profile becomes very similar to the \gls{TES} by the epoch $t_{-1}$.
The matching \gls{TES} has $\xi_s < \xi_{s,\mathrm{min}}$, however, suggesting that the core at this stage is stable everywhere (see \cref{eq:rhocmin} and related text).
As the core is further compressed due to the remaining inertia of the converging flows and the turbulent dissipation continues, $r_\mathrm{crit}$ shrinks and moves inside $r_\mathrm{tidal,max}$ by the epoch $t_0$, after which the central density and infall velocity start to increase dramatically.
We note that the measured velocity dispersions (\cref{fig:evolution_radial_profiles}(d)) are overall higher than the average scaling $c_s(r/r_{s,\mathrm{cloud}})^{0.5}$ expected from the initial condition.
This is due to a positive correlation between density and turbulent velocity.
More details will be presented in \cref{sec:linewidth_size}.

Although details differ, other cores evolve in a qualitatively similar manner to the selected core considered above: 1) cores initially form in regions where supersonic flows are converging; 2) due to low central density (and therefore low mass) and strong turbulence, cores have small $r_\mathrm{tidal,max}$ and large $r_\mathrm{crit}$ (which is often infinite) early in their formation (see \cref{fig:rcrit_rtidal}); 3) the central density gradually increases due to the converging flows while turbulence generally dissipates, such that $r_\mathrm{crit}$ decreases in time, while $r_\mathrm{tidal,max}$ grows slightly; 4) cores undergo runaway gravitational collapse roughly when $r_\mathrm{crit}$ drops below $r_\mathrm{tidal,max}$.
In \citetalias{paperII}, we will show that the structures of cores are quite consistent with \gls{TES} critical solutions at time when they initiate collapse.
In the following sections, we quantitatively investigate the aforementioned evolutionary sequence of cores.

\subsection{Definition of $t_\mathrm{crit}$ and Collapse Dynamics of Individual Cores}\label{sec:force_balance}

\begin{figure*}[htpb]
  \plotone{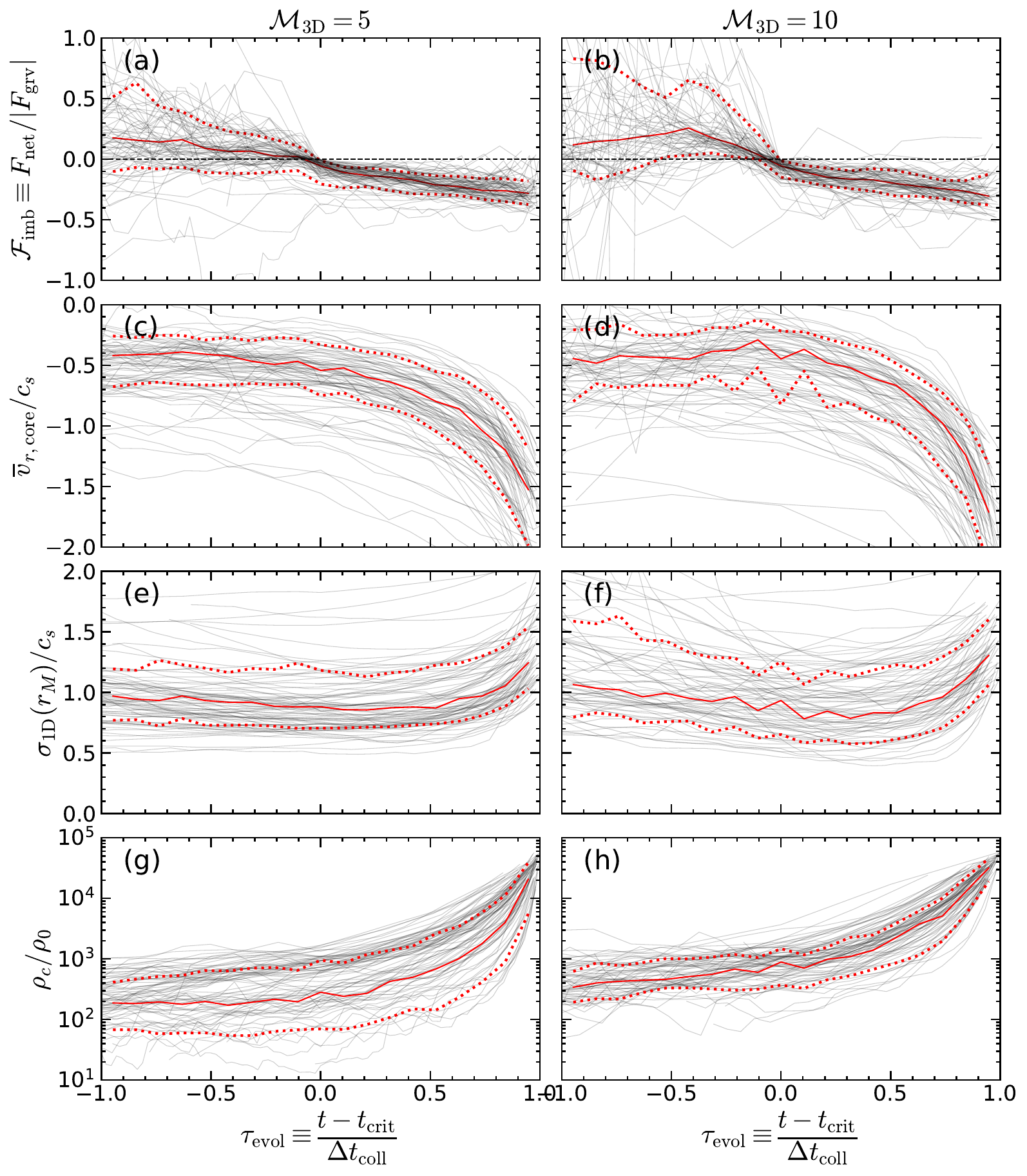}
  \caption{The evolutionary histories of individual cores in model \texttt{M5} (left column) and \texttt{M10} (right column).
    In each panel, the thin black lines are the tracks of individual cores, whereas the median and $\pm 34.1$th percentile values at each time bin are plotted by the red solid line and the red dotted lines, respectively.
    The abscissa is the normalized evolutionary time $\tau_\mathrm{evol}$ defined in \cref{eq:tau}.
    (a),(b): Fractional force imbalance $\mathcal{F}_\mathrm{imb}$ defined in \cref{eq:force_imbalance}.
    (c),(d): Mass-weighted average radial velocity $\overline{v}_{r,\mathrm{core}}$ defined in \cref{eq:vr_core}.
    Negative values indicate inflowing motion.
    (e),(f): Mass-weighted average turbulent velocity dispersion $\sigma_\mathrm{1D}$ defined in \cref{eq:sigma1d}, measured within $r_M$.
    (g),(h): Central density $\rho_c$, which approaches our sink formation threshold $\rho_\mathrm{thr} = 6\times 10^4\rho_0$ at $\tau_\mathrm{evol} = 1$.
    Some rare trajectories with large positive net forces after $t_\mathrm{crit}$ are caused by imperfect identification of $t_\mathrm{crit}$.
    We do not ``fix'' those edge cases because they do not affect the statistical results presented in this paper.
  }
  \label{fig:core_evolution}
\end{figure*}

If runaway collapse starts from radius $\sim r_\mathrm{crit}$ as predicted in \cref{sec:tes}, it is expected that the net force experienced by a core within $r_\mathrm{crit}$ will become negative when the collapse commences.
We therefore empirically identify the onset of collapse by finding the earliest time $t_\mathrm{crit}$ after which the net force integrated within $r_\mathrm{crit}$ remains negative until the end of the collapse.
That is,
\begin{equation}\label{eq:tcritdef}
    t_\mathrm{crit} \equiv \min\left\{t^* \mid F_\mathrm{net}(r_\mathrm{crit}) < 0\;\forall t\in (t^*, t_\mathrm{coll})\right\},
\end{equation}
where
\begin{equation}\label{eq:def_Fnet}
  F_\mathrm{net}(r) \equiv \int_0^{r} 4\pi r'^2 \left<\rho \right> f_\mathrm{net}\,dr'
\end{equation}
for $f_\mathrm{net}$ defined in \cref{eq:fnet}.

Properties of cores vary over time, but it is useful to define a characteristic core radius $R_\mathrm{core}$ and mass $M_\mathrm{core}$ by
the critical radius and mass within that radius at $t_\mathrm{crit}$, 
\begin{equation}\label{eq:Rcoredef}
  R_\mathrm{core} \equiv r_\mathrm{crit}(t = t_\mathrm{crit}),
\end{equation}
\begin{equation}\label{eq:Mcoredef}
  M_\mathrm{core} \equiv M_\mathrm{enc}(t = t_\mathrm{crit}; r = r_\mathrm{crit}).
\end{equation}
Here, the enclosed mass $M_\mathrm{enc}$ is computed using \cref{eq:menc} with $\left<\rho \right>$ directly measured from the simulations.
We exclude the cores having $R_\mathrm{core} < N_\mathrm{core,res}\Delta x$ with the fiducial choice $N_\mathrm{core,res} = 8$ from all our following analyses unless otherwise mentioned, because their internal turbulence is not resolved well enough to make their collapse dynamics reliable (see \cref{sec:resolution} for the related discussion of the resolution requirement).

In order to quantitatively investigate the dynamics of individual cores, we relate average infall speed to measured forces exerted on the core from a quasi-Lagrangian perspective.
For a given core mass $M_\mathrm{core}$
that is constant in time, we define the time-dependent, quasi-Lagrangian core radius $r_M(t)$ by requiring
\begin{equation}\label{eq:rMdef}
  M_\mathrm{enc}(r_M) = M_\mathrm{core}.
\end{equation}
Multiplying $\left<\rho \right>$ on both sides of \cref{eq:lagrangian_eom} and integrating over the sphere of radius $r_M$ leads to
\begin{equation}\label{eq:sphere_eom}
  M_\mathrm{core}\frac{d\overline{v}_{r,\mathrm{core}}}{dt} = F_\mathrm{net}(r_M),
\end{equation}
where
\begin{equation}\label{eq:vr_core}
\begin{split}
  \overline{v}_{r,\mathrm{core}} &\equiv \frac{\iiint_{r < r_M} \rho v_r d \mathcal{V}}{\iiint_{r < r_M} \rho d \mathcal{V}}\\
                                 &= M_\mathrm{core}^{-1} \int_0^{r_M} 4\pi r^2\left<\rho\right> \left<v_r \right>_\rho dr
\end{split}
\end{equation}
is the mass-weighted average radial velocity and $F_\mathrm{net}$ is the net integrated force defined in \cref{eq:def_Fnet}; see \autoref{app:angle-averaged-equations} for derivation.
\REV{We note that velocities are measured in the reference frame comoving with the local potential minimum, which defines the center of each core.}
We similarly define the thermal pressure gradient force $F_\mathrm{thm}$, the turbulent pressure gradient force $F_\mathrm{trb}$, the gravitational force $F_\mathrm{grv}$, the centrifugal force $F_\mathrm{cen}$, and the residual force due to the anisotropic turbulence $F_\mathrm{ani}$ by replacing $f_\mathrm{net}$ in \cref{eq:def_Fnet} with individual specific force components defined in \cref{eq:def_fthm,eq:def_ftrb,eq:def_fgrv,eq:def_fcen,eq:def_fani}.

Because the evolutionary histories of individual cores are offset in time, it is useful to introduce a normalized clock
\begin{equation}\label{eq:tau}
  \tau_\mathrm{evol} \equiv \frac{t - t_\mathrm{crit}}{\Delta t_\mathrm{coll}}
\end{equation}
for each core so that we can place cores on a common timeline.
Here,
\begin{equation}\label{eq:dtcoll}
  \Delta t_\mathrm{coll}\equiv t_\mathrm{coll} - t_\mathrm{crit}
\end{equation}
is the empirically measured duration of the collapse which varies from core to core (with values to be quantitatively presented in \citetalias{paperII}; this is generally comparable to twice the core's free-fall time at the central density, for the reasons discussed in \cref{sec:freefall_controlled}).
The evolutionary time $\tau_\mathrm{evol}$ is defined such that $\tau_\mathrm{evol} = 0$ and $1$ at the beginning ($t=t_\mathrm{crit}$) and at the end ($t=t_\mathrm{coll}$) of the collapse, respectively.

\cref{fig:core_evolution} summarizes the evolutionary history of individual cores by plotting the time variations of the fractional force imbalance defined by
\begin{equation}\label{eq:force_imbalance}
  \mathcal{F}_\mathrm{imb} \equiv \frac{F_\mathrm{net}(r_M)}{|F_\mathrm{grv}(r_M)|},
\end{equation}
the average infall velocity $\overline{v}_{r,\mathrm{core}}$ (\cref{eq:vr_core}), the average turbulent velocity dispersion $\sigma_\mathrm{1D}(r_M)$ and the central density $\rho_c$.
\cref{fig:core_evolution}(a) and (b) show that, before $t_\mathrm{crit}$, the net force exhibits significant temporal fluctuations as well as large core-to-core variations.
This is because cores form through chaotic and dynamic processes involving random collisions of supersonic flows.
The average net force before $t_\mathrm{crit}$ is slightly positive, indicating that the core-building flows are decelerated by the pressure gradients that they are setting up.
This deceleration additionally implies that it is the initial momentum (or ``inertia'') of the converging flows rather than gravitational collapse that builds a core around a stagnation point.
Over time, $\mathcal{F}_\mathrm{imb}$ evolves from small positive to small negative values and the turbulence level $\sigma_\mathrm{1D}$ slowly drops (\cref{fig:core_evolution}(e) and (f)).
Meanwhile, the radial velocity $\overline{v}_{r,\mathrm{core}}$ becomes increasingly negative and the central density $\rho_c$ increases;  these changes are slow prior to $t_\mathrm{crit}$, and rapid afterwards.

For $\tau_\mathrm{evol}>0$, cores are subject to a negative net force that accelerates the infall speed (\cref{fig:core_evolution}(c) and (d)).
This causes the central density to increase steeply after $\tau_\mathrm{evol} \approx 0.5$, growing by more than an order of magnitude at $\tau_\mathrm{evol} = 1$ compared to the value at $\tau_\mathrm{evol} = 0$ (\cref{fig:core_evolution}(g) and (h)).
Interestingly, however, the median fractional force imbalance during the runaway collapse (i.e., $0 < \tau_\mathrm{evol} < 1$) is only $\mathcal{F}_\mathrm{imb} = -0.2$.
Even at the end of the collapse when the imbalance is largest, the fractional imbalance is $\mathcal{F}_\mathrm{imb} =-0.28$, indicating that the collapse dynamics is far from gravitational free-fall, which corresponds to $\mathcal{F}_\mathrm{imb} = -1$.
Supersonic infall speeds produced by gravitational acceleration only appear near the end of the collapse ($\tau_\mathrm{evol} \gtrsim 0.7$), while most of the time the magnitude of $\overline{v}_{r,\mathrm{core}}$ remains subsonic.
\cref{fig:core_evolution}(c) and (d) show that subsonic infall motions are already present well before the onset of the collapse at $\tau_\mathrm{evol} =0 $, indicating that they are part of the core-building converging flows, and subsonic contraction persists through most of the evolution.

\cref{fig:core_evolution}(e) and (f) indicate that the turbulent kinetic energy of the core decays only quite gradually during the evolution, and actually increases during the final collapse (see also \cref{fig:evolution_radial_profiles}(d)).
This suggests that the turbulent dissipation may be partially balanced by adiabatic amplification due to core contraction \citep{robertson12}.
\REV{In \autoref{app:sigma_r}, we show that the increase of $\sigma_\mathrm{1D}$ is present even after subtracting the contribution from the bulk collapsing motion.}

From \cref{fig:core_evolution}(g),(h), there is only a slow increase in the central density of cores before $t_\mathrm{crit}$, consistent with the steady, subsonic infall speeds at those times. 
This behavior is similar to the \emph{hardening} stage observed in the simulations by \citet{collins24}.
Once the collapse sets in, the infall speeds are accelerated by the negative $F_\mathrm{net}$ and the central density $\rho_c$ rapidly increases.
It is interesting to compare $t_\mathrm{crit}$ with the ``singularity time'' $t_\mathrm{sing}$ of \citet{collins24}, where they defined $t_\mathrm{sing}$ as the time when $\partial\rho_c/\partial t$ exceeds a fixed threshold $10^5 \rho_0/t_{J,0}$ to identify the onset of the collapse.
We find $t_\mathrm{sing}$ generally occurs later than $t_\mathrm{crit}$, with larger delay observed for model \texttt{M5}: the difference between $t_\mathrm{sing}$ and $t_\mathrm{crit}$ in models \texttt{M5} and \texttt{M10} is $0.4\pm 0.36\Delta t_\mathrm{coll}$ and $0.25\pm 0.26 \Delta t_\mathrm{coll}$, respectively (cf. the model presented in \citet{collins24} has a Mach number $\mathcal{M}_\mathrm{3D} = 9$).

\begin{figure}[htpb]
  \epsscale{1.1}
  \plotone{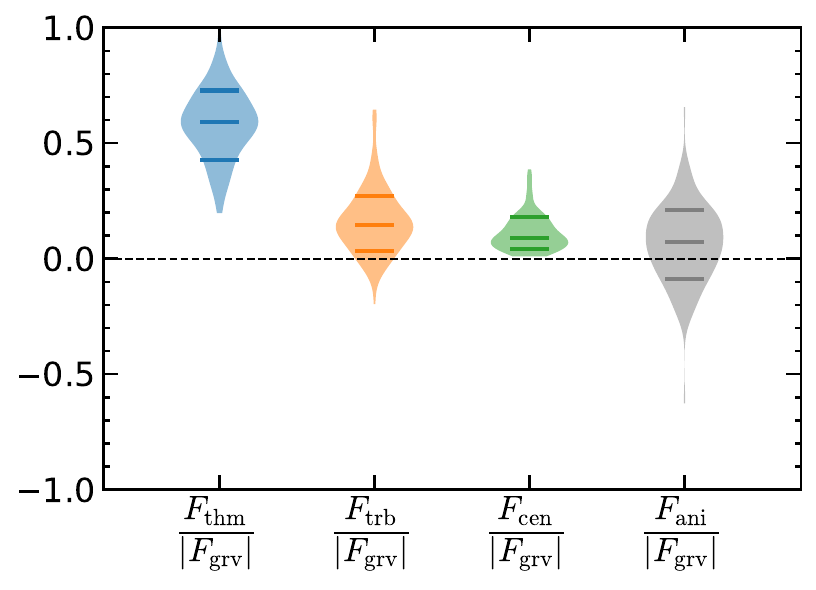}
  \caption{Distributions of the components of the normalized radial force at $t=t_\mathrm{crit}$ for the entire ensemble of resolved cores in models \texttt{M5} and \texttt{M10}. We show distributions for thermal (blue), turbulent (orange), centrifugal (green) terms, and the residual force due to anisotropic turbulence (gray); the median and $\pm 34.1$th percentiles are shown as horizontal line segments.}
  \label{fig:force_components}
\end{figure}

\cref{fig:force_components} plots the distribution of individual force components relative to the gravitational force measured at $t_\mathrm{crit}$.
It shows that gravity is mostly balanced by thermal pressure ($F_\mathrm{thm} \sim 0.59 F_\mathrm{grv}$), with a secondary contribution from the turbulent pressure $F_\mathrm{trb} \sim 0.15 F_\mathrm{grv}$.
The reason $F_\mathrm{trb}$ is generally smaller than $F_\mathrm{thm}$ despite the cores having transonic $\sigma_\mathrm{1D}$ is that $\left<\delta v_r^2 \right>_\rho$ increases with radius, making the $P_\mathrm{trb}$ profile shallower than $P_\mathrm{thm}$ for a given density gradient.
It is interesting to note that the turbulent pressure sometimes compresses a core rather than supporting it (i.e., $F_\mathrm{trb} < 0$), which can occur when the local power-law slope of the density profile is shallower than $-2p$.
The centrifugal force plays a relatively minor role in supporting cores, with $F_\mathrm{cen} \sim 0.09 F_\mathrm{grv}$.
The distribution of $F_\mathrm{ani}$ is centered slightly above the zero, with the median and standard deviations of $F_\mathrm{ani}/F_\mathrm{grv} = 0.07 \pm 0.16$.

\subsection{Critical Condition for Collapse}\label{sec:collapse_condition}

\begin{figure}[htpb]
  \epsscale{1.1}
  \plotone{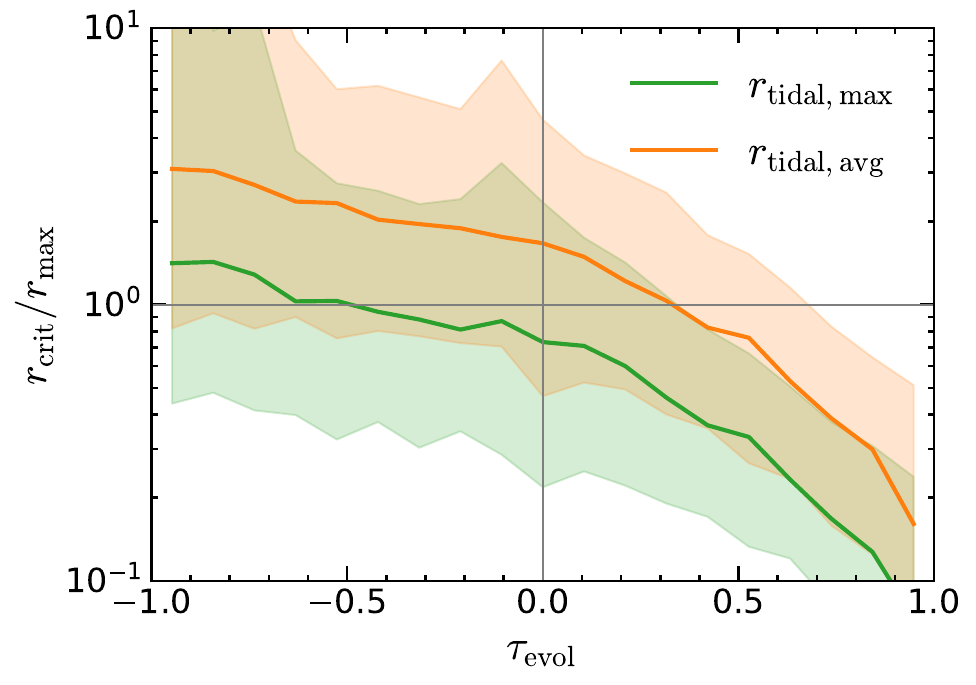}
  \caption{Time evolution of the ratio of the critical to the maximum radius, with the two alternative choices for the latter: $r_\mathrm{max} = r_\mathrm{tidal,max}$ (green) and $r_\mathrm{max} = r_\mathrm{tidal,avg}$ (orange).
  The solid line (for a given choice of $r_\mathrm{max}$) plots the median for the entire core sample, while the shades represent $\pm 34.1$th percentile ranges above and below the median for each time bin.}
  \label{fig:critical_conditions}
\end{figure}

Naive application of the \gls{TES} model leads to the prediction that collapse starts when the critical radius $r_\mathrm{crit}$ moves inside the ``maximum radius'' $r_\mathrm{max}$ such that the outer part of the core becomes unstable.
As already discussed in \cref{sec:intro}, however, cores generally do not have a sharp boundary but instead smoothly blend into the ambient cold \gls{ISM}, blurring the definition of $r_\mathrm{max}$.
From a dynamical point of view, $r_\mathrm{max}$ should reflect the structure of the gravitational field around a core, such that the gravity interior to $r_\mathrm{max}$ is primarily from the core and gravity outside is heavily influenced by neighboring structures.
\REV{A natural candidate for this quantity is the tidal radius, where \cref{fig:contours_schematics} illustrates two alternative definitions, $r_\mathrm{tidal,max}$ and $r_\mathrm{tidal,avg}$.
For all our prestellar cores, we find  $r_\mathrm{tidal,max}/r_\mathrm{tidal,avg} \sim 2\pm 0.6$.
To test whether these tidal radii indeed play a role as an effective maximum radius in determining} the onset of the collapse, we plot the ratios $r_\mathrm{crit}/r_\mathrm{tidal,max}$ and $r_\mathrm{crit}/r_\mathrm{tidal,avg}$ as a function of $\tau_\mathrm{evol}$ in \cref{fig:critical_conditions}.
This shows that the onset of the collapse (i.e., $\tau_\mathrm{evol} = 0$) as identified by direct force measurements coincides with the epoch when $r_\mathrm{crit}$ decreases to values roughly in between $r_\mathrm{tidal,max}$ and $r_\mathrm{tidal,avg}$, broadly consistent with the theoretical expectation.
The ratio of either $r_\mathrm{crit}/r_\mathrm{tidal,max}$ or $r_\mathrm{crit}/r_\mathrm{tidal,avg}$ involves a moderate level of scatter, however, indicating that 1) $r_\mathrm{max}$ is inherently a fuzzy quantity however it is defined, and/or 2) the density and velocity structure of simulated cores is sometimes not well explained by the idealized \gls{TES} model.

\begin{figure}[htpb]
  \epsscale{1.1}
  \plotone{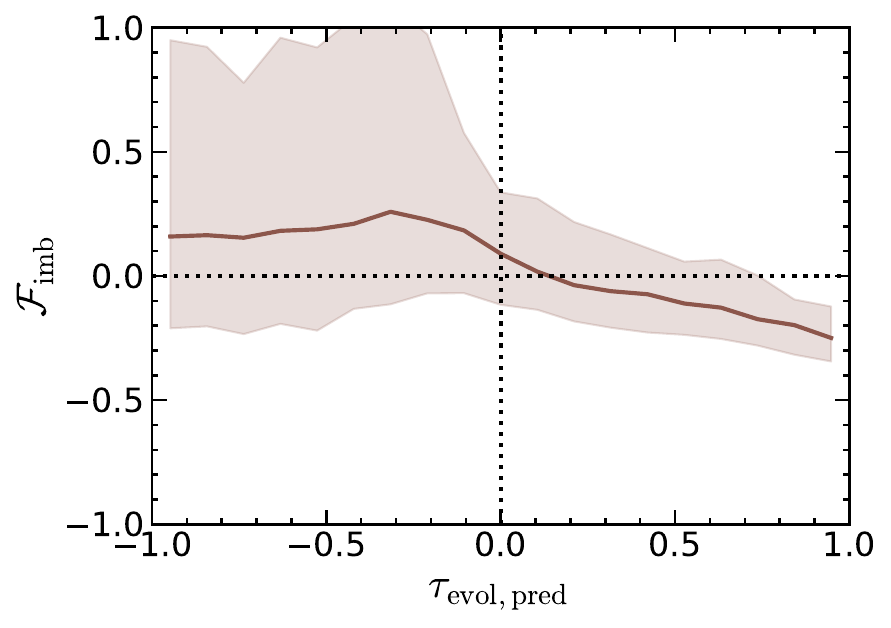}
  \caption{Temporal variations of the fractional net force $\mathcal{F}_\mathrm{imb}$ (\cref{eq:force_imbalance}) as a function of the predicted normalized time $\tau_\mathrm{evol,pred} \equiv (t - t_\mathrm{crit,pred})/(t_\mathrm{coll} - t_\mathrm{crit,pred})$, where $t_\mathrm{crit,pred}$ is the time when the conditions in \cref{eq:critical_condition} are first satisfied.
  The solid line plots the median $\mathcal{F}_\mathrm{imb}$ for the entire core sample, while the shades represent $\pm 34.1$th percentile ranges above and below the median for each time bin.
  On average, the distribution of the net force narrows and undergoes sign change from positive to negative at around $t_\mathrm{crit,pred}$, consistent with the theoretical prediction.}
  \label{fig:predicted_force_evolution}
\end{figure}

It is interesting to reverse the logic and examine how well the conditions (to be satisfied simultaneously)
\begin{subequations}\label{eq:critical_condition}
  \begin{align}
    r_\mathrm{crit} &\leq r_\mathrm{max}\\
    M_\mathrm{enc}(r_\mathrm{crit}) &\geq M_\mathrm{crit}
  \end{align}
\end{subequations}
predict the onset of the collapse.
Note that the second condition requires that the region inside $r_\mathrm{crit}$ actually encloses at least one critical mass, in order to suppress ``false alarms'' created by strong perturbations which temporarily raise $\rho_c$ without involving enough mass to render it unstable.
To test this predictor for the onset of collapse, we define the predicted critical time $t_\mathrm{crit,pred}$ as the earliest time that satisfies the conditions in \cref{eq:critical_condition}, for which we adopt $r_\mathrm{max} = 0.5(r_\mathrm{tidal,max} + r_\mathrm{tidal,avg})$.
\cref{fig:predicted_force_evolution} plots $\mathcal{F}_\mathrm{imb}$ versus $\tau_\mathrm{evol,pred} \equiv (t - t_\mathrm{crit,pred})/(t_\mathrm{coll} - t_\mathrm{crit,pred})$, showing that although \cref{eq:critical_condition} cannot precisely predict the onset of the collapse for individual cores, $t_\mathrm{crit,pred}$ based on these criteria statistically coincides with the epoch when the net force turns negative for the entire ensemble of cores.
Overall, the results shown in \cref{fig:critical_conditions,fig:predicted_force_evolution} support \cref{eq:critical_condition} as a critical condition for collapse.
We speculate that failed cores that do not satisfy \cref{eq:critical_condition} and disperse back into the ambient \gls{ISM} likely exist both in Nature and in our simulations (see \cref{sec:core_building} for related discussion).\footnote{Because our core tracking procedure starts from $t_\mathrm{coll}$, by construction it only selects prestellar cores that ultimately collapse.}

\subsection{Spatial Variations of Turbulent Scaling Relations and Correlation with Density}\label{sec:linewidth_size}

\begin{figure*}[htpb]\epsscale{1.2}
  \plotone{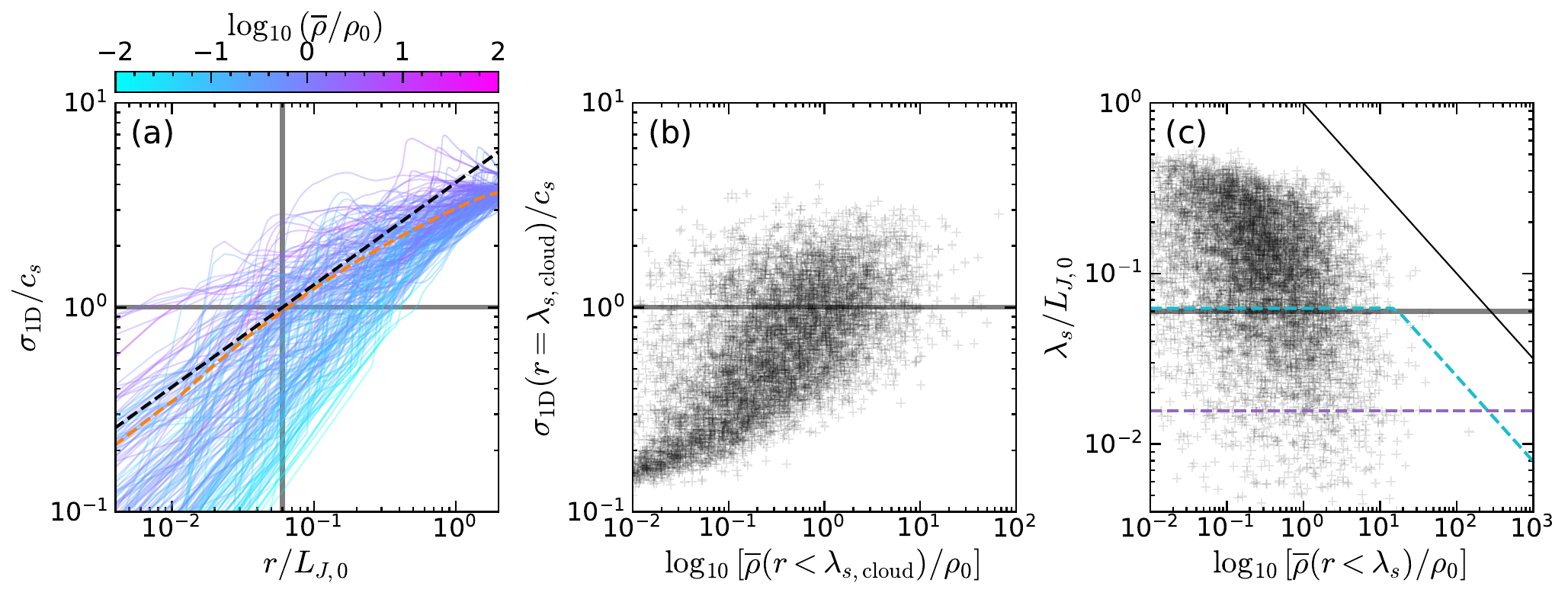}
  \caption{Linewidth-size relations for  model \texttt{M10} at the mean core forming time $t_\mathrm{cf} = 0.72 t_\mathrm{ff,0}$.
  Local linewidth-size relations are obtained by choosing a central location within the computational box, then calculating the one-dimensional velocity dispersion $\sigma_\mathrm{1D}$ using \cref{eq:sigma1d} as well as the mean density $\overline{\rho}$ as a function of the radius $r$ for a sphere centered on that location.
  (a) Local linewidth-size relations for 100 randomly chosen central positions, color-coded by $\overline{\rho}/\rho_0$.
  The orange dashed line plots the mean linewidth-size relation averaged over 1,000 randomly chosen locations at this time per simulation.
  The black dashed line plots the scaling $\sigma_\mathrm{1D} = c_s(r/\lambda_{s,\mathrm{cloud}})^{1/2}$ expected from the initial $k^{-2}$ power spectrum, with the cloud-average sonic scale $\lambda_{s,\mathrm{cloud}}$ from \cref{eq:lmb_sonic} marked by a vertical gray band.
  The intersection of colored lines with the horizontal gray band gives the actual sonic scale $\lambda_s$ on each linewidth-size curve.
  Decay of turbulence from its initial $\mathcal{M}_\mathrm{3D}=10$ level is clearly observed at large radii.
  (b) Velocity dispersion measured at $r=\lambda_{s,\mathrm{cloud}}$ (i.e., intersection points in panel (a) with vertical line) versus the mean density at that radius, demonstrating a positive correlation between the two quantities at a given size scale (also manifested by vertical color gradients in panel (a)).
  (c) Measured local sonic scale $\lambda_s$ (i.e., intersection points in panel (a) with horizontal line) versus the mean density at that radius.
  The horizontal gray band marks $\lambda_{s,\mathrm{cloud}}$.
  For reference, we also plot $L_J/L_{J,0}\equiv c_s[\pi /(G\overline{\rho})]^{1/2}/L_{J,0}$ with a black solid line, and mark $4\Delta x/L_{J,0}$ and $4\Delta x_\mathrm{AMR}/L_{J,0}$ (see \cref{sec:discussion_resolution}) with purple and cyan dashed lines, respectively.
  }
  \label{fig:linewidth_size}
\end{figure*}

As mentioned in \cref{sec:overall_evolution}, the level of turbulence at a given spatial scale varies significantly within the simulation.
In particular, while relations as expressed in \cref{eq:sigma1d} hold, there is no single value of $r_s$ or $\lambda_s$ that applies everywhere, meaning that the local value of $\lambda_s$ differs from the average value in the simulations, $\lambda_{s,\mathrm{cloud}}$ (\cref{eq:lmb_sonic}).
To illustrate the differences in the actual level of turbulence from the mean relationship, we calculate the \gls{RMS} velocity dispersion and the mean density averaged within a sphere as a function of the radius of the sphere, centered at the $1,000$ randomly selected locations in the computational domain for model \texttt{M10} at mean core formation time $t_\mathrm{cf} = 0.72 t_\mathrm{ff,0}$ defined as the ensemble average of $t_\mathrm{crit}$.
The linewidth-size relations for random centers are shown in \cref{fig:linewidth_size}(a), with coloring by density showing that velocity dispersion tends to be higher in regions with higher average density.

To see the relation between the density and velocity dispersion more clearly, in \cref{fig:linewidth_size}(b) we plot $\sigma_\mathrm{1D}$ measured at a fixed scale $r=\lambda_{s,\mathrm{cloud}}$ versus the average density at that scale, demonstrating a positive correlation between the density and velocity dispersion, with more than an order of magnitude variation.
\cref{fig:linewidth_size}(c) plots the actual sonic scale, at which the measured velocity dispersion equals the sound speed, versus the local average density at that scale, showing the \emph{local} sonic scale varies by almost two orders of magnitude.
Together, \cref{fig:linewidth_size} demonstrates that there are large differences from the mean linewidth-size relation at any given location, and resolving the sonic scale in most regions would require $\Delta x$ significantly smaller than $\lambda_{s,\mathrm{cloud}}$.
We will further discuss numerical resolution requirements in \cref{sec:discussion_resolution}.

\section{Discussion}\label{sec:discussion}

\subsection{Evolution to Collapse}\label{sec:core_building}

The positive net radial force during the core building stage ($t < t_\mathrm{crit}$) indicates that the formation of cores is not driven by top-town gravitational collapse.
Instead, most cores are built by inertial converging flows, which are decelerated either abruptly by shocks or gradually by pressure gradients.
When the converging flows are strong enough or maintained for a sufficient period of time, they succeed in assembling sufficient mass with large density contrast, satisfying \cref{eq:critical_condition}.
The region then becomes gravitationally unstable, leading to runaway collapse accelerated by a negative net radial force (\cref{fig:core_evolution,fig:predicted_force_evolution}).

\begin{figure}[htpb]
  \epsscale{1.1}
  \plotone{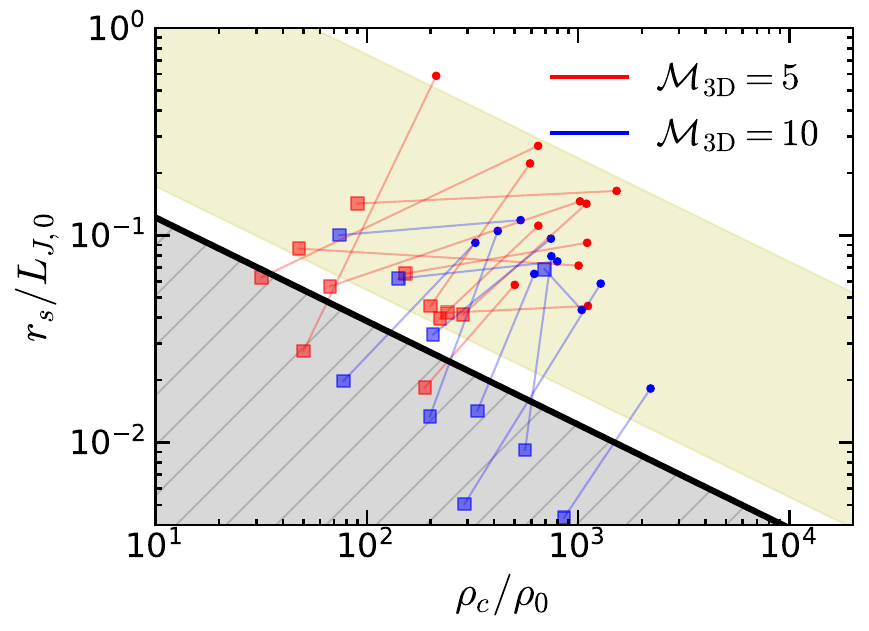}
  \caption{Local sonic radius versus central density measured for selected individual cores from model \texttt{M5} (red) and \texttt{M10} (blue) with black line and gray shading as in \cref{fig:tes_schematics}.
  Squares show pre-images of selected cores at three free-fall times before $t_\mathrm{crit}$, which are connected to corresponding $t_\mathrm{crit}$ cores by straight lines.
  Evolutionary trends are consistent with expectations that turbulence dissipates (increasing $r_s$) and stratification develops (increasing $\rho_c$).
  The yellow shaded band is the region where $\sigma_\mathrm{1D}$ is between $0.3$ (upper edge) and $2$ (lower edge) where most cores are found at the critical time (see \citetalias{paperII}).}
  \label{fig:rs_rhoc_trajectories}
\end{figure}

In \cref{fig:tes_schematics}, we provided a schematic indicating expected evolutionary trends for $r_s$ and $\rho_c$.
In \cref{fig:rs_rhoc_trajectories}, we indicate how the loci of actual cores vary from early to later times, by plotting $r_s$ and $\rho_c$ measured at the critical time $t_\mathrm{crit}$ as well as at earlier time $t_\mathrm{crit} - 3t_\mathrm{ff}(\overline{\rho}_\mathrm{crit})$, for some selected cores.
The lines connecting the two points show that the overall directions of evolution are consistent with our expectation in \cref{fig:tes_schematics}.
Some past images of cores live in the ``forbidden region'' where instability is completely suppressed by turbulence.
Others fall in the region where instability is allowed, but they generally have low $r_s$ and $\rho_c$ such that the critical conditions (\cref{eq:critical_condition}) are not easily satisfied.
Overall, \cref{fig:rs_rhoc_trajectories} indicates that cores start collapsing when the turbulence sufficiently dissipates and the central density becomes large enough due to the converging flows.
It also suggests that the density at which collapse starts is not unique, but rather depends on the local strength of turbulence parametrized by $r_s$.
In \citetalias{paperII}, we will present more detailed properties of critical cores and discuss them in the context of critical density.

While the present study focuses on cores in which collapse succeeds, there is likely a class of ``failed cores,'' 
in which the critical condition for collapse is never met, with the material that has been gathered instead dispersing back into the ambient medium.
A recent study by \citet{offner22} identified and tracked simulated cores using the density dendrogram.
They applied a clustering algorithm to a vector of measured physical properties of cores, finding the entire core dataset can be classified into three categories which they term ``turbulent'', ``coherent'', and ``pre/protostellar'' phases.
Their finding suggests that cores not only transition stochastically between these phases but also disperse entirely from any of these phases (see their Fig. 10).
This could happen, for example, when the core-building converging flows are not strong enough or are maintained for only a brief period of time; a core is hit by traveling shock waves.
Identifying these failed cores would require systematic analysis different from the present approach, in which we trace the prior history of cores that form sink particles.

Our critical conditions (\cref{eq:critical_condition}) suggest that the fate of a core depends not only on the local density but also on the local strength of turbulence and the gravitational potential terrain around the core.
When considering the effects of turbulence, almost all theories assume a single linewidth-size relation applied to all cores.
However, \cref{fig:linewidth_size} shows that both the slope and normalization of the \emph{local} linewidth-size relations significantly vary from region to region.
A complete theory should take into account 1) the effects of turbulence on the critical radius and mass of a core (e.g., from the \gls{TES} model), 2) the correlation between the density and turbulent velocities as well as spatial variations of the local linewidth-size relation (\cref{fig:linewidth_size}), and 3) geography of density and gravitational potential shaped by supersonic turbulence, which sets the tidal radius.
While the final task is especially challenging, a few different theoretical approaches to characterizing localized nonlinear density fluctuations \citep[e.g.,][]{padoan02,hennebelle08,hopkins12} could provide a starting point.

\subsection{Resolution Requirements for Turbulent Cores}\label{sec:discussion_resolution}

The analysis presented in this work indicates that the evolution of prestellar cores to a critical point, when instability and collapse commence, depends on the local strength of turbulence at earlier stages of evolution  (\cref{fig:rs_rhoc_trajectories}).
Moreover, perturbations introduced when typical gas densities are orders of magnitude below core values may be strongly enhanced by the time collapse becomes possible.
Resolving not only the final products -- dense cores -- but also the evolution of their turbulent, diffuse progenitors is therefore crucial in obtaining converged numerical results.
Many simulations with adaptive or moving mesh approaches, however, focus resolution primarily on the dense gas.
As a result, the turbulence in diffuse gas is not necessarily well resolved, with implications for dynamical consequences of this turbulence.
For example, some collapsing cores might instead have dispersed, if the turbulence in their progenitors had been better resolved.
Alternatively, if turbulence in diffuse gas is not resolved, perturbations it should introduce --- which would later lead to fragmentation --- could be missing.

As discussed in \cref{sec:resolution}, the cloud-scale average $\lambda_{s,\mathrm{cloud}}$ (\cref{eq:lmb_sonic}) is approximately twice the radius of a critical BE sphere (\cref{eq:rbe}) for cores at the 99th percentile of density, comparable to the post-shock value.
\cref{fig:linewidth_size}(d) shows that in practice, local values of $\lambda_s$ within our simulations span an order of magnitude above and below the cloud-average $\lambda_{s,\mathrm{cloud}}$.   
Thus, if we wish to ensure that turbulence is resolved everywhere at all stages of prestellar core evolution, we would need $\Delta x \ll \lambda_{s,\mathrm{cloud}} / 10$.
While strictly meeting this requirement would be quite challenging, it is still possible to \REV{marginally} resolve $\lambda_s$ in most of the gas.
For example, our resolution of $1024^3$ for model \texttt{M10} ensures that $95\%$ of the regions (or $90\%$ of the regions with density exceeding $\rho_0$) have $\lambda_s > 4 \Delta x$ (i.e., above the purple line in \cref{fig:linewidth_size}(c)).

In \cref{sec:resolution}, we also pointed out that setting the resolution to a constant fraction ($1/N_J \sim 1/4 \text{--} 1/30$) of the local Jeans length may not be enough to resolve the sonic scale.
To illustrate this, in \cref{fig:linewidth_size}(d) we indicate with a cyan dashed line the boundary defined by $4 \Delta x_\mathrm{AMR}$, where $\Delta x_\mathrm{AMR} \equiv \min(L_\mathrm{box}/256, L_J/16)$ is a ``typical'' resolution of \gls{AMR} simulations adopting a $256^3$ root grid and a Jeans refinement criterion with $N_J = 16$ (with higher root grid resolution and $N_J$, the horizontal and diagonal part of the boundary will shift down by corresponding factors).
Below this boundary, $\lambda_s$ would be resolved by fewer than 4 zones. 
For comparison, we also indicate (purple dashed line) a similar boundary defined by $4\Delta x$ for our numerical resolution.
The comparison shows that while \gls{AMR} simulations are better at resolving dense structures compared to uniform-grid simulations, the resolution of diffuse gas which is the \emph{progenitor} of the dense structures is lower.

To achieve $\Delta x = (1/4) (\lambda_{s,\mathrm{cloud}}/10)$ needed to resolve the sonic scale almost everywhere, the number of cells per dimension would need to satisfy \cref{eq:nmin_sonic} with $N_{s,\mathrm{res}} = 40$.
For a range $\mathcal{M}_\mathrm{3D} \sim 10\text{--}20$ typical of \glspl{GMC}, this implies that a uniform mesh simulation would need to exceed $2048^3\text{--}8192^3$ in order to resolve the sonic scale everywhere, as well as critical cores.
With the advent of GPU-accelerated codes, this is now within reach.
It will be very interesting to compare this kind of very high resolution uniform-grid simulation against \gls{AMR} simulations adopting traditional refinement criteria based on the Jeans length, and lagrangian (moving mesh) simulations in which a fixed mass is resolved.

\subsection{Free-fall vs. Controlled Collapse}\label{sec:freefall_controlled}
While theoretical work has emphasized that the collapse of centrally concentrated objects is qualitatively different from homogeneous free-fall collapse, the distinction is not always appreciated, perhaps because the quantitative difference in duration is modest.
Neglecting pressure, the time taken for a fluid element initially at rest at radius $r$ to reach the center is $\overline{t}_\mathrm{ff}(r) = [3\pi / (32G\overline{\rho}(r))]^{1/2}$ where $\overline{\rho}(r) = M_\mathrm{enc}(r) / (4\pi r^3/3)$ is the mean density within the radius $r$.
Because $\overline{\rho}(r)$ decreases outward for stratified objects, $\overline{t}_\mathrm{ff}(r)$ has a minimum at the center, $t_{\mathrm{ff},c} \equiv [3\pi / (32G\rho_c)]^{1/2}$,  
and increases outward.
Pressureless collapse therefore takes $\Delta t_\mathrm{coll} = t_{\mathrm{ff},c}$ to produce a singularity at the center of a stratified region, after which the remaining material accretes over time as set by $\bar{\rho}(r)$.

However, real core-building processes do not produce pressure-free systems, and collapse is not pressure-free.
More generally, the net force is given by $\mathcal{F}_\mathrm{imb} GM_\mathrm{enc}(r)/r^2$, with $-1 < \mathcal{F}_\mathrm{imb} <0$ for collapse modulated by pressure or other forces.
If $\mathcal{F}_\mathrm{imb}$ is constant, the duration is reduced to $\Delta t_\mathrm{coll} = |\mathcal{F}_\mathrm{imb}|^{-1/2}t_{\mathrm{ff},c}$.
Because $\Delta t_\mathrm{coll}$ depends only weakly on $|\mathcal{F}_\mathrm{imb}|$, the inclusion of pressure support does not dramatically delay the collapse.
For example, the \gls{LP} similarity solution has significant pressure support at the center, $\mathcal{F}_\mathrm{imb} = -0.4$, but the collapse timescale is only increased to $\Delta t_\mathrm{coll} = 1.58 t_{\mathrm{ff},c}$ \citep[see their Equation C10 and related text]{larson69}.
The prestellar cores in our simulations have $\mathcal{F}_\mathrm{imb} = -0.2$ on average during the collapse, such that $\Delta t_\mathrm{coll} = (0.2)^{-1/2} t_{\mathrm{ff},c} = 2.2 t_{\mathrm{ff},c}$, which is entirely consistent with the directly measured $\Delta t_\mathrm{coll} \approx 2t_{\mathrm{ff},c}$ (see \citetalias{paperII}).
We note that some other work \citep{murray17,cao23} has also presented evidence that non-gravitational forces remain large in the core-forming regions.

It is worth noting that the evidence for controlled collapse in our simulations shows that gravitational collapse is not primarily the result of fragmentation of filaments that have large mass per unit length ($\gg 2 c_s^2/G$), in which gravitational runaway is unimpeded.
While filaments like this could develop in a system with low virial parameters, we do not expect this to occur in realistic GMCs (see \autoref{app:initial_conditions}).

We also note that observed kinematics of prestellar cores are generally inconsistent with the inside-out collapse from the singular isothermal sphere, while they can be explained by a quasi-equilibrium collapse model starting from an idealized, isolated \gls{BE} sphere \citep{keto15,koumpia20}.
While our simulations indicate collapse indeed proceeds in a quasi-equilibrium with small net forces (\cref{fig:core_evolution}), they also highlight an important point that cores cannot be considered as isolated objects detached from their turbulent formation environment.
Both the internal turbulence and infall extending to large radii are prevalent in the simulated cores, and these must be taken into account when modeling the observed spectra.
Full forward modeling would therefore require coupling chemistry and radiation transfer within \gls{MHD} simulations.

\section{Summary and Conclusions}\label{sec:conclusion}

Dense cores are the immediate precursors of individual stars and stellar systems.
Because parent GMCs are highly turbulent and cores are part of complex larger scale structures, rather than being physically isolated, it is not obvious what physical conditions actually trigger the onset of gravitational collapse.
To investigate this question, we perform three-dimensional simulations of turbulent clouds and track the evolution of individual cores both prior to and subsequent to collapse (\cref{sec:simulations}).

We quantitatively investigate the dynamical evolution of individual cores by measuring the net radial force, mean velocity, velocity dispersion, and the central density as functions of time (\cref{fig:core_evolution}), identifying the onset of gravitational instability at time $t_\mathrm{crit}$, when the directly measured force $F_\mathrm{net}=0$ (\cref{sec:force_balance}).
For each core, at each time we compute a predicted critical radius $r_\mathrm{crit}$ (analogous to the Bonnor-Ebert critical radius, but allowing for turbulence; see \citetalias{moon24}).
We also compute tidal radii $r_\mathrm{tidal,max}$ and $r_\mathrm{tidal,avg}$, which depend on the gravitational potential terrain surrounding each core (see \cref{fig:contours_schematics,fig:evolution_projected_map}).
For each core, the critical radius decreases in time, while the tidal radius does not change systematically; we show that collapse occurs approximately when the critical radius becomes smaller than the tidal radius (\cref{sec:collapse_condition}).

Our main conclusions are as follows:
\begin{enumerate}
  \item
    The evolution of cores monitored in the simulations is broadly consistent with the four-stage descriptions of \citet{gong09} or \citet{collins24}.
    Cores are built in regions where turbulent flows locally converge, and initially the net radial force within cores is outward, with pressure forces and turbulence decelerating the inflow.
    When the central density becomes high enough and/or the turbulence dissipates sufficiently, the critical conditions (\cref{eq:critical_condition}) are met and the core becomes unstable (e.g., \cref{fig:evolution_radial_profiles,fig:core_evolution,fig:rs_rhoc_trajectories}).
    Runaway gravitational collapse accelerates inflow and leads to a dramatic increase in the central density (e.g., \cref{fig:core_evolution}).

  \item
    Cores do not have physical boundaries because they are formed dynamically out of a gaseous continuum.  Nevertheless, the hierarchical structure of the surrounding cloud creates an effective gravitational potential boundary that isolates each core (e.g., \cref{fig:contours_schematics,fig:evolution_projected_map}).
    We characterize the size of the gravitational potential ``pocket'' for each core based on tidal radii, which vary substantially from core to core (e.g., \cref{fig:rcrit_rtidal}(a)).
    Collapse does not occur until enough mass has collected locally for self-gravity to overwhelm the core's internal support.

  \item
    Overdense regions tend to be more turbulent than randomly sampled locations in our simulations, due to a correlation between the velocity dispersion and the average local density at a given scale (\cref{fig:linewidth_size}).
    We note that in theoretical models, for the purpose of quantifying turbulent support within cores it is often assumed \citep[e.g.][]{hennebelle08,hopkins12} that the velocity field samples from the large-scale power spectrum independent of the local density, which is not strictly consistent with the correlation we find.

  \item
    The fractional force imbalance within each core is defined as the net radial force divided by the gravitational force, $\mathcal{F}_\mathrm{imb}=F_\mathrm{net}/|F_\mathrm{grav}|$.
    $\mathcal{F}_\mathrm{imb}$ becomes negative at $t_\mathrm{crit}$ and grows in magnitude, but averages to only $\mathcal{F}_\mathrm{imb} = -0.2$ throughout the collapse, much smaller than the value would be for pressureless free-fall (\cref{fig:core_evolution}).
    Collapse of prestellar cores can therefore be viewed as a quasi-equilibrium process.
    There is no evidence in our simulations of unimpeded gravitational fragmentation, as would occur in filaments with mass per unit length far exceeding the critical value $2 c_s^2/G$.
    Internal pressure forces make core collapse highly non-homologous, such that by the time a singularity forms, only the very central part reaches $r=0$.

  \item
    Most cores in our simulations have subsonic inflows consistent with observed kinematics inferred from asymmetric molecular line profiles \citep[e.g.,][]{cwlee01}.
    These inflow motions are present from well before the onset of collapse during which the net radial force is mostly positive (\cref{fig:core_evolution}), suggesting that they represent the inertial, core-building converging flows (e.g., \citealt{cwlee01}; \citealt{gong09,gong11}; \citealt{padoan20}; \citealt{collins24}) rather than gravitational collapse.
    Supersonic infall speeds only appear near the end of the collapse and therefore would not be detected for most cores.
    The turbulent velocity dispersion averaged within the Lagrangian radius decreases only gradually with a slight upturn near the end of the collapse, possibly affected by adiabatic amplification \citep{robertson12,murray15}.
\end{enumerate}

Our work has several limitations.
First, the simulations presented in this work do not include magnetic fields.
While this choice is useful for comparison with the \gls{TES} model and therefore for testing the scenario proposed in \citetalias{moon24}, the effect of magnetic fields has to be taken into account to understand the dynamics of real cores.
We note, however, that previous studies generally find that the overall rate of star formation does not depend much on the cloud scale magnetic field strength unless an entire cloud is magnetically subcritical \citep[e.g.][]{2019FrASS...6....7K,2019FrASS...6....5H,jgkim21,2022MNRAS.515.4929G}, and the physical properties of cores at the time of collapse are also not strongly sensitive to the large-scale magnetic field strength, because this collapse cannot occur unless cores are supercritical \citep[e.g.][]{chen14,chen15,2018A&A...611A..24H}.
Secondly, the Eulerian nature of the simulations not only limits the time we can trace back the evolution of individual cores but also makes it challenging (although not impossible) to define a core as a single coherent entity through consecutive snapshots.
This limitation may be overcome by introducing Lagrangian tracer particles into simulations \citep{collins23,collins24}.
Finally, while we have demonstrated that collapse begins, as expected, when cores become critical in the sense $r_\mathrm{crit}<r_\mathrm{tidal}$, the tidal radius of each core is empirically identified rather than predicted in the simulation.
A very interesting question for future work will be to understand how the small-scale gravitational potential landscape may relate to global GMC properties (such as kinetic and magnetic energy levels relative to gravitational energy) and the spatio-temporal statistical characteristics of turbulence.

\begin{acknowledgments}
We thank the anonymous referee for a detailed review and constructive comments.
This work was supported in part by grant 510940 from the Simons Foundation to E.~C.\ Ostriker. 
S.M. thanks Alwin Mao for answering all the questions related to the structure finding algorithm and generously providing his Python implementation on which \texttt{GRID-dendro} is heavily based.
S.M. also thanks Kengo Tomida for providing the early version of the full multigrid gravity solver for \textit{Athena++}, and offering useful advice regarding the algorithm.

\end{acknowledgments}

\software{\textit{Athena++} \citep{stone20}, \texttt{xarray} \citep{hoyer2017xarray}}

\bibliographystyle{aasjournal}
\bibliography{mybib}

\appendix

\crefalias{section}{appsec}

\section{Angle-Averaged Equations of Motion}\label{app:angle-averaged-equations}
To derive \cref{eq:sphere_eom}, we start by taking time derivative of \cref{eq:vr_core},
\begin{equation}\label{eq:dpdt_1}
  M_\mathrm{core}\frac{d\overline{v}_{r,\mathrm{core}}}{dt} = \int_{0}^{r_M} 4\pi r^2 \frac{\partial}{\partial t} \left( \left<\rho \right> \left<v_r \right>_\rho \right) dr + 4\pi r_M^2 \left( \left<\rho \right> \left<v_r \right>_\rho^2 \right)(r=r_M),
\end{equation}
in which we have used the identity $\dot{r}_M = \left<v_r \right>_\rho (r = r_M)$ resulting from mass conservation.
Noting that the second term in the right hand side of \cref{eq:dpdt_1} can be equivalently expressed as $\int_0^{r_M} 4\pi (\partial/\partial r)(r^2 \left<\rho \right> \left<v_r \right>_\rho^2) dr$, \cref{eq:dpdt_1} can be rearranged into
\begin{equation}\label{eq:dpdt_2}
  M_\mathrm{core}\frac{d\overline{v}_{r,\mathrm{core}}}{dt} = \int_0^{r_M} 4\pi r^2 \left<\rho \right> \left( \frac{\partial \left<v_r \right>_\rho}{\partial t} + \left<v_r \right>_\rho \frac{\partial \left<v_r \right>_\rho}{\partial r} \right) dr + \int_0^{r_M} 4\pi r^2 \left<v_r \right>_\rho \left[ \frac{\partial \left<\rho \right>}{\partial t} + \frac{1}{r^2} \frac{\partial}{\partial r} \left( r^2 \left<\rho \right> \left<v_r \right>_\rho \right)  \right] dr.
\end{equation}
\REV{Here, the term in the parenthesis within the first integral is equal to $f_\mathrm{net}$ via \cref{eq:lagrangian_eom}, and the term in the square bracket within the second integral is identically zero, based on the angle-averaged continuity equation (see Equation (7) of \citetalias{moon24}).}
The right hand side of \cref{eq:dpdt_2} then reduces to \cref{eq:def_Fnet}, leading to \cref{eq:sphere_eom}.

\section{Decomposition of the Velocity Dispersion}\label{app:sigma_r}

\REV{
Prestellar cores acquire large collapsing velocities within them toward the end of their life (\cref{fig:core_evolution}(c),(d)), and this can contribute to the velocity dispersion $\sigma_\mathrm{1D}$ defined in \cref{eq:sigma1d} in addition to genuine ``turbulent'' motions.
To see this, it is convenient to define the mass-weighted volume-averaging operator
\begin{equation}\label{eq:vol_average}
    \llangle Q \rrangle_\rho \equiv \frac{\iiint \rho Q\,d\mathcal{V}}{\iiint \rho\, d\mathcal{V}} = \frac{\int \left<\rho\right>\left<Q\right>_\rho r^2dr}{\int \left<\rho\right> r^2dr}.
\end{equation}
such that $\sigma_\mathrm{1D}^2 = \llangle |\mathbf{v} - \mathbf{v}_\mathrm{com}|^2 \rrangle_\rho/3$.
Because the operator $\llangle \cdot \rrangle$ is linear, one can use the index notation and \cref{eq:def_delta_v} to write
\begin{equation}\label{eq:sigma_1d_decomposition}
\begin{split}
    \sigma_\mathrm{1D}^2 &= \frac{1}{3} \sum_i \left\llangle \left( \left< v_i\right>_\rho + \delta v_i - v_{\mathrm{com},i} \right)^2 \right\rrangle_\rho\\
    &= \frac{1}{3}\sum_i \left\llangle\left\langle v_i \right\rangle_\rho^2 \right\rrangle_\rho
    + \left\llangle  \delta v_i^2  \right\rrangle_\rho
    + \left\llangle  v_{\mathrm{com},i}^2  \right\rrangle_\rho
    + 2\left\llangle  \left\langle v_i \right\rangle_\rho \delta v_i  \right\rrangle_\rho
    - 2\left\llangle  \delta v_i v_{\mathrm{com},i} \right\rrangle_\rho
    - 2\left\llangle  \left\langle v_i \right\rangle_\rho v_{\mathrm{com},i} \right\rrangle_\rho.
\end{split}
\end{equation}
Here, one can use \cref{eq:def_delta_v,eq:vol_average} to show that the fourth term, $2\left\llangle  \left\langle v_i \right\rangle_\rho \delta v_i  \right\rrangle_\rho = 0$ by definition.
The last two terms can be combined into $-2\left\llangle v_i v_{\mathrm{com},i}\right\rrangle_\rho$, where we note that, in a curvilinear coordinate system, $v_{\mathrm{com},i}$ depends on position and therefore cannot be taken out of the angled brackets, even though $\mathbf{v}_\mathrm{com}$ is a constant vector.
To proceed, we utilize the coordinate transformation $q_i = \sum_j R_{ij} Q_j$ that relates general vector components $q_i$ to the Cartesian components $Q_j$.
For orthogonal coordinate systems, $R_{ij}R^T_{jk} = \delta_{ik}$.
Noting that the Cartesian components of the center of mass velocity, $V_{\mathrm{com},i}$, are constant, one can show that
\begin{equation}\label{eq:v_vcom}
    \sum_i \left\llangle v_i v_{\mathrm{com},i} \right\rrangle_\rho = \sum_{i,j} \left\llangle v_i R_{ij} V_{\mathrm{com},j} \right\rrangle_\rho = \sum_{i,j} \left\llangle R^T_{ji} v_i  \right\rrangle_\rho V_{\mathrm{com},j} = \sum_{j} \left\llangle V_j \right\rrangle_\rho V_{\mathrm{com},j} = \sum_j V_{\mathrm{com},j}^2 = |\mathbf{v}_\mathrm{com}|^2,
\end{equation}
and, similarly,
\begin{equation}\label{eq:vcom_sq}
    \sum_i \left\llangle v_{\mathrm{com},i}^2 \right\rrangle_\rho = \sum_{i,j,k} \left\llangle R_{ij}R_{ik} V_{\mathrm{com},j}V_{\mathrm{com},k} \right\rrangle_\rho = \sum_{i,j,k} \left\llangle R^T_{ji} R_{ik}  \right\rrangle_\rho V_{\mathrm{com},j}V_{\mathrm{com},k} = \sum_j V_{\mathrm{com},j}^2 = |\mathbf{v}_\mathrm{com}|^2.
\end{equation}
Substituting \cref{eq:v_vcom,eq:vcom_sq} into \cref{eq:sigma_1d_decomposition} yields
\begin{equation}\label{eq:sigma1d_expansion}
    \sigma_\mathrm{1D} = \sqrt{\sigma_\mathrm{1D,trb}^2 + \sigma_\mathrm{1D,blk}^2 - \frac{1}{3}|\mathbf{v}_\mathrm{com}|^2},
\end{equation}
where
\begin{equation}\label{eq:sigma_trb}
    \sigma_\mathrm{1D,trb} \equiv \frac{1}{\sqrt{3}}\left\llangle \sum_i \delta v_i^2 \right\rrangle_\rho^{1/2} = \frac{1}{\sqrt{3}} \left( \sum_i M_\mathrm{enc}^{-1} \int_0^{r} 4\pi r^2\left<\rho\right> \left<\delta v_i^2 \right>_\rho dr' \right)^{1/2}
\end{equation}
and
\begin{equation}\label{eq:sigma_blk}
    \sigma_\mathrm{1D,blk} \equiv \frac{1}{\sqrt{3}}\left\llangle \sum_i \left< v_i\right>_\rho^2 \right\rrangle_\rho^{1/2} = \frac{1}{\sqrt{3}} \left( \sum_i M_\mathrm{enc}^{-1} \int_0^{r} 4\pi r^2\left<\rho\right> \left<v_i \right>^2_\rho dr' \right)^{1/2}.
\end{equation}
We note that Equation (49) of \citetalias{moon24} is a special case of \cref{eq:sigma1d_expansion}, where $|\mathbf{v}_\mathrm{com}| = 0$ (i.e., the velocities are with respect to the center of momentum), $\sigma_\mathrm{1D,blk} = 0$, and the turbulence is isotropic.
While the expressions in \cref{eq:sigma1d_expansion,eq:sigma_trb,eq:sigma_blk} are valid in any orthogonal curvilinear coordinate system, the designation of $\sigma_\mathrm{1D,trb}$ and $\sigma_\mathrm{1D,blk}$ as the ``turbulent'' and ``bulk'' components, respectively, depends on choosing a coordinate system that aligns with the symmetry of the flow.
For example, while radial oscillations and rotational motions are correctly identified with nonzero $\sigma_\mathrm{1D,blk}$ in spherical coordinates via $\left< v_r\right>_\rho$ and $\left< v_\phi \right>_\rho$, respectively, adopting Cartesian coordinates would lead to $\sigma_\mathrm{1D,blk}=0$ as the velocity components are averaged out due to symmetry.
On the other hand, linear sloshing motion can be associated with nonzero $\sigma_\mathrm{1D,blk}$ in Cartesian coordinates but not in spherical coordinates.

Assuming the turbulence is statistically isotropic, one can use \cref{eq:linewidth_size} to perform the integral in \cref{eq:sigma_trb} for a given density profile.
For uniform density, this leads to $\sigma_\mathrm{1D,trb} = [3/(2p+3)]^{1/2}\left<\delta v_r^2\right>_\rho^{1/2}$.
For more general density profiles, we introduce an order unity factor $\eta_d$ to capture the degree of density concentration, such that
\begin{equation}
    \sigma_\mathrm{1D,trb} = \eta_d\left(\frac{3}{2p+3}\right)^{1/2}\left<\delta v_r^2\right>_\rho^{1/2}.
\end{equation}
We find $\eta_d \approx 0.9$ for \gls{TES} solutions with very weak dependence on $p$ and $\sigma_\mathrm{1D,trb}$ (e.g., the minimum and maximum values are $\eta_d = 0.85$ and $0.94$ for the range of $p = 0.3\text{--}0.7$ and $\sigma_\mathrm{1D} < 10$), and $\eta_d = [(1 + 2p/3) / (1 + 2p)]^{1/2}$ for $\rho \propto r^{-2}$ profile.

\begin{figure}[htpb]
  \plotone{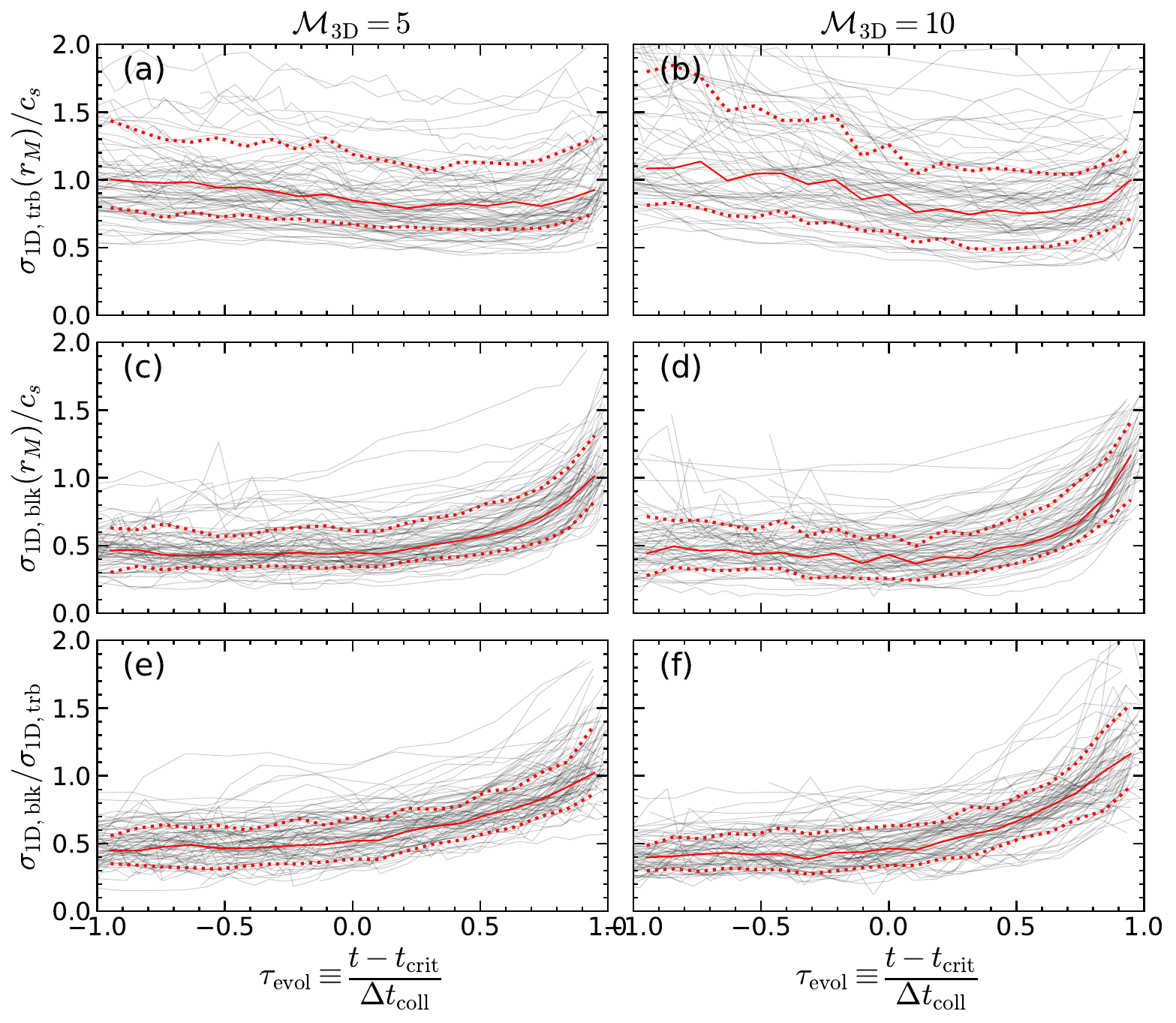}
  \caption{Time evolution of the turbulent ((a), (b)) and bulk ((c), (d)) components of the velocity dispersion, along with their ratio ((e), (f)).
  Black lines correspond to the trajectories of individual cores in model \texttt{M5} (left column) and \texttt{M10} (right column).
  Red solid and dotted lines plot the median and $\pm 34.1$th percentile values at each time bin.}
  \label{fig:sigma_trb_blk}
\end{figure}

\cref{fig:sigma_trb_blk} plots time evolution of $\sigma_\mathrm{1D,trb}$ and $\sigma_\mathrm{1D,blk}$ calculated using spherical coordinates, as well as their ratio, for individual cores in models \texttt{M5} and \texttt{M10}.
Comparison of \cref{fig:sigma_trb_blk} and \cref{fig:core_evolution}(e),(f) suggests that the steep upturn of $\sigma_\mathrm{1D}$ toward $\tau_\mathrm{evol} \to 1$ shown in the latter is largely due to increasing $\sigma_\mathrm{1D,blk}$.
However, \cref{fig:sigma_trb_blk} also shows that the turbulent velocity dispersion $\sigma_\mathrm{1D,trb}$ indeed increases when $\tau_\mathrm{evol} > 0.5$, which might be attributed to adiabatic amplification of turbulence \citep{robertson12}.

}

\section{Dependence on Initial Virial Parameter}\label{app:initial_conditions}

In our simulations, cores form and collapse within overdense structures shaped by velocity perturbations.
Because the crossing time of the perturbations increases with a scale as $l / \delta v(l) \propto l^{0.5}$, in our simulations with uniform initial density, density structures first appear at the smallest scales and then progressively grow at larger scales.
The largest structure develops when the peak and trough of the longest wavelength mode (i.e., $k = 2\pi / L_\mathrm{box}$) moving in opposite directions meet, which happens at roughly $t = 0.5t_\mathrm{flow,0}$ where
\begin{equation}
  t_\mathrm{flow,0} \equiv \frac{L_\mathrm{box}}{2\sigma_\mathrm{1D,box}}.
\end{equation}
is the flow-crossing time or dynamical time.
Using \cref{eq:lbox,eq:tff0}, the ratio between $t_\mathrm{flow,0}$ and $t_\mathrm{ff,0}$ can be expressed in terms of $\alpha_\mathrm{vir,box}$ as
\begin{equation}
  \frac{t_\mathrm{flow,0}}{t_\mathrm{ff,0}} \approx 1.03 \left(\frac{\alpha_\mathrm{vir,box}}{2}\right)^{-1/2}
\end{equation}
Our standard models have fixed $\alpha_\mathrm{vir,box} = 1.66$ and therefore $t_\mathrm{flow,0}$ and $t_\mathrm{ff,0}$ are comparable to each other.
To explore how the overall evolution changes depending on $\alpha_\mathrm{vir,box}$, we run two additional simulations named \texttt{M15L2} and \texttt{M3L4}, with high and low $\alpha_\mathrm{vir,box}$, respectively.

\begin{figure}[htpb]
  \plotone{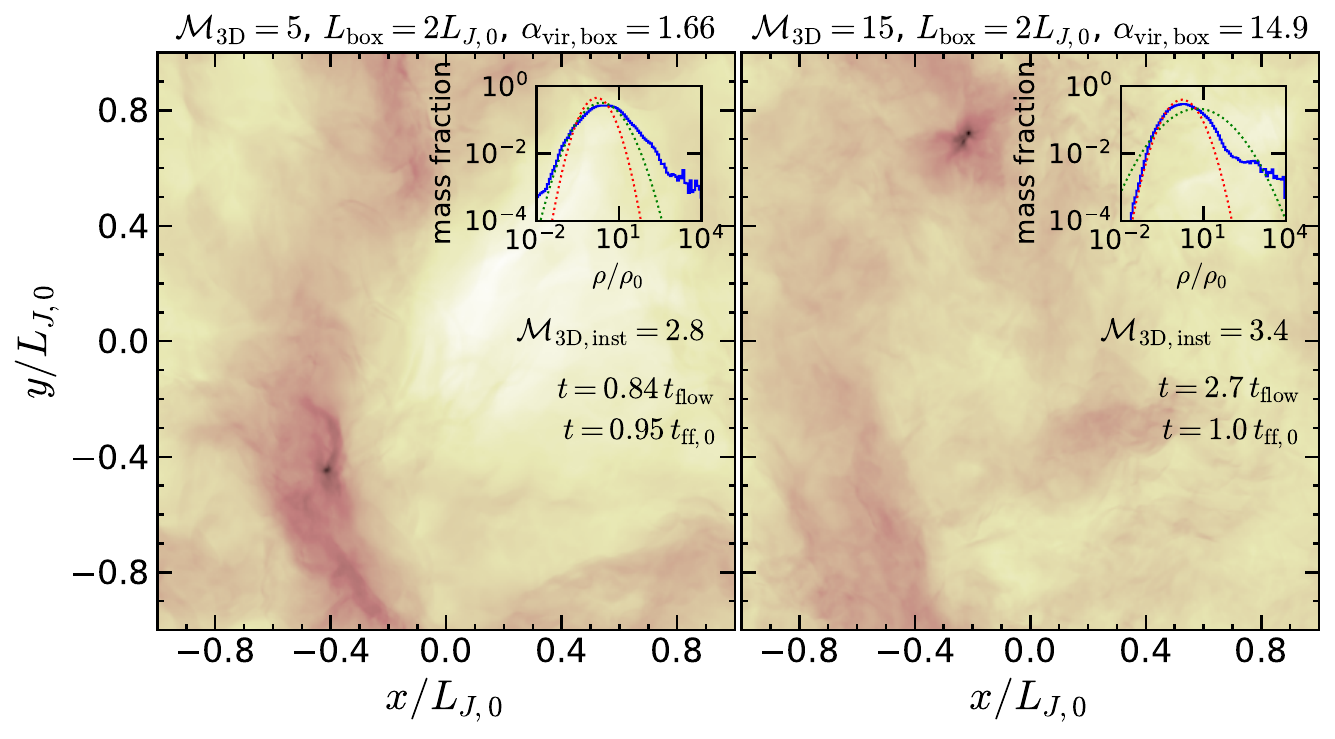}
  \caption{Density projections along the $z$-axis for models \texttt{M5} (\emph{left}) and the high-turbulence, low-gravity model \texttt{M15L2} (\emph{right}) at the respective times when the first sink particle forms (simulation time in units of the flow crossing time and free-fall time are given in each panel).
  Insets in each panel plot the density \gls{PDF} of each model as a solid line, with the log-normal distribution corresponding to the initial $\mathcal{M}_\mathrm{3D}$ and instantaneously measured Mach number $\mathcal{M}_\mathrm{3D,inst}$ overplotted as green and red dotted lines, respectively.}
  \label{fig:compare_at_tstar}
\end{figure}

The model \texttt{M15L2} has the box size $L_\mathrm{box} = 2L_{J,0}$ identical to model \texttt{M5}, but has higher initial Mach number $\mathcal{M}_\mathrm{3D} = 15$ such that $\alpha_\mathrm{vir,box} = 14.9$.
That is, this model initially has strong turbulence and weak gravity, with $t_\mathrm{flow,0}/t_\mathrm{ff,0}=0.38$.
\cref{fig:compare_at_tstar} illustrates the density distribution projected along the $z$-axis for model \texttt{M5} and \texttt{M15L2} (run with identical realization of the power spectrum of fluctuations), using the snapshot right before the formation of the first sink particle in each model.
In model \texttt{M5}, the first collapse occurs in the overdense region near the lower left quadrant at $t = 0.85 t_\mathrm{flow,0}$, which corresponds to $0.96 t_\mathrm{ff,0}$.
Although similar structure appears at $t = 0.85 t_\mathrm{flow,0}$ in model \texttt{M15L2} as well (because $n_\mathrm{seed}$ is the same), no local collapse occurs because strong turbulence makes $r_\mathrm{crit}$ quite large, such that it is not possible to satisfy 
the critical conditions for collapse (\cref{eq:critical_condition}). 
The first collapse eventually occurs in model \texttt{M15L2} at $t = 2.7 t_\mathrm{flow,0}$, corresponding to $1.0 t_\mathrm{ff,0}$, when the instantaneous Mach number has decreased to $\mathcal{M}_\mathrm{3D,inst} = 3.4$.
The time of first collapse relative to the free-fall time in each simulation is similar, and the turbulence has evolved to a similar level.
Thus, we see that if turbulence is initially strong compared to gravity (as is true in reality in the diffuse ISM), and then is allowed to decay without being replenished (as is true in certain locations within the ISM), the evolution ends up being similar to our standard models with initial $\alpha_\mathrm{vir,box}=1.66$.

Our second comparison model \texttt{M3L4} has box size $L_\mathrm{box} = 4L_{J,0}$ identical to model \texttt{M10}, but has lower initial Mach number $\mathcal{M}_\mathrm{3D} = 3$, such that initially $\alpha_\mathrm{vir,box} = 0.149$.
This model initially has weak turbulence and strong gravity, with $t_\mathrm{flow,0}/t_\mathrm{ff,0}=3.7$.
The initial conditions in model \texttt{M3L4}, with virial parameter well below unity, are unlike those found in the real ISM, and the evolution is qualitatively quite different from our standard models.

\begin{figure}[htpb]
  \plotone{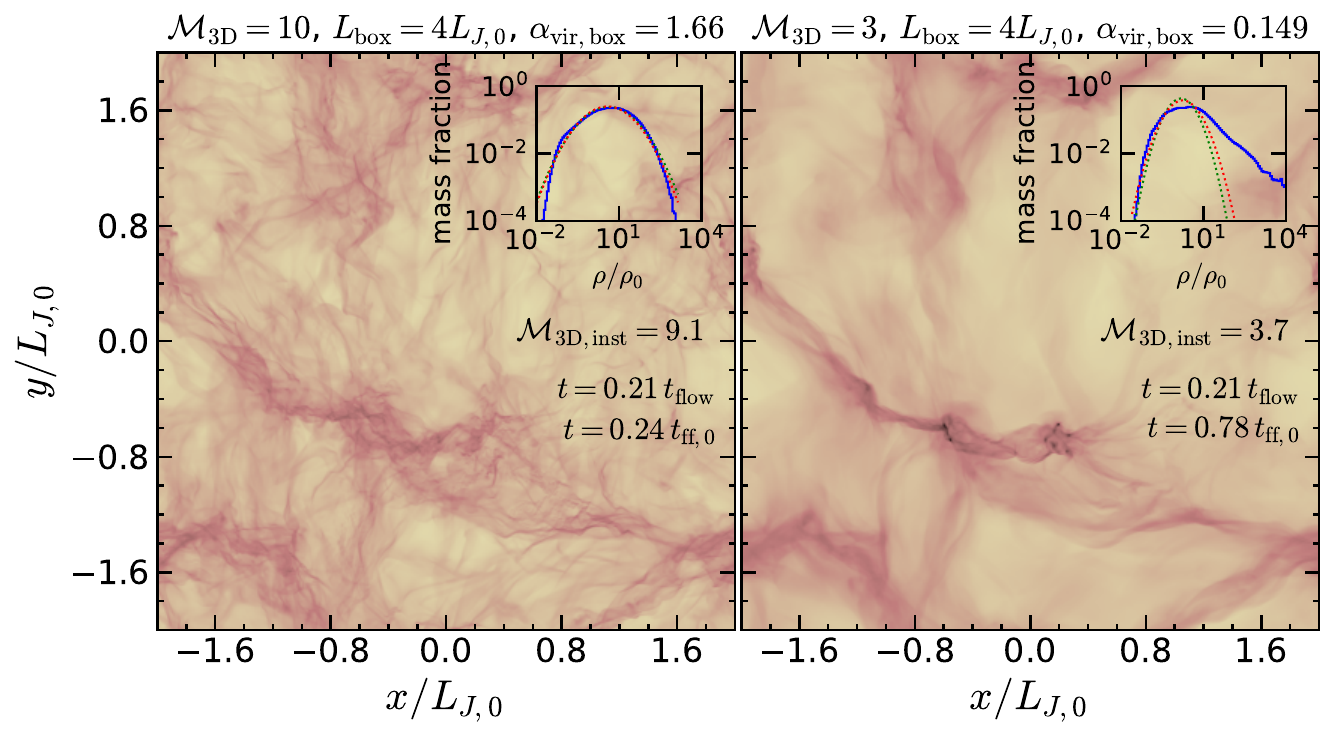}
  \caption{Density projections along the $z$-axis for models \texttt{M10} (\emph{left}) and the low-turbulence, high-gravity model \texttt{M3L4} (\emph{right}) at $t = 0.21 t_\mathrm{flow,0}$ (time in units of the flow crossing time and free-fall time is given in each panel).
  Insets show the density \gls{PDF} of each model as a solid line, with the log-normal distribution corresponding to the initial and instantaneously measured Mach numbers overplotted as green and red dotted lines.}
  \label{fig:compare_at_tflow}
\end{figure}

\cref{fig:compare_at_tflow} compares the projected density distribution between model \texttt{M10} and \texttt{M3L4} (again run with the identical perturbation power spectrum), both taken at $t = 0.21 t_\mathrm{flow,0}$.
At this time, no collapse has yet occurred in model \texttt{M10}, which has evolved only for $t=0.24 t_\mathrm{0,ff}$.
The stronger-gravity model \texttt{M3L4}, which has evolved for $t=0.78t_\mathrm{0,ff}$, has a number of locally collapsed regions.
Its density structure is, however, overall smoother due to lower turbulence, with prominent large-scale filaments.
The instantaneous Mach number has \emph{increased} to $\mathcal{M}_\mathrm{3D,inst} = 3.7$ due to conversion of gravitational energy to kinetic energy.
For this model, the density \gls{PDF} in the high-density regime follows a power-law, a characteristic of gravitational collapse.
This is because in model \texttt{M3L4}, $t_\mathrm{flow,0}$ is much larger than $t_\mathrm{ff,0}$ such that the emerging density structures are heavily influenced by gravity from the outset.

As noted above, the type of evolution shown for model \texttt{M3L4} is not expected for real GMCs, because they have virial parameters $\alpha_\mathrm{vir} > 1$ and abound with substructure at their birth.
In particular, if we consider the mass per unit length of a filament that has density comparable to the post-shock value for an isothermal shock, $\rho = {\cal M}_\mathrm{3D}^2\rho_0$, and cross-section equal to $r_{s,\mathrm{cloud}}^2$ for the sonic length given in \cref{eq:rs0} \REV{within a spherical cloud having virial parameter $\alpha_\mathrm{vir}= 5 \sigma_\mathrm{1D}^2 R_\mathrm{cloud}/(G M_\mathrm{cloud})$, we obtain 
\begin{equation}
  \mu_\mathrm{fil} = 2.01 \frac{c_s^2}{\alpha_\mathrm{vir}G}. 
\end{equation}
Thus, realistic GMCs with $\alpha_\mathrm{vir} \gtrsim 2 $} have sufficient turbulence that the mass per unit length of their internal quiescent substructures is expected to be below the critical value $2 c_s^2/G$, which is the maximum permitting an equilibrium \citep{ostriker64}.
In this case, overdense cores can form and collapse without the whole filament collapsing.

Only if an initially quiescent, thermally supported region suddenly cools down to very low temperature would the evolution resemble that of the low  $\alpha_\mathrm{vir}$ model \texttt{M3L4}.
This kind of evolution is perhaps more reminiscent of cosmological structure formation than structure formation in a turbulent GMC.

\end{document}